\documentclass[11pt,a4paper]{article}
\DeclareMathAlphabet{\scr}{U}{rsfs}{m}{n}

\pdfoutput=1
\usepackage{latexsym}
\usepackage{epsfig}
\usepackage[mathscr]{eucal}
\usepackage{amsfonts}
\usepackage{amscd}
\usepackage{cite}
\usepackage{array}   
\usepackage{amssymb}
\usepackage{colordvi}
\usepackage[centertags]{amsmath}
\usepackage{enumerate}
\usepackage{graphicx}
\usepackage{booktabs}
\usepackage{hyperref}
\usepackage{enumitem}
\setlist[description]{leftmargin=2\parindent,labelindent=\parindent}
\usepackage{theorem}
\usepackage{listings}
\usepackage{multirow}
\usepackage[framemethod=TikZ]{mdframed}
\usepackage{makecell}
\usepackage[footnotesize]{caption}
\usepackage{bbm}
\usepackage[labelfont=bf,textfont={sl,bf}]{subfig}
 \usepackage{slashed}
\usepackage{xcolor}
\xdefinecolor{mygray}{gray}{0.85}
\usepackage{ulem}
\usepackage{pifont}
\newcommand{\cmark}{\ding{51}}%
\newcommand{\xmark}{\ding{55}}%
\usepackage{mathtools}

\mdfsetup{%
   middlelinecolor=black,
   middlelinewidth=0.5pt,
   roundcorner=0pt,
   innertopmargin=-3pt,
   innerbottommargin=10pt
}

\newcommand{\newc}{\newcommand}
\newc{\be}{\begin{equation}}
\newc{\ee}{\end{equation}}
\newc{\bea}{\begin{eqnarray}}
\newc{\eea}{\end{eqnarray}}
\newc{\ol}{\overline}
\newc{\bs}{\boldsymbol}
\newc{\m}{\mathcal}
\newc{\lan}{\langle}
\newc{\ra}{\rangle}
\newc{\pa}{\partial}

\newcommand{\beq}{\begin{eqnarray}} 
\newcommand{\eeq}{\end{eqnarray}} 
\newcommand{\bpmatrix}{\begin{pmatrix}}
\newcommand{\epmatrix}{\end{pmatrix}}
\newcommand{\ba}{\begin{array}}
\newcommand{\ea}{\end{array}}

\newcommand{\figref}[1]{Fig.~\ref{#1}}
\renewcommand{\eqref}[1]{Eq.~(\ref{#1})}

\newcommand{\bc}{\begin{center}}
\newcommand{\ec}{\end{center}}

\renewcommand{\ol}{\text{1l}}

\newcommand{\s}{\newline \vspace*{-3.5mm}}

\begin{document}
\title{
\vspace*{-3cm}
\phantom{h} \hfill\mbox{\small KA-TP-28-2018}
\\[-1.1cm]
\phantom{h} \hfill\mbox{\small PSI-PR-18-10} 
\\[2cm]
\textbf{{\texttt{2HDECAY}} - A program for the Calculation of
  Electroweak One-Loop Corrections to Higgs Decays in the Two-Higgs-Doublet
  Model Including State-of-the-Art QCD Corrections}}

\date{}
\author{Marcel Krause$^{1\,}$\footnote{E-mail:
  \texttt{marcel.krause@kit.edu}}, Margarete M\"{u}hlleitner$^{1\,}$\footnote{E-mail:
  \texttt{margarete.muehlleitner@kit.edu}}, Michael Spira$^{2\,}$\footnote{E-mail:
  \texttt{michael.spira@psi.ch}}
\\[9mm]
{\small\it
$^1$Institute for Theoretical Physics, Karlsruhe Institute of Technology,} \\
{\small\it Wolfgang-Gaede-Str. 1, 76131 Karlsruhe, Germany.}\\[3mm]
{\small\it
$^2$Paul Scherrer Institute,} \\
{\small\it CH-5232 Villigen PSI, Switzerland.}\\[3mm]
}
\maketitle

\begin{abstract}
\noindent We present the program package {\texttt{2HDECAY}} for the
calculation of the partial decay widths and branching ratios of the Higgs
bosons of a general CP-conserving 2-Higgs doublet model
(2HDM). The tool includes the full electroweak one-loop corrections to
all two-body on-shell Higgs decays in the 2HDM that are not
loop-induced. It combines them with the state-of-the-art QCD
corrections that are already implemented in the
program {\texttt{HDECAY}}. For the 
renormalization of the electroweak sector an on-shell scheme is
implemented for most of the renormalization parameters. Exceptions are the
soft-$\mathbb{Z}_2$-breaking squared mass scale
$m_{12}^2$, where an $\overline{\text{MS}}$ condition is applied, as well as the 2HDM
mixing angles $\alpha$ and $\beta$, for which several different
renormalization schemes are implemented. The tool {\texttt{2HDECAY}}
can be used for phenomenological analyses of the branching ratios of
Higgs decays in the 2HDM. Furthermore, the separate output of the
electroweak contributions to the tree-level partial
decay widths for several different renormalization schemes, computed consistently with an automatic parameter conversion between the different schemes, allows for
an efficient analysis of the impact of the electroweak
corrections and the remaining theoretical error due to missing
higher-order corrections. The latest version of the program package
{\texttt{2HDECAY}} can be downloaded from the URL
\href{https://github.com/marcel-krause/2HDECAY}{https://github.com/marcel-krause/2HDECAY}.  
\end{abstract}
\thispagestyle{empty}
\vfill
\newpage

\section{Introduction}
\label{sec:Introduction}
The discovery of the Higgs particle, announced on 
4  July 2012 by the LHC experiments ATLAS \cite{Aad:2012tfa} and CMS
\cite{Chatrchyan:2012xdj} marked a milestone for 
particle physics. It structurally completed the Standard Model (SM)
providing us with a theory that remains weakly interacting all the
way up to the Planck scale. While the SM can successfully describe
numerous particle physics phenomena at the quantum level at highest
precision, it leaves open several questions. Among these are {\it
  e.g.}~the one for the nature of Dark Matter (DM), the baryon
asymmetry of the universe or the hierarchy problem. This calls for
physics beyond the SM (BSM). Models beyond the SM usually entail
enlarged Higgs sectors that can provide candidates for Dark Matter or
guarantee successful baryogenesis. Since the discovered Higgs boson
with a mass of 125.09~GeV \cite{Aad:2015zhl} behaves SM-like any BSM extension
has to make sure to contain a Higgs boson in its spectrum that is in
accordance with the LHC Higgs data. Moreover, the models have to be
tested against theoretical and further experimental
constraints from electroweak precision tests, $B$-physics, low-energy
observables and the negative searches for new particles that may be
predicted by some of the BSM theories. 

The lack of any direct sign of
new physics renders the investigation of the Higgs sector more and more
important. The precise investigation of the discovered Higgs boson may reveal
indirect signs of new physics through mixing with other Higgs bosons
in the spectrum, loop effects due to the additional Higgs bosons
and/or further new states predicted by the model, or decays into non-SM
states or Higgs bosons, including the possibility of invisible
decays. Due to the SM-like nature of the 125~GeV Higgs boson indirect
new physics effects on its properties are expected to be small. Moreover, different BSM
theories can lead to similar effects in the Higgs sector. In order not
to miss any indirect sign of new physics and to be able to identify the
underlying theory in case of discovery, highest precision in the prediction of the
observables and sophisticated experimental techniques are therefore
indispensable. The former calls for the inclusion of higher-order 
corrections at highest possible level, and theorists all over the
world have spent enormous efforts to improve the predictions
for Higgs observables \cite{deFlorian:2016spz}. 

Among the new physics models supersymmetric (SUSY) extensions
\cite{Golfand:1971iw,Volkov:1973ix,Wess:1974tw,Fayet:1974pd,Fayet:1976cr,Fayet:1976et,Fayet:1977yc,Nilles:1983ge,Haber:1984rc,Sohnius:1985qm,Gunion:1984yn,Gunion:1986nh,Gunion:1989we} 
certainly belong to the best motivated and most thoroughly
investigated models beyond the SM, and numerous higher-order 
predictions exist for the production and decay cross sections as well
as the Higgs potential parameters, {\it i.e.}~the masses and Higgs
self-couplings \cite{deFlorian:2016spz}. The Higgs sector of the
minimal supersymmetric extension (MSSM) \cite{Gunion:1989we,Martin:1997ns,Dawson:1997tz,Djouadi:2005gj} is a 2-Higgs doublet model
(2HDM) of type II \cite{Lee:1973iz,Branco:2011iw}. While due to
supersymmetry the MSSM Higgs potential parameters are given in terms
of gauge couplings this is not the 
case for general 2HDMs so that the 2HDM entails an interesting and
more diverse Higgs phenomenology and is also
affected differently by 
higher-order electroweak (EW) corrections. Moreover, 2HDMs allow for
successful baryogenesis
\cite{McLerran:1990zh,Turok:1990zg,Cohen:1991iu,Turok:1991uc,Funakubo:1993jg,Davies:1994id,Cline:1995dg,Cline:1996mga,Fromme:2006cm,Cline:2011mm,Dorsch:2013wja,Fuyuto:2015vna,Dorsch:2014qja,Dorsch:2016tab,Dorsch:2016nrg,Basler:2016obg,Dorsch:2017nza,Basler:2017uxn,Basler:2018cwe} and in their inert version provide a Dark
Matter candidate
\cite{Deshpande:1977rw,Barbieri:2006dq,LopezHonorez:2006gr,Dolle:2009fn,Honorez:2010re,Gustafsson:2012aj,Swiezewska:2012ej,Swiezewska:2012eh,Arhrib:2013ela,Klasen:2013btp,Abe:2014gua,Krawczyk:2013jta,Goudelis:2013uca,Chakrabarty:2015yia,Ilnicka:2015jba}.
 
The situation with respect to EW corrections in non-SUSY models is not
as advanced as for SUSY extensions. 
While the QCD corrections can be taken over to those
models with a minimum effort from the SM and the MSSM 
by applying appropriate changes, this is not the case
for the EW corrections. Moreover, some issues arise with
respect to renormalization. Thus, only recently a renormalization
procedure has been proposed by authors of this paper for the
mixing angles of the 2HDM that ensures
explicitly gauge-independent decay amplitudes,
\cite{Krause:2016gkg,Krause:2016oke}. Subsequent groups have confirmed
this in different Higgs channels
\cite{Denner:2016etu,Altenkamp:2017ldc,Altenkamp:2017kxk,Denner2018,Fox:2017hbw}. Moreover, in Ref.~\cite{Denner2018} four schemes have been proposed based on on-shell and symmetry-inspired renormalization conditions for the mixing angles (and by applying the background field method \cite{Zuber:1975sa,Zuber:1975gh,Boulware:1981,Abbott:1981,Abbott:1982,Hart:1983,Denner:1994xt}) and on $\overline{\mbox{MS}}$ prescriptions for the remaining new quartic Higgs couplings, and their features have been
  investigated in detail. The authors of Ref.~\cite{Kanemura:2017wtm} use an improved on-shell
  scheme that is essentially equivalent to the mixing angle
  renormalization scheme presented by our group in
  \cite{Krause:2016gkg,Krause:2016oke}. It has been applied to compute
  the renormalized one-loop Higgs boson couplings in the Higgs singlet
  model and the 2HDM and to implement these in the program
  {\texttt{H-COUP}} \cite{Kanemura:2017gbi}. In
  \cite{Kanemura:2018yai} the authors present, for these models, the 
  one-loop electroweak and QCD corrections to the Higgs decays into
  fermion and gauge boson pairs. The complete
  phenomenological analysis, however, requires  the corrections to all
  Higgs decays, as we present them here for the first time in the
  computer tool {\texttt{2HDECAY}}.

In \cite{Krause:2016xku} we completed the renormalization of the 2HDM and calculated
the higher-order corrections to Higgs-to-Higgs decays. We have applied
and extended this renormalization procedure in \cite{Krause:2017mal} to the
next-to-2HDM (N2HDM) which includes an additional real singlet. The
computation of the (N)2HDM EW corrections has shown that the corrections
can become very large for certain areas of the parameters space. There can be 
several reasons for this. The corrections can be parametrically enhanced
due to involved couplings that can be large
\cite{Kanemura:2002vm,Kanemura:2004mg,Krause:2016xku,Krause:2017mal}. This
is in particular the case for the  trilinear Higgs self-couplings that in
contrast to SUSY are not given in terms of the 
gauge couplings of the theory and that are so far only weakly
constrained by the LHC Higgs data. The corrections can be large
due to light particles in the loop in combination with not too small
couplings, {\it e.g.}~light Higgs states of the extended Higgs
sector. Also an inapt choice of the renormalization
scheme can artificially enhance loop corrections. Thus we found for our investigated
processes that process-dependent renormalization schemes or
$\overline{\mbox{MS}}$ renormalization of the scalar
  mixing angles can blow up the one-loop
corrections due to an insufficient cancellation of the large finite
contributions from wave function renormalization
constants \cite{Krause:2016oke,Krause:2016xku}. Moreover, 
counterterms can blow up in certain parameter 
regions because of small leading-order couplings, {\it e.g.}~in the 2HDM the coupling of
the heavy non-SM-like CP-even Higgs boson to gauge bosons, which in
the limit of a light SM-like CP-even Higgs boson is almost zero. The
same effects are observed for supersymmetric theories where a
badly chosen parameter set for the renormalization
can lead to very large counterterms and hence enhanced loop
corrections, {\it cf.}~Ref.~\cite{Belanger:2017rgu} for a recent analysis. 

This discussion shows that the renormalization of the EW corrections to BSM Higgs
observables is a highly non-trivial task. In addition, there may be no unique
best renormalization scheme for the whole parameter space of a
specific model, and the user has to decide which scheme to choose to
obtain trustworthy predictions. With the publication of the new tool {\tt
2HDECAY} we aim to give an answer to this problematic task. 

The program {\tt 2HDECAY} computes, for 17 different renormalization
schemes, the EW corrections to the Higgs decays of the 2HDM Higgs bosons
into all possible on-shell two-particle final states of the model that
are not loop-induced. It is 
combined with the widely used Fortran code {\tt HDECAY} version 6.52
\cite{DJOUADI199856,Djouadi:2018xqq} which provides the loop-corrected
decay widths and branching ratios for the 
SM, the MSSM and 2HDM incorporating the state-of-the-art higher-order QCD
corrections including also loop-induced and off-shell decays. Through the combination
of these corrections with the 
2HDM EW corrections {\tt 2HDECAY} becomes a code for the prediction of
the 2HDM Higgs boson decay widths at the presently highest possible level of
precision. Additionally, the separate output of the leading order (LO)
and next-to-leading order (NLO) EW corrected decay widths allows to
perform studies on the importance of the relative EW corrections (as function
of the parameter choices), comparisons with the relative EW corrections
within the MSSM or investigations on the most suitable renormalization
scheme for specific parameter regions. The comparison
  of the results for different renormalization schemes moreover
  permits to estimate the remaining theoretical error due to missing
  higher-order corrections. To that end, {\texttt{2HDECAY}} includes a parameter conversion routine which performs the automatic conversion of input parameters from one renormalization scheme to another for all 17 renormalization schemes that are implemented. With this tool we contribute to
the effort of improving the theory predictions for BSM Higgs physics
observables so that in combination with sophisticated experimental
techniques Higgs precision physics becomes possible and the gained 
insights may advance us in our understanding of the mechanism of
electroweak symmetry breaking (EWSB) and the true underlying theory.

The program package was developed and tested under 
{\texttt{Windows 10}}, {\texttt{openSUSE Leap 15.0}} and
{\texttt{macOS Sierra 10.12}}. It requires an up-to-date version of
{\texttt{Python 2}} or {\texttt{Python 3}} (tested with versions
{\texttt{2.7.14}} and {\texttt{3.5.0}}), the {\texttt{FORTRAN}}
compiler {\texttt{gfortran}} and the {\texttt{GNU C}} compilers
{\texttt{gcc}} (tested for compatibility with versions
{\texttt{6.4.0}} and {\texttt{7.3.1}}) and {\texttt{g++}}. The latest
version of the package can be downloaded from 
\begin{center}
	\href{https://github.com/marcel-krause/2HDECAY}{https://github.com/marcel-krause/2HDECAY} ~.
\end{center}

The paper is organized as follows. The subsequent
Sec.\,\ref{sec:EWQCD2HDMMain} forms the theoretical background for our
work. We briefly introduce the 2HDM, all
relevant parameters and particles and set our notation. 
We give a summary of all counterterms that are needed for the computation
of the EW corrections and state them explicitly. The relevant formulae
for the computation of the partial decay widths at one-loop level are
presented and the combination of the electroweak corrections with the
QCD corrections already implemented in {\texttt{HDECAY}} is described. In
Sec.\,\ref{sec:programDescriptionMain}, we introduce
{\texttt{2HDECAY}} in detail, describe the structure of the program
package and the input and output file formats. Additionally, we
provide installation and usage manuals. We conclude with a short
summary of our work in Sec.\,\ref{sec:summary}. As reference for the
user, we list exemplary input and output files in Appendices
\ref{sec:AppendixInputFile} and \ref{sec:AppendixOutputFile},
respectively. 

\section{One-Loop Electroweak and QCD Corrections in the 2HDM}
\label{sec:EWQCD2HDMMain}
In the following, we briefly set up our notation and introduce the
2HDM along with the input parameters used in our
parametrization. We give details on the EW one-loop
renormalization of the 2HDM. We discuss how the calculation of the
one-loop partial decay widths is performed. At the end of the section,
we explain how the EW corrections are combined with the existing
state-of-the-art QCD corrections already implemented in {\texttt{HDECAY}} and present the automatic parameter conversion routine that is implemented in {\texttt{2HDECAY}}. 

\subsection{Introduction of the 2HDM}
\label{sec:setupOfModel}
For our work, we consider a general CP-conserving 2HDM
\cite{Lee:1973iz,Branco:2011iw} with a global discrete $\mathbb{Z}_2$
symmetry that is softly broken. The model consists of two complex
$SU(2)_L$ doublets $\Phi _1$ and $\Phi _2$, both with hypercharge
$Y=+1$. The electroweak part of the 2HDM can be described by the
Lagrangian 
\begin{equation}
	\mathcal{L} ^\text{EW}_\text{2HDM} = \mathcal{L} _\text{YM} + \mathcal{L} _\text{F} + \mathcal{L} _\text{S} + \mathcal{L} _\text{Yuk} + {\mathcal{L}} _\text{GF} + \mathcal{L} _\text{FP} 
\label{eq:electroweakLagrangian}
\end{equation}
in terms of the Yang-Mills Lagrangian $\mathcal{L} _\text{YM}$ and the fermion
Lagrangian $\mathcal{L} _\text{F}$ containing the kinetic terms of the
gauge bosons and fermions and their interactions, the Higgs Lagrangian
$\mathcal{L} _\text{S}$, the Yukawa
Lagrangian $\mathcal{L}_{\text{yuk}}$ with the Higgs-fermion
interactions, the gauge-fixing and the Fadeev-Popov Lagrangian,
${\mathcal{L}} _\text{GF}$ and  $\mathcal{L}_\text{FP}$,
respectively. Explicit forms of  $\mathcal{L} _\text{YM}$ and
$\mathcal{L} _\text{F}$ can be found {\it
  e.g.}~in~\cite{Peskin:1995ev, Denner:1991kt} and of the general 2HDM
Yukawa Lagrangian {\it e.g.}~in \cite{Aoki:2009ha, Branco:2011iw}. We
do not give them explicitly here.
For the renormalization of the 2HDM, we follow the approach of
Ref.~\cite{Santos:1996vt} and apply the gauge-fixing only \textit{after}
the renormalization of the theory, {\it i.e.}~$\mathcal{L} _\text{GF}$
contains only fields that are already renormalized. For the purpose of
our work we do not present ${\mathcal{L}} _\text{GF}$ nor $\mathcal{L}_\text{FP}$ since 
their explicit forms are not needed in the following. 

The scalar Lagrangian $\mathcal{L} _\text{S}$ introduces the kinetic
terms of the Higgs doublets and their scalar potential. With the 
the covariant derivative  
\begin{equation}
D_\mu = \partial _\mu + \frac{i}{2} g \sum _{a=1}^3 \sigma ^a W_\mu ^a + \frac{i}{2} g{'} B_\mu 
\end{equation}
where $W_\mu ^a$ and $B_\mu$ are the gauge bosons of the $SU(2)_L$ and
$U(1)_Y$ respectively, $g$ and $g{'}$ are the corresponding coupling
constants of the gauge groups and $\sigma ^a$ are the Pauli matrices,
the scalar Lagrangian is given by
\begin{equation}
\mathcal{L} _S = \sum _{i=1}^2 (D_\mu \Phi _i)^\dagger (D^\mu \Phi _i) - V_\text{2HDM} ~.
\label{eq:scalarLagrangian}
\end{equation}
The scalar potential of the CP-conserving 2HDM reads
\cite{Branco:2011iw} 
\begin{equation}
\begin{split}
V_\text{2HDM} =&~ m_{11}^2 \left| \Phi _1 \right| ^2 + m_{22}^2 \left|
  \Phi _2 \right| ^2 - m_{12}^2 \left( \Phi _1 ^\dagger \Phi _2 +
  \textit{h.c.} \right) + \frac{\lambda _1}{2} \left( \Phi _1^\dagger
  \Phi _1 \right) ^2 + \frac{\lambda _2}{2} \left( \Phi _2^\dagger
  \Phi _2 \right) ^2 \\ 
&+ \lambda _3 \left( \Phi _1^\dagger \Phi _1 \right) \left( \Phi
  _2^\dagger \Phi _2 \right) + \lambda _4 \left( \Phi _1^\dagger \Phi
  _2 \right) \left( \Phi _2^\dagger \Phi _1 \right) + \frac{\lambda
  _5}{2} \left[ \left( \Phi _1^\dagger \Phi _2 \right) ^2 +
  \textit{h.c.} \right] ~. 
\end{split}
\label{eq:scalarPotential}
\end{equation}
Since we consider a CP-conserving model, the 2HDM potential can be
parametrized by three real-valued mass parameters $m_{11}$, $m_{22}$
and $m_{12}$ as well as five real-valued dimensionless coupling
constants $\lambda _i$ ($i=1,...,5$). For later convenience, we define
the frequently appearing combination of three of these coupling
constants as
\begin{equation}
\lambda _{345} \equiv \lambda _3 + \lambda _4 + \lambda _5 ~.
\end{equation}
For $m_{12}^2=0$, the potential $V_\text{2HDM}$ exhibits a discrete $\mathbb{Z}_2$
symmetry under the simultaneous field transformations $\Phi _1
\rightarrow - \Phi _1$ and $\Phi _2 \rightarrow \Phi
_2$. This
symmetry, implemented in the scalar potential in order to avoid
flavour-changing neutral currents (FCNC) at tree level, is softly
broken by a non-zero mass parameter $m_{12}$. 

After EWSB the neutral components of
the Higgs doublets develop vacuum expectation values (VEVs) which are
real in the CP-conserving case. After expanding about the real VEVs $v_1$
and $v_2$, the Higgs doublets $\Phi_i$ ($i=1,2$) can be expressed in
terms of the charged complex field $\omega_i^+$ and the real neutral
CP-even and CP-odd fields $\rho_i$ and $\eta_i$, respectively as
\begin{equation}
\Phi _1 = \begin{pmatrix} \omega ^+ _1 \\ \frac{v_1 + \rho _1 + i \eta
    _1 }{\sqrt{2}} \end{pmatrix} ~~~\text{and}~~~~ \Phi _2 = \begin{pmatrix}
  \omega ^+ _2 \\ \frac{v_2 + \rho _2 + i \eta
    _2}{\sqrt{2}} \end{pmatrix} 
\label{eq:vevexpansion}
\end{equation} 
where 
\begin{equation}
v^2 = v_1^2 + v_2^2 \approx (246.22~\text{GeV})^2
\label{eq:vevRelations} 
\end{equation}
is the SM VEV obtained from the Fermi constant
  $G_F$ and we define the ratio of the VEVs through the mixing angle $\beta$
as
\beq
\tan \beta = \frac{v_2}{v_1} 
\eeq
so that
\beq 
v_1 = v c_\beta \quad \mbox{and} \quad v_2 = v s_\beta~.
\eeq
Insertion of Eq.~(\ref{eq:vevexpansion}) in the kinetic part of the scalar
Lagrangian in Eq.~(\ref{eq:scalarLagrangian}) yields after rotation to
the mass eigenstates the tree-level relations for the masses of the
electroweak gauge bosons 
\begin{align}
	m_W^2 &= \frac{g^2v^2}{4}  \\
	m_Z^2 &= \frac{(g^2 + g{'}^2)v^2}{4}  \\
	m_\gamma ^2 &= 0 ~.
\end{align} 
The electromagnetic coupling constant $e$ is connected to the
fine-structure constant $\alpha _\text{em}$ and to the gauge boson
coupling constants through the tree-level relation
\begin{equation}
	e = \sqrt{4\pi \alpha _\text{em} } = \frac{g g{'}}{\sqrt{g^2 + g{'} ^2}} 
\label{eq:electromagneticCouplingDefinition}
\end{equation}
which allows to replace $g{'}$ in favor of $e$ or
$\alpha_\text{em}$. In our work, we use the fine-structure constant $\alpha
_\text{em}$ as an independent input. Alternatively, one could use the
tree-level relation to the Fermi constant 
\begin{equation}
G_F \equiv \frac{\sqrt{2} g^2}{8m_W^2} = \frac{\alpha _\text{em} \pi
}{\sqrt{2} m_W^2 \left( 1 - \frac{m_W^2}{m_Z^2} \right)} 
\label{eq:definitionFermiConstant}
\end{equation}
to replace one of the parameters of the electroweak sector in favor of
$G_F$. Since $G_F$ is used as an input value for {\texttt{HDECAY}}, we
present the formula here explicitly and explain the conversion between
the different parametrizations in Sec.\,\ref{sec:connectionHDECAY}. 

Inserting Eq.~(\ref{eq:vevexpansion}) in the scalar potential in
\eqref{eq:scalarPotential} leads to
\begin{equation}
\begin{split}
V_\text{2HDM} =&~ \frac{1}{2} \left( \rho _1 ~~ \rho _2 \right) M_\rho ^2 \begin{pmatrix} \rho _1 \\ \rho _2 \end{pmatrix} + \frac{1}{2} \left( \eta _1 ~~ \eta _2 \right) M_\eta ^2 \begin{pmatrix} \eta _1 \\ \eta _2 \end{pmatrix} + \frac{1}{2} \left( \omega ^\pm _1 ~~ \omega ^\pm _2 \right) M_\omega ^2 \begin{pmatrix} \omega ^\pm _1 \\ \omega ^\pm _2 \end{pmatrix} \\
& + T_1 \rho _1 + T_2 \rho _2 + ~~ \cdots
\end{split}
\label{eq:scalarPotentialMultilinearFields}
\end{equation}
where the terms $T_1$ and $T_2$ and the matrices $M_\omega ^2$,
$M_\rho ^2$ and $M_\eta ^2$ are defined below. By requiring the
VEVs of \eqref{eq:vevexpansion} to
represent the minimum of the potential, the minimum conditions for the
potential can be expressed as 
\begin{equation}
\frac{\partial V_\text{2HDM}}{\partial \Phi _i} \Bigg| _{\left\langle \Phi _j \right\rangle} = 0~.
\end{equation}
This is equivalent to the statement that the two terms linear in the
CP-even fields $\rho _1$ and $\rho _2$, the tadpole terms, 
\begin{align}
\frac{T_1}{v_1} &\equiv m_{11}^2 - m_{12}^2 \frac{v_2}{v_1} + \frac{v_1^2 \lambda _1}{2} + \frac{v_2 ^2 \lambda _{345}}{2} \label{eq:tadpoleCondition1}  \\
\frac{T_2}{v_2} &\equiv m_{22}^2 - m_{12}^2 \frac{v_1}{v_2} + \frac{v_2^2 \lambda _2}{2} + \frac{v_1 ^2 \lambda _{345}}{2} \label{eq:tadpoleCondition2}
\end{align}
have to vanish at tree level:
\begin{equation}
	T_1 = T_2 = 0 ~~~ (\text{at tree level}) ~.
\label{eq:tadpoleVanishAtTreelevel}
\end{equation}
The tadpole equations can be solved for $m_{11}^2$ and $m_{22}^2$ in
order to replace these two parameters by the tadpole parameters $T_1$
and $T_2$. 

The terms bilinear in the fields given in
\eqref{eq:scalarPotentialMultilinearFields} define the scalar mass
matrices 
\begin{align}
M_\rho ^2 &\equiv \begin{pmatrix}
m_{12}^2 \frac{v_2}{v_1} + \lambda _1 v_1^2 & -m_{12}^2 + \lambda _{345} v_1 v_2 \\ -m_{12}^2 + \lambda _{345} v_1 v_2 & m_{12}^2 \frac{v_1}{v_2} + \lambda _2 v_2^2
\end{pmatrix} + \begin{pmatrix}
\frac{T_1}{v_1} & 0 \\ 0 & \frac{T_2}{v_2} 
\end{pmatrix}  \label{eq:massMatrices1} \\
M_\eta ^2 &\equiv \left( \frac{m_{12}^2}{v_1v_2} - \lambda _5 \right) \begin{pmatrix}
v_2^2 & -v_1 v_2 \\ -v_1 v_2 & v_1 ^2 
\end{pmatrix} + \begin{pmatrix}
\frac{T_1}{v_1} & 0 \\ 0 & \frac{T_2}{v_2} 
\end{pmatrix}  \\
M_\omega ^2 &\equiv \left( \frac{m_{12}^2}{v_1v_2} - \frac{\lambda _4 + \lambda _5}{2} \right) \begin{pmatrix}
v_2^2 & -v_1 v_2 \\ -v_1 v_2 & v_1 ^2 
\end{pmatrix} + \begin{pmatrix}
\frac{T_1}{v_1} & 0 \\ 0 & \frac{T_2}{v_2} 
\end{pmatrix}  \label{eq:massMatrices3}
\end{align}
where Eqs.\,(\ref{eq:tadpoleCondition1}) and
(\ref{eq:tadpoleCondition2}) have already been applied to replace the
parameters $m_{11}^2$ and $m_{22}^2$ in favor of $T_1$ and
$T_2$. Keeping the latter explicitly in the expressions of the
mass matrices is crucial for the correct renormalization of the scalar
sector, as explained in Sec.\,\ref{sec:renormalization2HDM}. By means
of two mixing angles $\alpha$ and $\beta$ which define the rotation
matrices\footnote{Here and in the following, we use the short-hand
  notation $s_x \equiv \sin (x)$, $c_x \equiv \cos (x)$, $t_x \equiv
  \tan (x)$ for convenience.} 
\begin{equation}
	R (x) \equiv \begin{pmatrix} c_x & - s_x \\ s_x & c_x \end{pmatrix} 
\end{equation}
the fields $\omega ^+ _i$, $\rho _i$ and $\eta _i$ are rotated to the mass basis according to 
\begin{align}
	\begin{pmatrix} \rho _1 \\ \rho _2 \end{pmatrix} &= R(\alpha ) \begin{pmatrix} H \\ h \end{pmatrix}  \label{eq:rotationCPEven} \\
	\begin{pmatrix} \eta _1 \\ \eta _2 \end{pmatrix} &= R(\beta ) \begin{pmatrix} G^0 \\ A \end{pmatrix}  \\
	\begin{pmatrix} \omega ^\pm _1 \\ \omega ^\pm _2 \end{pmatrix} &= R(\beta ) \begin{pmatrix} G^\pm \\ H^\pm \label{eq:rotationCharged} \end{pmatrix} 
\end{align}
with the two CP-even Higgs bosons $h$ and $H$, the CP-odd Higgs boson
$A$, the CP-odd Goldstone boson $G^0$ and the charged Higgs bosons
$H^\pm$ as well as the charged Goldstone bosons $G^\pm$. In the mass
basis, the diagonal mass matrices are given by
\begin{align}
	D_\rho ^2 &\equiv \begin{pmatrix} m_H^2 & 0 \\ 0 & m_h^2 \end{pmatrix}  \\
	D_\eta ^2 &\equiv \begin{pmatrix} m_{G^0}^2 & 0 \\ 0 & m_A^2 \end{pmatrix}  \\
	D_\omega ^2 &\equiv \begin{pmatrix} m_{G^\pm}^2 & 0 \\ 0 & m_{H^\pm}^2 \end{pmatrix} 
\end{align}
with the diagonal entries representing the squared masses of the
respective particles. The Goldstone bosons are massless,
\begin{equation}
	m_{G^0}^2 = m_{G^\pm}^2 = 0~.
\end{equation}
%
%
The squared masses expressed in terms of the potential parameters and
the mixing angle $\alpha$ can be cast into the form~\cite{Kanemura:2004mg}  
\begin{align}
m_H^2 &= c_{\alpha - \beta}^2 M_{11}^2 + s_{2(\alpha - \beta )} M_{12}^2 + s_{\alpha - \beta } ^2 M_{22}^2  \label{eq:parameterTransformationInteractionToMass1} \\
m_h^2 &= s_{\alpha - \beta}^2 M_{11}^2 - s_{2(\alpha - \beta )} M_{12}^2 + c_{\alpha - \beta } ^2 M_{22}^2  \\
m_A^2 &= \frac{m_{12}^2}{s_\beta c_\beta} - v^2 \lambda _5  \\
m_{H^\pm } ^2 &= \frac{m_{12}^2}{s_\beta c_\beta} - \frac{v^2}{2}
\left( \lambda _4 + \lambda _5 \right) \\
t_{2(\alpha - \beta ) } &= \frac{2M_{12}^2}{M_{11}^2 - M_{22}^2}  \label{eq:parameterTransformationInteractionToMass5}
\end{align}
where we have introduced
\begin{align}
M_{11}^2 &\equiv v^2 \left( c_\beta ^4 \lambda _1 + s_\beta ^4 \lambda _2 + 2 s_\beta ^2 c_\beta ^2 \lambda _{345} \right)  \\
M_{12}^2 &\equiv s_\beta c_\beta v^2 \left( - c_\beta ^2 \lambda _1 + s_\beta ^2 \lambda _2 + c_{2\beta } \lambda _{345} \right)  \\
M_{22}^2 &\equiv \frac{m_{12}^2}{s_\beta c_\beta } + \frac{v^2}{8} \left( 1 - c_{4\beta } \right) \left( \lambda _1 + \lambda _2 - 2\lambda _{345} \right) ~.
\end{align}
Inverting these relations, the quartic couplings $\lambda _i$
($i=1,...,5$) can be expressed in terms of the mass
parameters $m_h^2$, $m_H^2$, $m_A^2$, $m_{H^\pm}^2$ and the CP-even
mixing angle $\alpha$ as \cite{Kanemura:2004mg} 
\begin{align}
\lambda _1 &= \frac{1}{v^2 c_\beta ^2} \left( c_\alpha ^2 m_H^2 + s_\alpha ^2 m_h^2 - \frac{s_\beta}{c_\beta} m_{12}^2 \right)  \label{eq:parameterTransformationMassToInteraction1} \\
\lambda _2 &= \frac{1}{v^2 s_\beta ^2} \left( s_\alpha ^2 m_H^2 + c_\alpha ^2 m_h^2 - \frac{c_\beta}{s_\beta} m_{12}^2 \right)  \\
\lambda _3 &= \frac{2m_{H^\pm}^2}{v^2} + \frac{s_{2\alpha}}{s_{2\beta} v^2} \left( m_H^2 - m_h^2\right) - \frac{m_{12}^2}{s_\beta c_\beta v^2}  \\
\lambda _4 &= \frac{1}{v^2} \left( m_A^2 - 2m_{H^\pm}^2 + \frac{m_{12}^2}{s_\beta c_\beta} \right)  \\
\lambda _5 &= \frac{1}{v^2} \left( \frac{m_{12}^2}{s_\beta c_\beta} - m_A^2 \right) ~.  \label{eq:parameterTransformationMassToInteraction5} 
\end{align}
In order to avoid tree-level FCNC currents, as introduced by the most
general 2HDM Yukawa Lagrangian, one type of fermions is allowed to
couple only to one Higgs doublet by imposing a global $\mathbb{Z}_2$
symmetry under which $\Phi_{1,2} \to \mp \Phi_{1,2}$. Depending on the
$\mathbb{Z}_2$ charge assignments, there are four phenomenologically
different types of 2HDMs summarized in  Tab.\,\ref{tab:yukawaDefinitions}.
\begin{table}[tb]
\centering
  \begin{tabular}{ c c c c }
    \hline
    & $u$-type & $d$-type & leptons \\ \hline
    I &  $\Phi _2$ & $\Phi _2$ & $\Phi _2$ \\
    II & $\Phi _2$ & $\Phi _1$ & $\Phi _1$ \\
    lepton-specific & $\Phi _2$ & $\Phi _2$ & $\Phi _1$ \\
    flipped & $\Phi _2$ & $\Phi _1$ & $\Phi _2$ \\
    \hline
  \end{tabular}
    \caption{The four Yukawa types of the $\mathbb{Z}_2$-symmetric
      2HDM defined by the Higgs doublet that couples to each kind of fermions.}
   \label{tab:yukawaDefinitions}
\end{table}
For the four 2HDM types considered in this work, all Yukawa couplings can be
parametrized through six different Yukawa coupling parameters $Y_i$
($i=1,...,6$) whose values for the different types are presented
in Tab.\,\ref{tab:yukawaCouplings}. They are introduced here for later
convenience. 
\begin{table}[tb]
\centering
  \begin{tabular}{ c c c c c c c }
    \hline
    2HDM type & $Y_1$ & $Y_2$ & $Y_3$ & $Y_4$ & $Y_5$ & $Y_6$ \\ \hline
    I  & $\frac{c_\alpha }{s_\beta }$ & $\frac{s_\alpha }{s_\beta }$ & $-\frac{1}{t_\beta}$ & $\frac{c_\alpha }{s_\beta }$ & $\frac{s_\alpha }{s_\beta }$ & $-\frac{1}{t_\beta}$ \\
    II & $-\frac{s_\alpha }{c_\beta }$ & $\frac{c_\alpha }{c_\beta }$ & $t_\beta $ & $-\frac{s_\alpha }{c_\beta }$ & $\frac{c_\alpha }{c_\beta }$ & $t_\beta $ \\
    lepton-specific  & $\frac{c_\alpha }{s_\beta }$ & $\frac{s_\alpha }{s_\beta }$ & $-\frac{1}{t_\beta}$ & $-\frac{s_\alpha }{c_\beta }$ & $\frac{c_\alpha }{c_\beta }$ & $t_\beta $ \\
    flipped  & $-\frac{s_\alpha }{c_\beta }$ & $\frac{c_\alpha }{c_\beta }$ & $t_\beta $ & $\frac{c_\alpha }{s_\beta }$ & $\frac{s_\alpha }{s_\beta }$ & $-\frac{1}{t_\beta}$ \\
    \hline
  \end{tabular}
    \caption{Parametrization of the Yukawa coupling parameters in
      terms of six parameters $Y_i$ ($i=1,...,6$)
      for each 2HDM type.}
   \label{tab:yukawaCouplings}
\end{table}

We conclude this section with an overview over the full set of
independent parameters that is used as input for the computations in
{\texttt{2HDECAY}}. Additionally to the parameters defined by
${\mathcal L}_{\text{2HDM}}^{\text{EW}}$, {\texttt{HDECAY}} requires
the electromagnetic coupling constant
  $\alpha_{\text{em}}$ in the Thomson limit for the calculation of
  the loop-induced decays into a photon pair and into $Z\gamma$, the
  strong coupling constant 
$\alpha _s$ for the loop-induced decay into gluons
and the QCD corrections as well as the total decay widths of
the $W$ and $Z$ bosons, $\Gamma _W$ and $\Gamma _Z$, for the
computation of the off-shell decays into
massive gauge boson final states. In the mass basis of the
scalar sector, the set of independent parameters is given by 
\begin{equation}
\{ G_F, \alpha _s , \Gamma _W , \Gamma _Z , \alpha _\text{em} , m_W ,
m_Z , m_f, V_{ij} , t_\beta , m_{12}^2 , \alpha , m_h , m_H , m_A ,
m_{H^\pm} \} ~.
\label{eq:inputSetMassBase} 
\end{equation}
Here $m_f$ denote the fermion masses of the strange, charm, bottom and
top quarks and of the $\mu$ and $\tau$ leptons
($f=s,c,b,t,\mu,\tau$). All other fermion 
masses are assumed to be zero in {\texttt{HDECAY}} and will also be
assumed to be zero in our computation of the EW corrections to the
decay widths. The fermion and gauge boson masses are defined in
accordance with the recommendations of the LHC Higgs cross section
working group \cite{Denner:2047636}. The $V_{ij}$ denote the CKM
mixing matrix elements. All {\texttt{HDECAY}} decay widths are
computed in terms of the Fermi constant $G_F$  
except for processes involving on-shell external photon vertices that are 
expressed by $\alpha_\text{em}$ in the Thomson limit. In the
computation of the EW corrections, however, we require the on-shell
masses $m_W$ and $m_Z$ and the electromagnetic coupling at 
the $Z$ boson mass scale,
$\alpha_{\text{em}}(m_Z^2)$ (not to be confused with the mixing angle
$\alpha$ in the Higgs sector), as input parameters for our
renormalization conditions. We will come back to this
  point later.

Alternatively, the original parametrization of the scalar
potential in the interaction basis can be
used\footnote{{\texttt{HDECAY}} internally translates the parameters
  from the interaction 
  to the mass basis, in terms of which the decay widths are
  implemented.}. In this case, the set of 
independent parameters is given by 
\begin{equation}
\{ G_F, \alpha _s , \Gamma _W , \Gamma _Z , \alpha _\text{em} , m_W , m_Z ,
m_f, V_{ij} , t_\beta , m_{12}^2 , \lambda _1 , \lambda _2 , \lambda _3
, \lambda _4 , \lambda _5 \} ~.
\label{eq:inputSetInteractionBase}
\end{equation}
However, we want to emphasize that the automatic parameter conversion routine in {\texttt{2HDECAY}} is only performed when the parameters are given in the mass basis of \eqref{eq:inputSetMassBase}.

Actually, also the tadpole parameters $T_1$ and $T_2$ should be
included in the two sets as independent parameters of the Higgs
potential. However, as described in
Sec.\,\ref{sec:renormalization2HDM}, the treatment of the minimum of
the Higgs potential at higher orders requires special care, and in an
alternative treatment of the minimum conditions, the tadpole
parameters disappear as independent parameters. In any case, after the
renormalization procedure is completely performed, the tadpole
parameters vanish again and hence, do not count as input parameters
for {\texttt{2HDECAY}}. 

\subsection{Renormalization}
\label{sec:renormalization2HDM}
We focus on the calculation of EW one-loop corrections to decay widths
of Higgs particles in the 2HDM. Since the higher-order (HO)
corrections of these decay widths are in general ultraviolet (UV)-divergent, a
proper regularization and renormalization of the UV divergences is
required. In the following, we briefly present the definition of the
counterterms (CTs) needed for the calculation of the EW one-loop
corrections. For a thorough derivation and presentation of the
gauge-independent renormalization of the 2HDM, we refer the reader to
\cite{Krause:2016gkg, Krause:2016oke, Krause:2016xku}.\footnote{See
  also Refs.~\cite{Denner:2016etu,Altenkamp:2017ldc,Denner2018}
  for a discussion of the renormalization of the 2HDM. For recent
  works discussing gauge-independent renormalization within multi-Higgs
  models, see \cite{Fox:2017hbw,Grimus:2018rte}.}

All input parameters that are renormalized for the calculation of the
EW corrections (apart from the mixing angles $\alpha $ and
$\beta$ and the soft-$\mathbb{Z}_2$-breaking scale $m_{12}$) are
renormalized in the on-shell (OS) scheme. For the physical fields,
we employ the conditions that any mixing of fields with the same
quantum numbers shall vanish on the mass shell of the respective
particles and that the fields are normalized by fixing the
residue of their corresponding propagators at their poles to unity. Mass
CTs are fixed through the condition that the masses are defined as the
real parts of the poles of the renormalized propagators. These OS
conditions suffice to renormalize most of the parameters of the 2HDM
necessary for our work. The renormalization of the mixing angles
$\alpha$ and $\beta$ follows an OS-motivated approach, as discussed in
Sec.\,\ref{sec:renormalizationMixingAngles}, while $m_{12}$ is
renormalized via an $\overline{\text{MS}}$ condition as discussed in
Sec.\,\ref{sec:renormalizationSoftm12Squared}. 

\subsubsection{Renormalization of the Tadpoles}
\label{sec:renormalizationTadpoles}
As shown for the 2HDM for the first time in \cite{Krause:2016gkg,
  Krause:2016oke}, the proper treatment of the
tadpole terms at one-loop order is crucial for the 
gauge-independent definition of the CTs of the mixing angles $\alpha$
and $\beta$. This allows for the calculation of one-loop partial decay widths with
a manifestly gauge-independent relation between input variables and
the physical observable.
In the following, we briefly repeat the different renormalization
conditions for the tadpoles that can be employed in the 2HDM. 

The \textit{standard tadpole scheme} is a commonly used
renormalization scheme for the tadpoles ({\it cf.}~{\it e.g.}~\cite{Denner:1991kt}
for the SM or \cite{Kanemura:2004mg, Kanemura:2015mxa} for the
2HDM). While the tadpole parameters vanish at tree level, as stated in
\eqref{eq:tadpoleVanishAtTreelevel}, they are in general non-vanishing
at higher orders in perturbation theory. Since the tadpole terms,
being the terms linear in the Higgs potential, define the minimum
of the potential, it is necessary to employ a renormalization of the
tadpoles in such a way that the ground state of the potential still
represents the minimum at higher orders. In the standard tadpole
scheme, this condition is imposed on the loop-corrected potential. By
replacing the tree-level tadpole terms at one-loop order with the
physical ({\it i.e.}~renormalized) tadpole terms and the tadpole CTs $\delta
T _i$, 
\begin{equation}
	T_i ~\rightarrow ~ T_i + \delta T_i ~~~ (i = 1,2) 
\end{equation}
the correct minimum of the loop-corrected potential
is obtained by demanding the renormalized tadpole
terms $T_i$ to vanish. This directly connects the tadpole CTs $\delta
T_i$ with the corresponding one-loop tadpole diagrams, 
\begin{equation}
i\delta T_{H/h} = \mathord{ \left(\rule{0cm}{30px}\right. \vcenter{
    \hbox{ \includegraphics[height=57px , trim = 16.6mm 12mm 16.6mm
      10mm, clip]{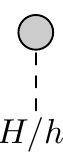} } }
  \left.\rule{0cm}{30px}\right)  } 
\label{eq:tadpoleCountertermDefinition}
\end{equation}
where we switched the tadpole terms from the interaction basis to the
mass basis by means of the rotation matrix $R(\alpha )$, as indicated
in \eqref{eq:rotationCPEven}. Since the tadpole terms explicitly
appear in the mass matrices in
Eqs.\,(\ref{eq:massMatrices1})-(\ref{eq:massMatrices3}), their CTs
explicitly appear in the mass matrices at one-loop
order. The rotation from the interaction to the mass basis yields
nine tadpole CTs in total which depend
on the two tadpole CTs $\delta T_{H/h}$ defined by the one-loop
tadpole diagrams in \eqref{eq:tadpoleCountertermDefinition}: 
\begin{mdframed}[frametitle={Renormalization of the tadpoles (standard scheme)},frametitlerule=true,frametitlebackgroundcolor=black!14,frametitlerulewidth=0.6pt]\begin{align}
\delta T_{HH} &= \frac{c _\alpha ^3 s _\beta + s _\alpha ^3 c _\beta }{vs _\beta c _\beta } \delta T_{H} - \frac{s_{2\alpha} s_{\beta - \alpha} }{vs_{2\beta} } \delta T_{h}  \label{eq:RenormalizationRadpolesTadpoleCountertermDeltaTH0H0ExplicitForm} \\
\delta T_{Hh} &= -\frac{s _{2\alpha} s _{\beta - \alpha} }{vs_{2\beta}} \delta T_{H} + \frac{s _{2\alpha} c _{\beta - \alpha} }{vs_{2\beta}} \delta T_{h}  \\
\delta T_{hh} &= \frac{s_{2\alpha} c_{\beta - \alpha} }{vs_{2\beta}} \delta T_{H} - \frac{s_\alpha ^3 s _\beta - c _\alpha ^3 c _\beta }{vs_\beta c_\beta } \delta T_{h}  \\
\delta T_{G^0G^0} &= \frac{c_{\beta -\alpha} }{v} \delta T_{H} + \frac{s _{\beta - \alpha} }{v} \delta T_{h}  \label{eq:RenormalizationRadpolesTadpoleCountertermDeltaTG0G0ExplicitForm} \\
\delta T_{G^0A} &= -\frac{s_{\beta - \alpha} }{v} \delta T_{H} + \frac{c _{\beta - \alpha} }{v} \delta T_{h}  \\
\delta T_{AA} &= \frac{c_\alpha s_\beta ^3 + s _\alpha c_\beta ^3 }{vs_\beta c_\beta } \delta T_{H} - \frac{s_\alpha s_\beta ^3 - c_\alpha c_\beta ^3 }{vs_\beta c_\beta } \delta T_{h}  \label{eq:RenormalizationRadpolesTadpoleCountertermDeltaTA0A0ExplicitForm} \\
\delta T_{G^\pm G^\pm } &= \frac{c_{\beta -\alpha} }{v}
                          \delta T_{H} + \frac{s _{\beta -
                          \alpha} }{v} \delta T_{h}  \\ 
\delta T_{G^\pm H^\pm } &= -\frac{s_{\beta - \alpha} }{v}
                          \delta T_{H} + \frac{c _{\beta -
                          \alpha} }{v} \delta T_{h}  \\ 
\delta T_{H^\pm H^\pm } &= \frac{c_\alpha s_\beta ^3
                          + s _\alpha c_\beta ^3
                          }{vs_\beta c_\beta }
                          \delta T_{H} - \frac{s_\alpha
                          s_\beta ^3 - c_\alpha
                          c_\beta ^3 }{vs_\beta
                          c_\beta } \delta
                          T_{h}  
\label{eq:RenormalizationRadpolesTadpoleCountertermDeltaTHpHpExplicitForm}
\end{align}\end{mdframed}
Since the minimum of the potential is defined through the
loop-corrected scalar potential, which in general is a gauge-dependent
quantity, the CTs defined through this minimum ({\it e.g.}~the CTs of
the scalar or gauge boson masses) become manifestly gauge-dependent
themselves. This is no problem as long as all gauge dependences
arising in a fixed-order calculation cancel against each other. In the
2HDM, however, an improper renormalization condition for the mixing
angle CTs within the standard tadpole scheme can lead to uncanceled
gauge dependences in the calculation of partial decay widths. This is
discussed in more detail in
Sec.\,\ref{sec:renormalizationMixingAngles}. Apart from the appearance
of the tadpole diagrams in
Eqs.\,(\ref{eq:RenormalizationRadpolesTadpoleCountertermDeltaTH0H0ExplicitForm})-(\ref{eq:RenormalizationRadpolesTadpoleCountertermDeltaTHpHpExplicitForm}),
and subsequently in the CTs and the wave function renormalization
constants (WFRCs) defined through these, the renormalization condition
in \eqref{eq:tadpoleCountertermDefinition} ensures that all other
appearances of tadpoles are canceled in the one-loop calculation,
{\it i.e.}~tadpole diagrams in the self-energies or vertex corrections
do not have to be taken into account. 
	
\begin{figure}[t!]
\centering
\includegraphics[width=\linewidth, trim=2.2cm 2.2cm 0cm 2.2cm, clip]{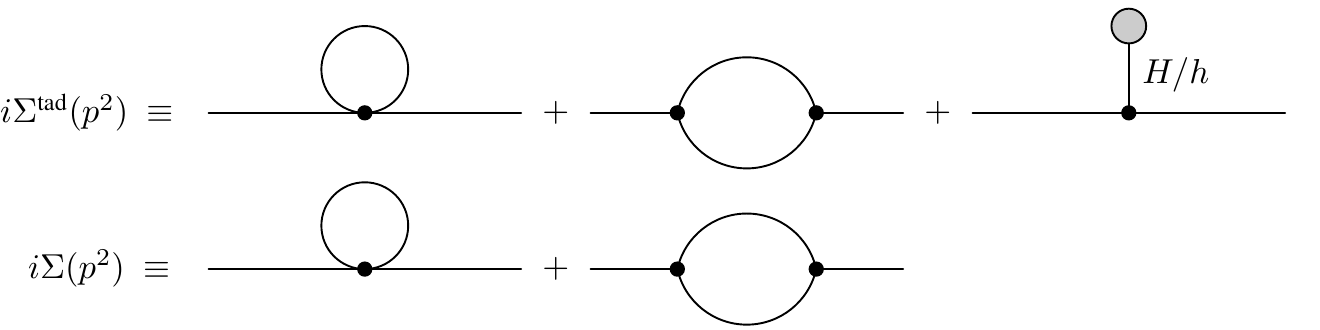}
\caption{Generic definition of the self-energies $\Sigma$ and $\Sigma
  ^\text{tad}$ as function of the external momentum $p^2$
  used in our CT definitions of the 2HDM. While $\Sigma$
  is the textbook definition of the one-particle irreducible
  self-energy, the self-energy $\Sigma ^\text{tad}$ additionally
  contains tadpole diagrams, indicated by the gray
  blob. For the actual calculation, the full particle content of the 2HDM has to be
  inserted into the self-energy topologies depicted here.} 
\label{fig:definitionOfSelfenergies}
\end{figure} 
An alternative treatment of the tadpole renormalization was proposed
by J. Fleischer and F. Jegerlehner in the SM
\cite{PhysRevD.23.2001}. It was applied to the extended scalar sector
of the 2HDM for the first time in \cite{Krause:2016gkg,
  Krause:2016oke} and is called  \textit{alternative (FJ) tadpole
  scheme} in the following. In this alternative approach, the VEVs
$v_{1,2}$ are considered as the fundamental quantities instead of the
tadpole terms. The \textit{proper} VEVs are the renormalized all-order
VEVs of the Higgs fields which represent the true ground state of the
theory and which are connected to the particle masses and the
couplings of the electroweak sector. Since the alternative approach
relies on the minimization of the gauge-independent tree-level scalar
potential, the mass CTs defined in this framework become manifestly
gauge-independent quantities by themselves. Moreover, the alternative
tadpole scheme connects the all-order renormalized VEVs directly to
the corresponding tree-level VEVs. Since the tadpoles are not the
fundamental quantities of the Higgs minimum in this framework, they do
not receive CTs. Instead, CTs for the VEVs are introduced by replacing
the VEVs with the renormalized VEVs and their CTs, 
\begin{equation}
	v_i ~ \rightarrow ~ v_i + \delta v_i
\end{equation}
and by fixing the latter in such a way that it is ensured that the
renormalized VEVs represent the proper tree-level minima to all
orders. At one-loop level, this leads to the following connection
between the VEV CTs in the interaction basis and the one-loop tadpole
diagrams in the mass basis,
\begin{equation}
\delta v_1 = \frac{-i c_\alpha }{m_{H}^2} \mathord{
  \left(\rule{0cm}{30px}\right. \vcenter{
    \hbox{ \includegraphics[height=57px , trim = 18mm 12mm 17.4mm
      10mm, clip]{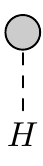} } }
  \left.\rule{0cm}{30px}\right)  } - \frac{-i s_\alpha }{m_{h}^2}
\mathord{ \left(\rule{0cm}{30px}\right. \vcenter{
    \hbox{ \includegraphics[height=57px , trim = 18mm 12mm 17.4mm
      10mm, clip]{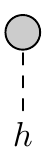} } }
  \left.\rule{0cm}{30px}\right)  } ~~~\text{and} ~~~ \delta v_2 = \frac{-i
  s_\alpha }{m_{H}^2} \mathord{
  \left(\rule{0cm}{30px}\right. \vcenter{
    \hbox{ \includegraphics[height=57px , trim = 18mm 12mm 17.4mm
      10mm, clip]{TadpoleDiagramHiggsBasisHH.pdf} } }
  \left.\rule{0cm}{30px}\right)  } + \frac{-i c_\alpha }{m_{h}^2}
\mathord{ \left(\rule{0cm}{30px}\right. \vcenter{
    \hbox{ \includegraphics[height=57px , trim = 18mm 12mm 17.4mm
      10mm, clip]{TadpoleDiagramHiggsBasish0.pdf} } }
  \left.\rule{0cm}{30px}\right)  } ~. 
\label{eq:vevCountertermDefinition} 
\end{equation}
The renormalization of the VEVs in the alternative tadpole scheme
effectively shifts the VEVs by tadpole contributions. As a
consequence, tadpole diagrams have to be considered wherever they can
appear in the 2HDM. For the self-energies, this means that the
fundamental self-energies used to define the CTs are the ones defined
as $\Sigma ^\text{tad}$ in \figref{fig:definitionOfSelfenergies}
instead of the usual one-particle irreducible self-energies
$\Sigma$. Additionally, tadpole diagrams have to be considered in the
calculation of the one-loop vertex corrections to the Higgs decays. In 
summary, the renormalization of the tadpoles in the alternative scheme
leads to the following conditions:

\begin{mdframed}[frametitle={Renormalization of the tadpoles (alternative FJ scheme)},frametitlerule=true,frametitlebackgroundcolor=black!14,frametitlerulewidth=0.6pt]\begin{align}
\delta T_{ij} &= 0   \\
\Sigma (p^2) ~ &\rightarrow ~\Sigma ^\text{tad} (p^2)   \\
\text{Tadpole diagrams have to be }&\text{considered in the vertex
                                     corrections.} 
\nonumber
\end{align}\end{mdframed}

\subsubsection{Renormalization of the Gauge Sector}
\label{sec:renormalizationGaugeSector}
For the renormalization of the gauge sector, we introduce CTs and
WFRCs for all parameters and fields of the electroweak sector of the
2HDM by applying the shifts 
\begin{align}
m_W^2 ~ &\rightarrow ~ m_W^2 + \delta m_W^2  \\
m_Z^2 ~ &\rightarrow ~ m_Z^2 + \delta m_Z^2  \\
\alpha _\text{em} ~ &\rightarrow ~ \alpha _\text{em} + \delta \alpha _\text{em} \equiv \alpha _\text{em} + 2 \alpha _\text{em} \delta Z_e \\	
W^\pm _\mu ~ &\rightarrow ~ \left( 1 + \frac{\delta Z_{WW} }{2} \right) W^\pm _\mu \\
\begin{pmatrix} Z \\ \gamma \end{pmatrix} ~ &\rightarrow ~ \begin{pmatrix} 1 + \frac{\delta Z_{ZZ} }{2} & \frac{\delta Z_{Z \gamma }}{2} \\ \frac{\delta Z_{\gamma Z }}{2} & 1 + \frac{\delta Z_{\gamma \gamma }}{2} \end{pmatrix} \begin{pmatrix} Z \\ \gamma \end{pmatrix} 
\;
\end{align}
where for convenience, we additionally introduced the shift
\begin{equation}
	e ~ \rightarrow ~ e\,( 1 + \delta Z_e ) 
\end{equation}
for the electromagnetic coupling constant by using
\eqref{eq:electromagneticCouplingDefinition}. Applying OS
conditions to the gauge sector of the 2HDM leads to equivalent
expressions for the CTs as derived in Ref.~\cite{Denner:1991kt} for the
SM\footnote{In contrast to Ref.~\cite{Denner:1991kt}, however, we choose a
  different sign for the $SU(2)_L$ term of the covariant derivative,
  which subsequently leads to a different sign in front of the second
  term of \eqref{eq:RenormalizationGaugeSectorExplicitFormDeltaZe}.},
for the standard and alternative tadpole scheme, respectively,
\begin{mdframed}[frametitle={Renormalization of the gauge sector (standard scheme)},frametitlerule=true,frametitlebackgroundcolor=black!14,frametitlerulewidth=0.6pt]\begin{align}
\delta m_W^2 &= \textrm{Re} \left[ \Sigma _{WW} ^{T} \left( m_W ^2 \right) \right]  \\
\delta m_Z^2 &= \textrm{Re} \left[ \Sigma _{ZZ} ^{T} \left( m_{Z} ^2 \right) \right] 
\end{align}\end{mdframed}
\begin{mdframed}[frametitle={Renormalization of the gauge sector (alternative FJ scheme)},frametitlerule=true,frametitlebackgroundcolor=black!14,frametitlerulewidth=0.6pt]\begin{align}
\delta m_W^2 &= \textrm{Re} \left[ \Sigma _{WW} ^{\textrm{tad},T} \left( m_W ^2 \right) \right]  \\
\delta m_Z^2 &= \textrm{Re} \left[ \Sigma _{ZZ} ^{\textrm{tad},T} \left( m_{Z} ^2 \right) \right]  
\end{align}\end{mdframed}
The WFRCs are the same in both tadpole schemes, 
\begin{mdframed}[frametitle={Renormalization of the gauge sector
    (standard and alternative FJ scheme)},frametitlerule=true,frametitlebackgroundcolor=black!14,frametitlerulewidth=0.6pt,nobreak=true]\begin{align}
\delta Z_e (m_Z^2) &= \frac{1}{2} \left. \frac{\partial \Sigma ^T _{\gamma \gamma } \left( p^2 \right) }{\partial p^2 } \right| _{p^2 = 0} + \frac{s_W }{c_W } \frac{\Sigma ^T _{\gamma Z} \left( 0\right) }{m_Z ^2 } - \frac{1}{2} \Delta \alpha (m_Z^2) \label{eq:RenormalizationGaugeSectorExplicitFormDeltaZe} \\
\delta Z_{WW} &= - \textrm{Re} \left[ \frac{\partial \Sigma ^T _{WW} \left( p^2 \right) }{\partial p^2 } \right] _{p^2 = m_W^2 }  \label{eq:RenormalizationGaugeSectorExplicitFormDeltaWW} \\
\begin{pmatrix} \delta Z_{ZZ} & \delta Z _{Z \gamma } \\ \delta Z _{\gamma Z } & \delta Z_{ \gamma \gamma } \end{pmatrix} &= \begin{pmatrix} - \textrm{Re} \left[ \frac{\partial \Sigma ^T _{ZZ} \left( p^2 \right) }{\partial p^2 } \right]  _{p^2 = m_Z^2 } & \frac{2}{m_Z ^2 } \Sigma ^{T} _{Z \gamma } \left( 0 \right)  \\ -\frac{2}{m_Z ^2 } \textrm{Re} \left[  \Sigma ^{T} _{Z \gamma } \left( m_Z ^2 \right)  \right] & - \textrm{Re} \left[ \frac{\partial \Sigma ^T _{\gamma \gamma } \left( p^2 \right) }{\partial p^2 } \right] _{p^2 = 0} \end{pmatrix}  \label{eq:RenormalizationGaugeSectorExplicitFormDeltaZZ} 
\end{align}\end{mdframed}
The superscript $T$ indicates that only the transverse parts of the
self-energies are taken into account. The CT for the
electromagnetic coupling $\delta Z_e (m_Z^2)$ is defined at the scale
of the $Z$ boson mass instead of the Thomson limit. For this, the additional term
\begin{equation}
\Delta \alpha (m_Z^2) = \frac{\partial \Sigma _{\gamma \gamma} ^{\text{light},T} (p^2) }{\partial p^2} \Bigg| _{p^2 = 0} - \frac{\Sigma ^T _{\gamma \gamma } (m_Z^2) }{m_Z^2}
\label{eq:lightContributions}
\end{equation}
is required, where the transverse photon self-energy $\Sigma _{\gamma \gamma}
^{\text{light},T} (p^2)$ in \eqref{eq:lightContributions} contains
solely light fermion contributions ({\it i.e.}~contributions from all
fermions apart from the $t$ quark). This ensures that the results of
our EW one-loop computations are independent of large
logarithms due to light fermion contributions \cite{Denner:1991kt}. 

For later convenience, we additionally introduce the shift of the weak coupling constant
\begin{equation}
g ~ \rightarrow ~ g + \delta g ~.
\end{equation}
Since $g$ is not an independent parameter in our approach, {\it
  cf.}~\eqref{eq:electromagneticCouplingDefinition}, the CT $\delta g$
is not independent either and can be expressed through the other CTs
derived in this subsection as 
\begin{equation}
	\frac{\delta g}{g} = \delta Z_e (m_Z^2) + \frac{1}{2( m_Z^2 - m_W^2)} \left( \delta m_W^2 - \frac{m_W^2}{m_Z^2} \delta m_Z^2 \right) ~.
\end{equation}

\subsubsection{Renormalization of the Scalar Sector}
\label{sec:renormalizationScalarSector}
In the scalar sector of the 2HDM, the masses and fields of the scalar particles are shifted as
\begin{align}
m_{H}^2 ~ &\rightarrow ~ m_{H}^2 + \delta m_{H}^2  \\
m_{h}^2 ~ &\rightarrow ~ m_{h}^2 + \delta m_{h}^2  \\
m_{A}^2 ~ &\rightarrow ~ m_{A}^2 + \delta m_{A}^2  \\
m_{H^\pm }^2 ~ &\rightarrow ~ m_{H^\pm }^2 + \delta m_{H^\pm }^2  \\
\begin{pmatrix}  H \\ h \end{pmatrix} ~ &\rightarrow ~ \begin{pmatrix} 1 + \frac{\delta Z _{{H} {H}}}{2} & \frac{\delta Z _{{H} {h}}}{2}  \\ \frac{\delta Z _{{h} {H}}}{2}  & 1 + \frac{\delta Z _{{h} {h}}}{2} \end{pmatrix} \renewcommand*{\arraystretch}{1.6} \begin{pmatrix}  H \\ h \end{pmatrix}  \label{RenormalizationOnShellLabelSectionScalarSectorFieldRenormalizationConstantsCPEvenHiggses} \\
\begin{pmatrix}  G^0 \\ A \end{pmatrix} ~ &\rightarrow ~ \begin{pmatrix} 1 + \frac{\delta Z _{{G}^0 {G}^0}}{2}  & \frac{\delta Z _{{G}^0 {A}}}{2}  \\ \frac{\delta Z _{{A} {G}^0}}{2}  & 1 + \frac{\delta Z _{{A} {A}}}{2} \renewcommand*{\arraystretch}{1.6} \end{pmatrix} \begin{pmatrix}  G^0 \\ A \end{pmatrix} \label{RenormalizationOnShellLabelSectionScalarSectorFieldRenormalizationConstantsCPOddHiggses} \\
\begin{pmatrix}  G^\pm \\ H^\pm \end{pmatrix} ~ &\rightarrow ~ \begin{pmatrix} 1 + \frac{\delta Z _{{G}^\pm {G}^\pm }}{2}  & \frac{\delta Z _{{G}^\pm {H}^\pm }}{2}  \\ \frac{\delta Z _{{H}^\pm {G}^\pm }}{2}  & 1 + \frac{\delta Z _{{H}^\pm {H}^\pm }}{2} \renewcommand*{\arraystretch}{1.6} \end{pmatrix} \begin{pmatrix}  G^\pm \\ H^\pm \end{pmatrix} \label{RenormalizationOnShellLabelSectionScalarSectorFieldRenormalizationConstantsChargedHiggses} ~.
\end{align}
Applying OS renormalization conditions leads to the following CT
definitions \cite{Krause:2016gkg}, 
\begin{mdframed}[frametitle={Renormalization of the scalar sector (standard scheme)},frametitlerule=true,frametitlebackgroundcolor=black!14,frametitlerulewidth=0.6pt,nobreak=true]\begin{align}
\delta Z_{Hh} &= \frac{2}{m_{H}^2 - m_{h}^2} \textrm{Re} \Big[ \Sigma _{Hh} (m_{h}^2) - \delta T_{Hh} \Big]  \label{RenormalizationScalarFieldsMassesExplicitFormWaveFunctionRenormalizationConstantH0h0} \\
\delta Z_{hH} &= -\frac{2}{m_{H}^2 - m_{h}^2} \textrm{Re} \Big[ \Sigma _{Hh} (m_{H}^2) - \delta T_{Hh} \Big]   \label{RenormalizationScalarFieldsMassesExplicitFormWaveFunctionRenormalizationConstanth0H0} \\
\delta Z_{G^0A} &= -\frac{2}{m_{A}^2} \textrm{Re} \Big[ \Sigma _{G^0A} (m_{A}^2) - \delta T_{G^0A} \Big]   \\
\delta Z_{AG^0} &= \frac{2}{m_{A}^2} \textrm{Re} \Big[ \Sigma _{G^0A} (0) - \delta T_{G^0A} \Big]  \\
\delta Z_{G^\pm H^\pm} &= -\frac{2}{m_{H^\pm}^2} \textrm{Re} \Big[ \Sigma _{G^\pm H^\pm} (m_{H^\pm}^2) - \delta T_{G^\pm H^\pm} \Big]  \\
\delta Z_{H^\pm G^\pm} &= \frac{2}{m_{H^\pm}^2} \textrm{Re} \Big[ \Sigma _{G^\pm H^\pm} (0) - \delta T_{G^\pm H^\pm} \Big]  \\
\delta m_{H}^2 &= \textrm{Re} \Big[ \Sigma _{HH} (m_{H}^2) - \delta T_{HH} \Big]  \\
\delta m_{h}^2 &= \textrm{Re} \Big[ \Sigma _{hh} (m_{h}^2) - \delta T_{hh} \Big]   \\
\delta m_{A}^2 &= \textrm{Re} \Big[ \Sigma _{AA} (m_{A}^2) - \delta T_{AA} \Big]  \\
\delta m_{H^\pm }^2 &= \textrm{Re} \Big[ \Sigma _{H^\pm H^\pm } (m_{H^\pm }^2) - \delta T_{H^\pm H^\pm } \Big] 
\end{align}\end{mdframed}
\begin{mdframed}[frametitle={Renormalization of the scalar sector (alternative FJ  scheme)},frametitlerule=true,frametitlebackgroundcolor=black!14,frametitlerulewidth=0.6pt,nobreak=true]\begin{align}
\delta Z_{Hh} &= \frac{2}{m_{H}^2 - m_{h}^2} \textrm{Re} \Big[ \Sigma ^\textrm{tad} _{Hh} (m_{h}^2) \Big]  \label{RenormalizationScalarFieldsMassesExplicitFormWaveFunctionRenormalizationConstantH0h0Alt} \\
\delta Z_{hH} &= -\frac{2}{m_{H}^2 - m_{h}^2} \textrm{Re} \Big[ \Sigma ^\textrm{tad} _{Hh} (m_{H}^2) \Big]  \label{RenormalizationScalarFieldsMassesExplicitFormWaveFunctionRenormalizationConstanth0H0Alt} \\
\delta Z_{G^0A} &= -\frac{2}{m_{A}^2} \textrm{Re} \Big[ \Sigma ^\textrm{tad} _{G^0A} (m_{A}^2) \Big]  \\
\delta Z_{AG^0} &= \frac{2}{m_{A}^2} \textrm{Re} \Big[ \Sigma ^\textrm{tad} _{G^0A} (0) \Big]  \\
\delta Z_{G^\pm H^\pm} &= -\frac{2}{m_{H^\pm}^2} \textrm{Re} \Big[ \Sigma ^\textrm{tad} _{G^\pm H^\pm} (m_{H^\pm}^2) \Big]  \\
\delta Z_{H^\pm G^\pm} &= \frac{2}{m_{H^\pm}^2} \textrm{Re} \Big[ \Sigma ^\textrm{tad} _{G^\pm H^\pm} (0) \Big]  \\
\delta m_{H}^2 &= \textrm{Re} \Big[ \Sigma ^\textrm{tad} _{HH} (m_{H}^2) \Big]   \label{RenormalizationScalarFieldsMassesExplicitFormMassCountertermH0} \\
\delta m_{h}^2 &= \textrm{Re} \Big[ \Sigma ^\textrm{tad} _{hh} (m_{h}^2) \Big]  \label{RenormalizationScalarFieldsMassesExplicitFormMassCountertermh0} \\
	\delta m_{A}^2 &= \textrm{Re} \Big[ \Sigma ^\textrm{tad} _{AA} (m_{A}^2) \Big]  \\
	\delta m_{H^\pm }^2 &= \textrm{Re} \Big[ \Sigma ^\textrm{tad} _{H^\pm H^\pm } (m_{H^\pm }^2) \Big]  \label{RenormalizationScalarFieldsMassesExplicitFormMassCountertermHp}
\end{align}\end{mdframed}
\begin{mdframed}[frametitle={Renormalization of the scalar sector
    (standard and alternative FJ scheme)},frametitlerule=true,frametitlebackgroundcolor=black!14,frametitlerulewidth=0.6pt,nobreak=true]\begin{align}
\delta Z_{HH} &= - \textrm{Re} \left[ \frac{\partial \Sigma _{H H } \left( p^2 \right) }{\partial p^2 } \right] _{p^2 = m_{H} ^2}  \label{RenormalizationScalarFieldsMassesExplicitFormWaveFunctionRenormalizationConstantH0H0} \\
\delta Z_{hh} &= - \textrm{Re} \left[ \frac{\partial \Sigma _{h h } \left( p^2 \right) }{\partial p^2 } \right] _{p^2 = m_{h} ^2} \\
\delta Z_{G^0G^0} &= - \textrm{Re} \left[ \frac{\partial \Sigma _{G^0 G^0 } \left( p^2 \right) }{\partial p^2 } \right] _{p^2 = 0} \\
\delta Z_{AA} &= - \textrm{Re} \left[ \frac{\partial \Sigma _{A A } \left( p^2 \right) }{\partial p^2 } \right] _{p^2 = m_{A} ^2}  \\
\delta Z_{G^\pm G^\pm } &= - \textrm{Re} \left[ \frac{\partial \Sigma _{G^\pm G^\pm } \left( p^2 \right) }{\partial p^2 } \right] _{p^2 = 0} \\
\delta Z_{H^\pm H^\pm } &= - \textrm{Re} \left[ \frac{\partial \Sigma _{H^\pm H^\pm } \left( p^2 \right) }{\partial p^2 } \right] _{p^2 = m_{H^\pm} ^2}  \label{RenormalizationScalarFieldsMassesExplicitFormWaveFunctionRenormalizationConstantHpHp}
\end{align}\end{mdframed}
with the tadpole CTs in the standard scheme defined in Eqs.\,(\ref{eq:RenormalizationRadpolesTadpoleCountertermDeltaTH0H0ExplicitForm})-(\ref{eq:RenormalizationRadpolesTadpoleCountertermDeltaTHpHpExplicitForm}).

\subsubsection{Renormalization of the Scalar Mixing Angles}
\label{sec:renormalizationMixingAngles}
In the following, we describe the renormalization of the scalar mixing
angles $\alpha$ and $\beta$ in the 2HDM. In our approach, we perform
the rotation from the interaction to the mass basis,
{\it cf.}~Eqs.\,(\ref{eq:rotationCPEven})-(\ref{eq:rotationCharged}), before
renormalization so that the mixing angles need to be renormalized. 
At one-loop level, the bare mixing angles are replaced by their
renormalized values and counterterms as
\begin{align}
\alpha ~ &\rightarrow ~ \alpha + \delta \alpha  \\
\beta ~ &\rightarrow ~ \beta + \delta \beta ~.
\end{align}
The renormalization of the mixing angles in the 2HDM is a non-trivial
task and several different schemes have been proposed in the
literature. In the following, we only briefly present the definition
of the mixing angle CTs in all different schemes that are implemented
in {\texttt{2HDECAY}} and refer to \cite{Krause:2016gkg,
  Krause:2016oke} for details on the derivation of these
schemes. \\

\textbf{$\overline{\text{MS}}$ scheme.} It was shown in \cite{Lorenz:2015, Krause:2016gkg} that an
$\overline{\text{MS}}$ condition for $\delta \alpha$ and $\delta
\beta$ can lead to one-loop corrections that
are orders of magnitude larger than the LO result\footnote{In \cite{Denner:2016etu},
    an $\overline{\text{MS}}$ condition for the scalar mixing angles
    in certain processes led to corrections that are numerically
    well-behaving due to a partial cancellation of large contributions
    from tadpoles. In the decays considered in our work, an
    $\overline{\text{MS}}$ condition of $\delta \alpha$ and $\delta
    \beta$ in general leads to very large corrections, however.}. We implemented this scheme in {\texttt{2HDECAY}} for reference, as the $\overline{\text{MS}}$ CTs contain only the UV divergences of the CTs, but no finite parts $\left. \delta \alpha \right| _\text{fin}$ and $\left. \delta \beta \right| _\text{fin}$. After having checked for UV finiteness of the full decay width, the CTs of the mixing angles $\alpha$ and $\beta$ are effectively set to zero in {\texttt{2HDECAY}} in this scheme for the numerical evaluation of the partial decay widths.
\begin{mdframed}[frametitle={ Renormalization of $\delta \alpha$ and $\delta \beta$: $\overline{\text{MS}}$ scheme (both schemes) },frametitlerule=true,frametitlebackgroundcolor=black!14,frametitlerulewidth=0.6pt,nobreak=true]\begin{align}
\left. \delta \alpha \right| _\text{fin} &= 0 \\
\left. \delta \beta \right| _\text{fin} &= 0 
\end{align}\end{mdframed}
The $\overline{\text{MS}}$ CTs of $\alpha$ and
  $\beta$ depend on the renormalization scale $\mu_R$. The user has to
  specifiy in the input file the scale at which $\alpha$ and $\beta$
  are understood to be given when the $\overline{\text{MS}}$ renormalization scheme is
  chosen. The one-loop corrected decay widths that contain these CTs,
  then additionally depend on the renormalization scale of $\alpha$ and
  $\beta$. The scale at which the decays are evaluated is also defined 
  by the user in the input file and should be chosen appropriately in
  order to avoid the appearance of large logarithms in the EW one-loop
  corrections. In case this scale differs from the scale of the
  $\overline{\text{MS}}$ mixing angles 
  $\alpha$ and $\beta$, the automatic parameter conversion
  routine converts $\alpha$ and $\beta$ to the scale of the
  loop-corrected decay widths, as further described in
  Sec.\,\ref{sec:ParameterConversion}. For the
  conversion, the UV-divergent terms for the CTs $\delta \alpha$ and
  $\delta \beta$ are needed, \textit{i.e.}\,the terms proportional to
  $1/\varepsilon$. These UV-divergent terms are presented analytically
  for both the standard and alternative tadpole scheme in
  Ref.\,\cite{Altenkamp:2017ldc}. We cross-checked these terms
  analytically in an independent calculation. \\

\textbf{KOSY scheme.} The KOSY scheme (denoted by the authors' initials)
was suggested in \cite{Kanemura:2004mg}. It combines the standard
tadpole scheme with the definition of the counterterms 
through off-diagonal wave function renormalization constants.
As shown in \cite{Krause:2016gkg,Krause:2016oke}, the KOSY scheme not
only implies a gauge-dependent definition of the mixing angle CTs but also
leads to explicitly gauge-dependent decay amplitudes. The CTs are
derived by temporarily switching from the mass to the gauge
basis. Since $\beta$ diagonalizes both the charged and CP-odd
sector not all scalar fields can be defined OS at the same time,
unless a systematic modification of the $SU(2)$
  relations is performed which we do not do here. 
We implemented two different CT definitions where $\delta \beta$ is defined
through the CP-odd or the charged sectors, indicated by superscripts $o$
and $c$, respectively. The KOSY scheme is implemented in
{\texttt{2HDECAY}} both in the standard and in the alternative FJ
scheme as a benchmark scheme for comparison with other schemes, but
for actual computations, we do not recommend to use it due to the
explicit gauge dependence of the decay amplitudes. In the KOSY
scheme, the mixing angle CTs are defined as 
\begin{mdframed}[frametitle={Renormalization of $\delta \alpha$ and $\delta \beta$: KOSY scheme (standard scheme)},frametitlerule=true,frametitlebackgroundcolor=black!14,frametitlerulewidth=0.6pt,nobreak=true]\begin{align}
\delta \alpha &= \frac{1}{2(m_H^2 - m_h^2)} \text{Re} \left[ \Sigma _{Hh} (m_H^2) + \Sigma _{Hh} (m_h^2) - 2\delta T_{Hh} \right] \\
\delta \beta ^o &= -\frac{1}{2m_A^2} \text{Re} \left[ \Sigma _{G^0A} (m_A^2) + \Sigma _{G^0A} (0) - 2\delta T_{G^0A} \right] \\
\delta \beta ^c &= -\frac{1}{2m_{H^\pm}^2} \text{Re} \left[ \Sigma _{G^\pm H^\pm} (m_{H^\pm}^2) + \Sigma _{G^\pm H^\pm} (0) - 2\delta T_{G^\pm H^\pm} \right] 
\end{align}\end{mdframed}
\begin{mdframed}[frametitle={Renormalization of $\delta \alpha$ and $\delta \beta$: KOSY scheme (alternative FJ scheme)},frametitlerule=true,frametitlebackgroundcolor=black!14,frametitlerulewidth=0.6pt,nobreak=true]\begin{align}
\delta \alpha &= \frac{1}{2(m_H^2 - m_h^2)} \text{Re} \left[ \Sigma ^\text{tad} _{Hh} (m_H^2) + \Sigma ^\text{tad} _{Hh} (m_h^2) \right] \\
\delta \beta ^o &= -\frac{1}{2m_A^2} \text{Re} \left[ \Sigma ^\text{tad} _{G^0A} (m_A^2) + \Sigma ^\text{tad} _{G^0A} (0) \right]  \\
\delta \beta ^c &= -\frac{1}{2m_{H^\pm}^2} \text{Re} \left[ \Sigma ^\text{tad} _{G^\pm H^\pm} (m_{H^\pm}^2) + \Sigma ^\text{tad} _{G^\pm H^\pm} (0) \right] 
\end{align}\end{mdframed}
\vspace*{0.14cm}

\textbf{$p_{*}$-pinched scheme.} One possibility to avoid
gauge-parameter-dependent mixing angle CTs was
suggested in \cite{Krause:2016gkg, Krause:2016oke}. The main idea is
to maintain the OS-based definition of $\delta \alpha$ and $\delta
\beta$ of the KOSY scheme, but instead of using the usual
gauge-dependent off-diagonal WFRCs, the WFRCs are defined through
pinched self-energies in the alternative FJ scheme by applying the pinch
technique (PT) \cite{Binosi:2004qe, Binosi:2009qm, Cornwall:1989gv,
  Papavassiliou:1989zd, Degrassi:1992ue, Papavassiliou:1994pr,
  Watson:1994tn, Papavassiliou:1995fq}. As worked out for the
2HDM for the first time in \cite{Krause:2016gkg, Krause:2016oke}, the
pinched scalar self-energies are equivalent to the usual scalar
self-energies in the alternative FJ scheme, evaluated in Feynman-'t
Hooft gauge ($\xi =1$), up to additional UV-finite self-energy contributions
$\Sigma ^\text{add} _{ij} (p^2)$. The mixing angle
CTs depend on the scale where the pinched self-energies are
evaluated. In the $p_{*}$-pinched scheme, we follow the approach of
\cite{Espinosa:2002cd} in the MSSM, where the self-energies $\Sigma
^\text{tad} _{ij} (p^2)$ are evaluated at the scale 
\begin{equation}
	p_{*}^2 \equiv \frac{m_i^2 + m_j^2}{2} ~.
\end{equation}
At this scale, the additional contributions $\Sigma ^\text{add} _{ij}
(p^2)$ vanish. Using the $p_{*}$-pinched scheme at
one-loop level yields explicitly gauge-parameter-independent partial
decay widths. The mixing angle CTs are defined as 
\begin{mdframed}[frametitle={Renormalization of $\delta \alpha$ and $\delta \beta$: $p_{*}$-pinched scheme (alternative FJ scheme)},frametitlerule=true,frametitlebackgroundcolor=black!14,frametitlerulewidth=0.6pt,nobreak=true]\begin{align}
\delta \alpha &= \frac{1}{m_H^2 - m_h^2} \text{Re} \left[ \Sigma ^\text{tad} _{Hh} \left( \frac{m_H^2 + m_h^2}{2} \right) \right] _{\xi = 1}  \\
\delta \beta ^o &= -\frac{1}{m_A^2} \text{Re} \left[ \Sigma ^\text{tad} _{G^0A} \left( \frac{m_A^2 }{2} \right) \right] _{\xi = 1}  \\
\delta \beta ^c &= -\frac{1}{m_{H^\pm}^2} \text{Re} \left[ \Sigma ^\text{tad} _{G^\pm H^\pm} \left( \frac{m_{H^\pm}^2 }{2} \right) \right] _{\xi = 1} 
\end{align}\end{mdframed}

\textbf{OS-pinched scheme.} In order to allow for the analysis of the effects
of different scale choices of the mixing angle CTs, we implemented 
another OS-motivated scale choice, which is called the OS-pinched
scheme. Here, the additional terms do not vanish and are given by
\cite{Krause:2016gkg} 
\begin{align}
\Sigma ^\textrm{add} _{H h } (p^2) &= \frac{\alpha _\text{em} m_Z^2 s_{\beta - \alpha} c_{\beta - \alpha} }{8 \pi m_W^2 \left( 1 - \frac{m_W^2}{m_Z^2} \right) } \left( p^2 - \frac{ m_{H}^2 + m_{h}^2}{2} \right) \bigg\{ \left[ B_0( p^2; m_Z^2, m_{A}^2) - B_0( p^2; m_Z^2, m_{Z}^2) \right] \nonumber \\
&\hspace*{0.4cm} + 2\frac{m_W^2}{m_Z^2} \left[ B_0( p^2; m_W^2, m_{H^\pm }^2) - B_0( p^2; m_W^2, m_{W}^2) \right] \bigg\} \label{eq:RenormalizationScalarAnglesAdditionalTermCPEven} \\
\Sigma ^\textrm{add} _{G^0 A } (p^2) &= \frac{\alpha _\text{em} m_Z^2 s_{\beta - \alpha} c_{\beta - \alpha}}{8 \pi m_W^2 \left( 1 - \frac{m_W^2}{m_Z^2} \right) } \left( p^2 - \frac{m_{A}^2}{2} \right) \left[ B_0( p^2; m_Z^2, m_{H}^2) - B_0( p^2; m_Z^2, m_{h}^2) \right] \label{eq:RenormalizationScalarAnglesAdditionalTermCPOdd} \\
\Sigma ^\textrm{add} _{G^\pm H^\pm } (p^2) &= \frac{\alpha _\text{em} s_{\beta - \alpha} c_{\beta - \alpha}}{4 \pi \left( 1 - \frac{m_W^2}{m_Z^2} \right) } \left( p^2 - \frac{m_{H^\pm }^2}{2} \right) \left[ B_0( p^2; m_W^2, m_{H}^2) - B_0( p^2; m_W^2, m_{h}^2) \right] ~. \label{eq:RenormalizationScalarAnglesAdditionalTermCharged} 
\end{align}
The mixing angle CTs in the OS-pinched scheme are then defined as
\begin{mdframed}[frametitle={Renormalization of $\delta \alpha$ and $\delta \beta$: OS-pinched scheme (alternative FJ scheme)},frametitlerule=true,frametitlebackgroundcolor=black!14,frametitlerulewidth=0.6pt,nobreak=true]\begin{align}
\delta \alpha &= \frac{\textrm{Re} \Big[ \left[ \Sigma ^\textrm{tad} _{Hh} (m_{H}^2) + \Sigma ^\textrm{tad} _{Hh} (m_{h}^2) \right] _{\xi = 1} + \Sigma ^\textrm{add} _{Hh} (m_{H}^2) + \Sigma ^\textrm{add} _{Hh} (m_{h}^2) \Big]}{2\left( m_{H}^2 - m_{h}^2\right) }   \label{eq:RenormalizationScalarAnglesDeltaAlphaOSPinchedResult} \\
\delta \beta ^o &= -\frac{\textrm{Re} \Big[ \left[ \Sigma ^\textrm{tad} _{G^0A} (m_{A}^2) + \Sigma ^\textrm{tad} _{G^0A} (0) \right] _{\xi = 1} + \Sigma ^\textrm{add} _{G^0A} (m_{A}^2) + \Sigma ^\textrm{add} _{G^0A} (0) \Big]}{2 m_{A}^2}   \label{eq:RenormalizationScalarAnglesDeltaBeta1OSPinchedResult} \\
\delta \beta ^c &= -\frac{\textrm{Re} \Big[ \left[ \Sigma ^\textrm{tad} _{G^\pm H^\pm } (m_{H^\pm }^2) + \Sigma ^\textrm{tad} _{G^\pm H^\pm } (0) \right] _{\xi = 1} + \Sigma ^\textrm{add} _{G^\pm H^\pm } (m_{H^\pm }^2) + \Sigma ^\textrm{add} _{G^\pm H^\pm } (0) \Big]}{2 m_{H^\pm }^2}   \label{eq:RenormalizationScalarAnglesDeltaBeta2OSPinchedResult}
\end{align}\end{mdframed}

\textbf{Process-dependent schemes.} The definition of the mixing angle
CTs through observables, like {\it e.g.}~partial decay widths of Higgs
bosons, was proposed for the MSSM in \cite{Coarasa:1996qa,
  Freitas:2002um} and for the 2HDM in \cite{Santos:1996hs}. This
scheme leads to explicitly gauge-independent partial
decay widths per construction. Moreover, the connection of the mixing angle CTs with
physical observables allows for a more physical interpretation of the
unphysical mixing angles $\alpha $ and $\beta$. However, as
it was shown in \cite{Krause:2016gkg, Krause:2016oke}, 
process-dependent schemes can in general lead to very large one-loop
corrections. We implemented three different process-dependent schemes
for $\delta \alpha$ and $\delta \beta$ in
{\texttt{2HDECAY}}. The schemes differ in the processes that are used
for the definition of the CTs. In all cases we have chosen leptonic
Higgs boson decays. For these, the QED corrections can be
separated in a UV-finite way from the rest of the EW corrections and therefore be
excluded from the counterterm
definition. This is necessary to avoid
the appearance of infrared (IR) divergences in the CTs \cite{Freitas:2002um}. The
NLO corrections to the partial decay widths of the leptonic decay
of a Higgs particle $\phi _i$ into a pair of leptons $f_j$, $f_k$ can
then be cast into the form 
\begin{equation}
\Gamma _{\phi _i f_j f_k}^{\text{NLO,weak}} = \Gamma _{\phi _i f_j f_k}^{\text{LO}} \left( 1 + 2\text{Re} \left[ \mathcal{F}_{\phi _i f_j f_k}^\text{VC} + \mathcal{F}_{\phi _i f_j f_k}^\text{CT} \right] \right) 
\end{equation} 
where $\mathcal{F}_{\phi _i f_j f_k}^\text{VC}$ and $\mathcal{F}_{\phi
  _i f_j f_k}^\text{CT}$ are the form factors of the vertex
corrections and the CT, respectively, and the superscript weak
indicates that in the vertex corrections IR-divergent QED
contributions are excluded. The form factor $\mathcal{F}_{\phi _i f_j
  f_k}^\text{CT}$ contains either $\delta \alpha$ or $\delta \beta$ or
both simultaneously as well as other CTs that are fixed as described
in the other subsections of
Sec.\,\ref{sec:renormalization2HDM}. Employing the renormalization
condition 
\begin{equation}
\Gamma ^\text{LO} _{\phi _i f_j f_k} \equiv \Gamma ^\text{NLO,weak} _{\phi _i f_j f_k}
\end{equation}
for two different decays then allows for a process-dependent
definition of the mixing angle CTs. For more details on the
calculation of the CTs in process-dependent schemes in the 2HDM, we
refer to \cite{Krause:2016gkg, Krause:2016oke}. In {\texttt{2HDECAY}},
we have chosen the following three different combinations of processes
as definition for the CTs,
\begin{enumerate}
\item $\delta \beta$ is first defined by $A \rightarrow \tau ^+ \tau ^-$ and $\delta \alpha$ is subsequently defined by $H \rightarrow \tau ^+ \tau ^-$.
\item $\delta \beta$ is first defined by $A \rightarrow \tau ^+ \tau ^-$ and $\delta \alpha$ is subsequently defined by $h \rightarrow \tau ^+ \tau ^-$.
\item $\delta \beta$ and $\delta \alpha$ are simultaneously defined by $H \rightarrow \tau ^+ \tau ^-$ and $h \rightarrow \tau ^+ \tau ^-$.
\end{enumerate}
Employing these renormalization conditions yields the following
definitions of the mixing angle CTs\footnote{While the definition of
  the CTs is generically the same for both tadpole schemes, their actual analytic
  forms differ in both schemes since some of the CTs used in the
  definition differ in the two schemes, as well. However, when choosing a process-dependent scheme for the mixing angle CTs, the full partial decay width is independent of the chosen tadpole scheme, which was checked explicitly by us. Therefore, in {\texttt{2HDECAY}} we implemented the process-dependent schemes in the alternative tadpole scheme, only. }: 
\begin{mdframed}[frametitle={Renormalization of $\delta \alpha$ and $\delta \beta$: process-dependent scheme 1 (both schemes)},frametitlerule=true,frametitlebackgroundcolor=black!14,frametitlerulewidth=0.6pt,nobreak=true]\begin{align}
\delta \alpha &= \frac{- Y_5}{Y_4} \bigg[ \mathcal{F}^\textrm{VC}_{H \tau \tau } + \frac{\delta g}{g} + \frac{\delta m_\tau }{m_\tau } - \frac{\delta m_W^2}{2m_W^2} + Y_6 \delta \beta + \frac{\delta Z_{HH}}{2} + \frac{Y_4}{Y_5} \frac{\delta Z_{hH}}{2} + \frac{\delta Z^\textrm{L} _{\tau \tau}}{2} \\
&\hspace*{1.4cm} + \frac{\delta Z^\textrm{R} _{\tau \tau}}{2} \bigg]   \nonumber \\
\delta \beta &= \frac{- Y_6}{1+Y_6^2} \bigg[ \mathcal{F}^\textrm{VC}_{A \tau \tau } + \frac{\delta g}{g} + \frac{\delta m_\tau }{m_\tau } - \frac{\delta m_W^2}{2m_W^2} + \frac{\delta Z_{AA}}{2} - \frac{1}{Y_6} \frac{\delta Z_{G^0A}}{2} + \frac{\delta Z^\textrm{L} _{\tau \tau}}{2} + \frac{\delta Z^\textrm{R} _{\tau \tau}}{2} \bigg]  
\end{align}\end{mdframed}
\begin{mdframed}[frametitle={Renormalization of $\delta \alpha$ and $\delta \beta$: process-dependent scheme 2 (both schemes)},frametitlerule=true,frametitlebackgroundcolor=black!14,frametitlerulewidth=0.6pt,nobreak=true]\begin{align}
\delta \alpha &= \frac{Y_4}{Y_5} \bigg[ \mathcal{F}^\textrm{VC}_{h \tau \tau } + \frac{\delta g}{g} + \frac{\delta m_\tau }{m_\tau } - \frac{\delta m_W^2}{2m_W^2} + Y_6 \delta \beta + \frac{\delta Z_{hh}}{2} + \frac{Y_5}{Y_4} \frac{\delta Z_{Hh}}{2} + \frac{\delta Z^\textrm{L} _{\tau \tau}}{2} \\
&\hspace*{1.1cm} + \frac{\delta Z^\textrm{R} _{\tau \tau}}{2} \bigg]  \nonumber \\
\delta \beta &= \frac{- Y_6}{1+Y_6^2} \bigg[ \mathcal{F}^\textrm{VC}_{A \tau \tau } + \frac{\delta g}{g} + \frac{\delta m_\tau }{m_\tau } - \frac{\delta m_W^2}{2m_W^2} + \frac{\delta Z_{AA}}{2} - \frac{1}{Y_6} \frac{\delta Z_{G^0A}}{2} + \frac{\delta Z^\textrm{L} _{\tau \tau}}{2} + \frac{\delta Z^\textrm{R} _{\tau \tau}}{2} \bigg] 
\end{align}\end{mdframed}
\begin{mdframed}[frametitle={Renormalization of $\delta \alpha$ and $\delta \beta$: process-dependent scheme 3 (both schemes)},frametitlerule=true,frametitlebackgroundcolor=black!14,frametitlerulewidth=0.6pt,nobreak=true]\begin{align}
\delta \alpha &= \frac{Y_4 Y_5}{Y_4^2 + Y_5^2} \bigg[ \mathcal{F}^\textrm{VC}_{h \tau \tau } - \mathcal{F}^\textrm{VC}_{H \tau \tau } + \frac{\delta Z_{hh}}{2} - \frac{\delta Z_{HH}}{2} + \frac{Y_5}{Y_4} \frac{\delta Z_{Hh}}{2} - \frac{Y_4}{Y_5} \frac{\delta Z_{hH}}{2} \bigg]   \\
\delta \beta &= \frac{- 1}{Y_6(Y_4^2+Y_5^2)} \bigg[ (Y_4^2 + Y_5^2) \left( \frac{\delta g}{g} + \frac{\delta m_\tau }{m_\tau } - \frac{\delta m_W^2}{2m_W^2} + \frac{\delta Z^\textrm{L} _{\tau \tau}}{2} + \frac{\delta Z^\textrm{R} _{\tau \tau}}{2} \right) \\
&\hspace*{0.4cm}+ Y_4Y_5 \left( \frac{\delta Z_{Hh}}{2} + \frac{\delta Z_{hH}}{2} \right) + Y_4^2 \left( \frac{\delta Z_{hh}}{2} + \mathcal{F}^\textrm{VC}_{h \tau \tau } \right) + Y_5^2 \left( \frac{\delta Z_{HH}}{2} + \mathcal{F}^\textrm{VC}_{H \tau \tau } \right) \bigg]  \nonumber 
\end{align}\end{mdframed}
Note that for the process-dependent schemes, decays have to be chosen
that are experimentally accessible. This may not be the case for
certain parameter configurations, in which case the user has to
choose, if possible, the decay combination that leads to large enough 
decay widths to be measurable. \\

\textbf{Physical (on-shell) schemes.} In order to exploit the advantages of process-dependent schemes, \textit{i.e.}\,gauge independence of the mixing angle CTs that are defined within these schemes, while simultaneously avoiding possible drawbacks, \textit{e.g.}\,potentially large NLO corrections, the mixing angle CTs can be defined through certain observables or combinations of $S$ matrix elements in such a way that the CTs of all other parameters of the theory do not contribute to the mixing angle CTs. Such a scheme was proposed for the quark mixing within the SM in \cite{Denner:2004bm} and for the mixing angle CTs in the 2HDM in \cite{Denner2018}, where the derivation of the scheme is presented in detail. Here, we only recapitulate the key ideas and state the relevant formulae. For the sole purpose of renormalizing the mixing angles, two right-handed fermion singlets $\nu _{1\text{R}}$ and $\nu _{2\text{R}}$ are added to the 2HDM Lagrangian. An additional discrete $\mathbb{Z}_2$ symmetry is imposed under which the singlets transform as
\begin{align}
\nu _{1\text{R}} &\longrightarrow - \nu _{1\text{R}} \\
\nu _{2\text{R}} &\longrightarrow \nu _{2\text{R}} 
\end{align}
which prevents lepton generation mixing. The two singlets are coupled via Yukawa couplings $y_{\nu _1}$ and $y_{\nu _2}$ to two arbitrary left-handed lepton doublets of the 2HDM, giving rise to two massive Dirac neutrinos $\nu _1$ and $\nu _2$. The CT of the mixing angle $\alpha$ can then be defined by demanding that the ratio of the decay amplitudes of the decays $H \rightarrow \nu _i \bar{\nu} _i$ and $h \rightarrow \nu _i \bar{\nu} _i$ (for either $i=1$ or $i=2$) is the same at tree level and at NLO. Taking the ratio of the decay amplitudes has the advantage that other CTs apart from some WFRCs and the mixing angle CTs cancel against each other. For the CT of the mixing angle $\beta$, analogous conditions are imposed, involving additionally the decay of the pseudoscalar Higgs boson $A$ into the pair of massive neutrinos in the ratios of the LO and NLO decay amplitudes. In all cases, the mixing angle CTs are then given as functions of the scalar WFRCs as well as the genuine one-loop vertex corrections to the decays of the scalar particles into the pair of massive neutrinos, namely $\delta _{H\nu _i \bar{\nu} _i}$, $\delta _{h\nu _i \bar{\nu} _i}$ and $\delta _{A\nu _i \bar{\nu} _i}$, as given in Ref.\,\cite{Denner2018}. In this reference, three combinations of ratios of decay amplitudes were chosen to define three different renormalization schemes for the mixing angle CTs in the physical (on-shell) scheme:
\begin{itemize}
	\item ``OS1'' scheme: $\mathcal{A} _{H_1 \rightarrow \nu _1 \bar{\nu }_1} / \mathcal{A} _{H_2 \rightarrow \nu _1 \bar{\nu }_1} $ for $\delta \alpha$ and $\mathcal{A} _{A \rightarrow \nu _1 \bar{\nu }_1} / \mathcal{A} _{H_1 \rightarrow \nu _1 \bar{\nu }_1} $ for $\delta \beta$
	\item ``OS2'' scheme: $\mathcal{A} _{H_1 \rightarrow \nu _2 \bar{\nu }_2} / \mathcal{A} _{H_2 \rightarrow \nu _2 \bar{\nu }_2} $ for $\delta \alpha$ and $\mathcal{A} _{A \rightarrow \nu _2 \bar{\nu }_2} / \mathcal{A} _{H_1 \rightarrow \nu _2 \bar{\nu }_2} $ for $\delta \beta$
	\item ``OS12'' scheme: $\mathcal{A} _{H_1 \rightarrow \nu _2 \bar{\nu }_2} / \mathcal{A} _{H_2 \rightarrow \nu _2 \bar{\nu }_2} $ for $\delta \alpha$ and a specific combination of all possible decay amplitudes $\mathcal{A} _{H_i \rightarrow \nu _j \bar{\nu }_j}$ and $\mathcal{A} _{A \rightarrow \nu _j \bar{\nu }_j}$ ($i,j=1,2$) for $\delta \beta$ .
\end{itemize}
All three of these schemes were implemented in
{\texttt{2HDECAY}}\footnote{As for the process-dependent schemes
  before, the generic form of the CTs is valid for both the standard
  and alternative tadpole scheme, while the actual analytic
  expressions differ between the schemes. Since the full partial decay
  width is again independent of the tadpole scheme when using the
  physical (on-shell) scheme, we implemented these schemes in the
  alternative tadpole scheme,
  only.}\textsuperscript{,}\footnote{Note
      that the CTs of the physical on-shell schemes are defined in
      \cite{Denner2018} in the framework of the complex mass scheme
      \cite{Denner:2005fg, Denner:2006ic} while in {\texttt{2HDECAY}},
      we take the real parts of the self-energies through which these
      CTs are defined. These different definitions can lead to
      different finite parts in the one-loop partial decay
      widths. These differences are formally of
        next-to-next-to leading order.} according to the following definitions of the mixing angle CTs:
\begin{mdframed}[frametitle={ Renormalization of $\delta \alpha$ and $\delta \beta$: physical (on-shell) scheme OS1 (both schemes) },frametitlerule=true,frametitlebackgroundcolor=black!14,frametitlerulewidth=0.6pt,nobreak=true]\begin{align}
\delta \alpha &= s_\alpha c_\alpha \left( \delta _{H \nu _1 \bar{\nu} _1} - \delta _{h \nu _1 \bar{\nu} _1} \right) + s_\alpha c_\alpha \frac{\delta Z _{HH} - \delta Z_{hh}}{2} + \frac{c_\alpha ^2 \delta Z_{Hh} - s_\alpha ^2 \delta Z_{hH}}{2}  \\
\delta \beta &= t_\beta \Bigg[ c_\alpha ^2 \delta _{H\nu _1 \bar{\nu} _1} + s_\alpha ^2 \delta _{h\nu _1 \bar{\nu} _1} - \delta _{A \nu _1 \bar{\nu} _1} + \frac{ c_\alpha ^2 \delta Z_{HH} + s_\alpha ^2 \delta Z_{hh} -\delta Z_{AA}}{2} \\
&\hspace*{1.24cm} - s_\alpha c_\alpha \frac{\delta Z_{Hh} + \delta Z_{hH}}{2} \Bigg] + \frac{\delta Z_{G^0A}}{2}  \nonumber 
\end{align}\end{mdframed}
\begin{mdframed}[frametitle={ Renormalization of $\delta \alpha$ and $\delta \beta$: physical (on-shell) scheme OS2 (both schemes) },frametitlerule=true,frametitlebackgroundcolor=black!14,frametitlerulewidth=0.6pt,nobreak=true]\begin{align}
\delta \alpha &= s_\alpha c_\alpha \left( \delta _{h \nu _2 \bar{\nu} _2} - \delta _{H \nu _2 \bar{\nu} _2} \right) + s_\alpha c_\alpha \frac{\delta Z _{hh} - \delta Z_{HH}}{2} + \frac{s_\alpha ^2 \delta Z_{Hh} - c_\alpha ^2 \delta Z_{hH}}{2}  \\
\delta \beta &= \frac{1}{t_\beta } \Bigg[ \delta _{A \nu _2 \bar{\nu} _2} - s_\alpha ^2 \delta _{H\nu _2 \bar{\nu} _2} - c_\alpha ^2 \delta _{h\nu _2 \bar{\nu} _2}  + \frac{ \delta Z_{AA} - s_\alpha ^2 \delta Z_{HH} - c_\alpha ^2 \delta Z_{hh} }{2} \\
&\hspace*{1.24cm} - s_\alpha c_\alpha \frac{\delta Z_{Hh} + \delta Z_{hH}}{2} \Bigg] + \frac{\delta Z_{G^0A}}{2}  \nonumber 
\end{align}\end{mdframed}
\begin{mdframed}[frametitle={ Renormalization of $\delta \alpha$ and $\delta \beta$: physical (on-shell) scheme OS12 (both schemes) },frametitlerule=true,frametitlebackgroundcolor=black!14,frametitlerulewidth=0.6pt,nobreak=true]\begin{align}
\delta \alpha &= s_\alpha c_\alpha \left( \delta _{h \nu _2 \bar{\nu} _2} - \delta _{H \nu _2 \bar{\nu} _2} \right) + s_\alpha c_\alpha \frac{\delta Z _{hh} - \delta Z_{HH}}{2} + \frac{s_\alpha ^2 \delta Z_{Hh} - c_\alpha ^2 \delta Z_{hH}}{2}  \\
\delta \beta &= s_\beta c_\beta \left[ c_{2\alpha} \frac{\delta Z_{HH} - \delta Z_{hh}}{2} - s_{2\alpha} \frac{\delta Z_{Hh} + \delta Z_{hH}}{2} \right] + \frac{\delta Z_{G^0A}}{2} \\
&\hspace*{0.4cm} + s_\beta c_\beta \left[ \delta _{A\nu _2 \bar{\nu } _2} - \delta _{A\nu _1 \bar{\nu } _1} + c_\alpha ^2 \delta _{H\nu _1 \bar{\nu } _1} - s_\alpha ^2 \delta _{H\nu _2 \bar{\nu } _2} + s_\alpha ^2 \delta _{h\nu _1 \bar{\nu } _1} - c_\alpha ^2 \delta _{h\nu _2 \bar{\nu } _2} \right]  \nonumber 
\end{align}\end{mdframed}
For $y _{\nu _i} \rightarrow 0$ ($i=1,2$), the two Dirac neutrinos become massless again, the right-handed neutrino singlets decouple and the original 2HDM Lagrangian is recovered. The vertex corrections $\delta _{H\nu _i \bar{\nu} _i}$, $\delta _{h\nu _i \bar{\nu} _i}$ and $\delta _{A\nu _i \bar{\nu} _i}$ are non-vanishing in this limit, however, so that the mixing angle CTs can still be defined through these processes. The mixing angle CTs defined in these physical (on-shell) schemes are manifestly gauge-independent. \\

\textbf{Rigid symmetry scheme.} The renormalization of mixing matrix elements, \textit{e.g.}\,of $\alpha$ and $\beta$ for the scalar sector of the 2HDM, can be connected to the renormalization of the WFRCs by using the rigid symmetry of the Lagrangian. More specifically, it is possible to renormalize the fields and dimensionless parameters of the unbroken gauge theory and to connect the renormalization of the mixing matrix elements of \textit{e.g.}\,the scalar sector through a field rotation from the symmetric to the broken phase of the theory. Such a scheme was applied for the renormalization of the SM in \cite{Bohm:1986rj}. In \cite{Denner2018}, the scheme was applied to the scalar mixing angles of the 2HDM within the framework of the background field method (BFM) \cite{Zuber:1975sa,Zuber:1975gh,Boulware:1981,Abbott:1981,Abbott:1982,Hart:1983,Denner:1994xt}, which allows to formulate the mixing angle CTs as functions of the WFRCs $\delta Z_{\hat{H} \hat{h}}$ and $\delta Z_{\hat{h} \hat{H}}
$ in the alternative tadpole scheme, where the hat denotes that the fields are given in the BFM framework. These WFRCs differ from the ones used in the non-BFM framework, \textit{i.e.}\,$\delta Z_{Hh}$ and $\delta Z_{hH}
$ as given by Eqs.\,(\ref{RenormalizationScalarFieldsMassesExplicitFormWaveFunctionRenormalizationConstantH0h0Alt}) and (\ref{RenormalizationScalarFieldsMassesExplicitFormWaveFunctionRenormalizationConstanth0H0Alt}), by some additional term as presented in App.\,B of \cite{Denner2018} which coincides with the additional term given in \eqref{eq:RenormalizationScalarAnglesAdditionalTermCPEven} derived by means of the PT. The scalar self-energies involved in defining the mixing angle CTs are evaluated in a specifically chosen gauge, \textit{e.g.}\, the Feynman-'t Hooft gauge. This leads to the following definition of the CTs according to \cite{Denner2018} which is implemented in {\texttt{2HDECAY}},
\begin{mdframed}[frametitle={ Renormalization of $\delta \alpha$ and $\delta \beta$: BFMS scheme (alternative FJ scheme) },frametitlerule=true,frametitlebackgroundcolor=black!14,frametitlerulewidth=0.6pt,nobreak=true]\begin{align}
\delta \alpha &= \frac{\textrm{Re} \Big[ \left[ \Sigma ^\textrm{tad} _{Hh} (m_{H}^2) + \Sigma ^\textrm{tad} _{Hh} (m_{h}^2) \right] _{\xi = 1} + \Sigma ^\textrm{add} _{Hh} (m_{H}^2) + \Sigma ^\textrm{add} _{Hh} (m_{h}^2) \Big]}{2\left( m_{H}^2 - m_{h}^2\right) }  \\
\delta \beta &= \frac{s_{2\beta}}{s_{2\alpha}} \frac{\textrm{Re} \Big[ \left[ \Sigma ^\textrm{tad} _{Hh} (m_{h}^2) - \Sigma ^\textrm{tad} _{Hh} (m_{H}^2) \right] _{\xi = 1} + \Sigma ^\textrm{add} _{Hh} (m_{h}^2) - \Sigma ^\textrm{add} _{Hh} (m_{H}^2) \Big]}{2\left( m_{H}^2 - m_{h}^2\right) } \\
&\hspace*{0.4cm} + \frac{e}{2m_W \sqrt{ 1 - \frac{m_W^2}{m_Z^2} } } \left[ s_{\beta - \alpha} \frac{\delta T_H}{m_H^2} - c_{\beta - \alpha} \frac{\delta T_h}{m_h^2} \right] \nonumber 
\end{align}\end{mdframed}
where we replaced the BFM WFRCs with the corresponding self-energies and additional terms. As mentioned in \cite{Denner2018}, we want to remark that the definition of $\delta \alpha$ in the BFMS scheme coincides with the definition in the OS-pinched scheme of \eqref{eq:RenormalizationScalarAnglesDeltaAlphaOSPinchedResult}, while the definition of $\delta \beta$ in the BFMS scheme is different from the one in the OS-pinched scheme.

\subsubsection{Renormalization of the Fermion Sector}
\label{sec:renormalizationFermionSector}
The masses $m_f$, where $f$ generically stands for any fermion of the
2HDM, the CKM matrix elements $V_{ij}$ ($i,j=1,2,3$), the Yukawa coupling parameters
$Y_k$ ($k=1,...,6)$ and the fields of the fermion sector are replaced by the
renormalized quantities and the respective CTs and WFRCs as 
\begin{align}
m_f ~ &\rightarrow ~ m_f + \delta m_f  \\
V_{ij} ~&\rightarrow ~ V_{ij} + \delta V_{ij}  \\
Y_k ~& \rightarrow ~ Y_k + \delta Y_k  \\
f_i^L ~& \rightarrow ~ \left(\delta _{ij} + \frac{\delta Z_{ij}^{f,L} }{2} \right) f_j^L  \\
f_i^R ~& \rightarrow ~ \left(\delta _{ij} + \frac{\delta Z_{ij}^{f,R} }{2} \right) f_j^R  
\end{align}
where we use Einstein's sum convention in the last two lines. The
superscripts $L$ and $R$ denote the left- and right-chiral component
of the fermion fields, respectively. The Yukawa coupling parameters
$Y_i$ are not independent input parameters, but functions of $\alpha$ and
$\beta$, {\it cf.}~Tab.~\ref{tab:yukawaCouplings}. Their one-loop
counterterms are therefore given in terms of $\delta \alpha$
and $\delta \beta$ defined in
Sec.\,\ref{sec:renormalizationMixingAngles} by the following formulae
which are independent of the 2HDM type, 
\begin{align}
\delta Y_1 &= Y_1 \left( -\frac{Y_2}{Y_1} \delta \alpha + Y_3 \delta \beta \right)  \\
\delta Y_2 &= Y_2 \left( \frac{Y_1}{Y_2} \delta \alpha + Y_3 \delta \beta \right)  \\
\delta Y_3 &= \left( 1+ Y_3 ^2 \right) \delta \beta  \\
\delta Y_4 &= Y_4 \left( -\frac{Y_5}{Y_4} \delta \alpha + Y_6 \delta \beta \right)  \\
\delta Y_5 &= Y_5 \left( \frac{Y_4}{Y_5} \delta \alpha + Y_6 \delta \beta \right)  \\
\delta Y_6 &= \left( 1+ Y_6 ^2 \right) \delta \beta ~. 
\end{align}
Before presenting the renormalization conditions of the 
mass CTs and WFRCs, we shortly discuss the
renormalization of the CKM matrix. In \cite{Denner:1991kt} the
renormalization of the CKM matrix is connected to the renormalization
of the fields, which in turn are renormalized in an OS approach,
leading to the definition ($i,j,k=1,2,3$)
\begin{equation}
	\delta V_{ij} = \frac{1}{4} \left[ \left( \delta Z ^{u,L} _{ik} - \delta Z ^{u,L \dagger} _{ik} \right) V_{kj} - V_{ik} \left( \delta Z ^{d,L} _{kj} - \delta Z ^{d,L \dagger} _{kj} \right)  \right]
\label{eq:CKMCTdefinition}
\end{equation}
where the superscripts $u$ and $d$ denote up-type and down-type
quarks, respectively. This definition of the CKM matrix CTs leads to 
uncanceled explicit gauge dependences when used in the calculation of
EW one-loop corrections, however, \cite{Gambino:1998ec,
  Barroso:2000is, Kniehl:2000rb, Pilaftsis:2002nc, Yamada:2001px,
  Diener:2001qt}. Since the CKM matrix is approximately a unit matrix
\cite{1674-1137-38-9-090001}, the 
numerical effect of this gauge dependence is typically very small, but
the definition nevertheless introduces uncanceled explicit gauge dependences
into the partial decay widths, which should be avoided. In our work,
we follow the approach of 
Ref.~\cite{Yamada:2001px} and use pinched fermion self-energies for the
definition of the CKM matrix CT. An analytic analysis shows that this
is equivalent with defining the CTs in \eqref{eq:CKMCTdefinition} in
the Feynman-'t Hooft gauge. 

Apart from the CKM matrix CT, all other CTs of the fermion sector are
defined through OS conditions. The resulting forms of the CTs are
analogous to the ones presented in \cite{Denner:1991kt} and given by 
\begin{mdframed}[frametitle={Renormalization of the fermion sector (standard scheme)},frametitlerule=true,frametitlebackgroundcolor=black!14,frametitlerulewidth=0.6pt,nobreak=true]\begin{align}
	\delta m_{f, i} &= \frac{m_{f,i}}{2} \text{Re} \left( \Sigma _{ii}^{f,L} (m_{f,i}^2) + \Sigma _{ii}^{f,R} (m_{f,i}^2) + 2\Sigma _{ii}^{f,S} (m_{f,i}^2) \right)  \\
	\delta Z^{f,L}_{ij} &= \frac{2}{m_{f,i}^2 - m_{f,j}^2} \text{Re} \bigg[ m_{f,j}^2 \Sigma _{ij}^{f,L} (m_{f,j}^2) + m_{f,i} m_{f,j} \Sigma _{ij} ^{f,R} (m_{f,j}^2) \\
	&\hspace*{3.2cm} + (m_{f,i}^2 + m_{f,j}^2) \Sigma _{ij}^{f,S} (m_{f,j}^2) \bigg] ~~~~~~ (i\neq j)  \nonumber \\
	\delta Z^{f,R}_{ij} &= \frac{2}{m_{f,i}^2 - m_{f,j}^2} \text{Re} \bigg[ m_{f,j}^2 \Sigma _{ij}^{f,R} (m_{f,j}^2) + m_{f,i} m_{f,j} \Sigma _{ij} ^{f,L} (m_{f,j}^2)  \\
	&\hspace*{3.2cm} + 2 m_{f,i}m_{f,j} \Sigma _{ij}^{f,S} (m_{f,j}^2) \bigg] ~~~~~~ (i\neq j)   \nonumber 
\end{align}\end{mdframed} 
\begin{mdframed}[frametitle={Renormalization of the fermion sector (alternative FJ scheme)},frametitlerule=true,frametitlebackgroundcolor=black!14,frametitlerulewidth=0.6pt,nobreak=true]\begin{align}
\delta m_{f, i} &= \frac{m_{f,i}}{2} \text{Re} \left( \Sigma _{ii}^{f,L} (m_{f,i}^2) + \Sigma _{ii}^{f,R} (m_{f,i}^2) + 2\Sigma _{ii}^{\text{tad},f,S} (m_{f,i}^2) \right)  \\
\delta Z^{f,L}_{ij} &= \frac{2}{m_{f,i}^2 - m_{f,j}^2} \text{Re} \bigg[ m_{f,j}^2 \Sigma _{ij}^{f,L} (m_{f,j}^2) + m_{f,i} m_{f,j} \Sigma _{ij} ^{f,R} (m_{f,j}^2) \\
&\hspace*{3.2cm} + (m_{f,i}^2 + m_{f,j}^2) \Sigma _{ij}^{\text{tad},f,S} (m_{f,j}^2) \bigg] ~~~~~~ (i\neq j)  \nonumber \\
\delta Z^{f,R}_{ij} &= \frac{2}{m_{f,i}^2 - m_{f,j}^2} \text{Re} \bigg[ m_{f,j}^2 \Sigma _{ij}^{f,R} (m_{f,j}^2) + m_{f,i} m_{f,j} \Sigma _{ij} ^{f,L} (m_{f,j}^2) \\
&\hspace*{3.2cm} + 2 m_{f,i}m_{f,j} \Sigma _{ij}^{\text{tad},f,S} (m_{f,j}^2) \bigg] ~~~~~~ (i\neq j)  \nonumber  \label{RenormalizationFermionSectorExplicitFormMassCountertermTauAlternativeTadpoleScheme}
\end{align}\end{mdframed}
\begin{mdframed}[frametitle={Renormalization of the fermion sector
    (standard and alternative FJ scheme)},frametitlerule=true,frametitlebackgroundcolor=black!14,frametitlerulewidth=0.6pt,nobreak=true]\begin{align}
\delta V_{ij} &= \frac{1}{4} \left[ \left( \delta Z ^{u,L} _{ik} - \delta Z ^{u,L \dagger} _{ik} \right) V_{kj} - V_{ik} \left( \delta Z ^{d,L} _{kj} - \delta Z ^{d,L \dagger} _{kj} \right)  \right] _{\xi = 1}  \\
\delta Z^{f,L} _{ii} &= - \textrm{Re} \Big[ \Sigma ^{f,L} _{ii} (m_{f,i} ^2) \Big] - m_{f,i} ^2 \textrm{Re} \left[ \frac{\partial \Sigma ^{f,L} _{ii} (p ^2)}{\partial p^2} + \frac{\partial \Sigma ^{f,R} _{ii} (p ^2)}{\partial p^2} + 2\frac{\partial \Sigma ^{f,S} _{ii} (p ^2)}{\partial p^2} \right] _{p^2 = m_{f,i} ^2}   \raisetag{2.2\baselineskip} \\
\delta Z^{f,R} _{ii} &= - \textrm{Re} \Big[ \Sigma ^{f,R} _{ii} (m_{f,i} ^2) \Big] - m_{f,i} ^2 \textrm{Re} \left[ \frac{\partial \Sigma ^{f,L} _{ii} (p ^2)}{\partial p^2} + \frac{\partial \Sigma ^{f,R} _{ii} (p ^2)}{\partial p^2} + 2\frac{\partial \Sigma ^{f,S} _{ii} (p ^2)}{\partial p^2} \right] _{p^2 = m_{f,i} ^2}   \raisetag{2.2\baselineskip}
\end{align}\end{mdframed}
where as before, the superscripts $L$ and $R$ denote the left- and
right-chiral parts of the self-energies, while the superscript $S$
denotes the scalar part. 

\subsubsection{Renormalization of the Soft-$\mathbb{Z}_2$-Breaking Parameter $m_{12}^2$} 
\label{sec:renormalizationSoftm12Squared}
The last remaining parameter of the 2HDM that needs to be renormalized
is the soft-$\mathbb{Z}_2$-breaking parameter $m_{12}^2$. As before,
we replace the bare parameter by the renormalized one and its
corresponding CT, 
\begin{equation}
m_{12}^2 ~\rightarrow ~ m_{12}^2 + \delta m_{12}^2 ~.
\end{equation}
In order to fix $\delta m_{12}^2$ in a physical way, one could use a
process-dependent scheme analogous to
Sec.\,\ref{sec:renormalizationMixingAngles} for the scalar mixing
angles. Since $m_{12}^2$ only appears in trilinear Higgs couplings,  a
Higgs-to-Higgs decay width would have to be chosen as observable that
fixes the CT. However, as discussed in \cite{Krause:2016xku}, a
process-dependent definition of $\delta m_{12}^2$ can lead to very
large one-loop corrections in Higgs-to-Higgs decays. We therefore 
employ an $\overline{\text{MS}}$ condition in {\texttt{2HDECAY}} to
fix the CT. This is done by calculating the 
off-shell decay process $h \rightarrow hh$ at one-loop order and by
extracting all UV-divergent terms. This fixes the CT of $m_{12}^2$ to 
\begin{mdframed}[frametitle={Renormalization of $m_{12}^2$ (standard
    and alternative FJ scheme)},frametitlerule=true,frametitlebackgroundcolor=black!14,frametitlerulewidth=0.6pt,nobreak=true]
\begin{align}
\delta m_{12}^2 &= \frac{\alpha _\text{em} m_{12}^2 }{16\pi m_W^2 \left( 1 - \frac{m_W^2}{m_Z^2} \right)} \Big[ \frac{8m_{12}^2}{s_{2\beta }} - 2m_{H^\pm }^2 - m_A^2 + \frac{s_{2\alpha }}{s_{2\beta }} (m_H^2 - m_h^2) - 3(2m_W^2 + m_Z^2) \nonumber  \\
&\hspace*{0.3cm} + \sum _u 3 m_u^2 \frac{1}{s_\beta ^2} - \sum _d 6 m_d^2  Y_3 \left( -Y_3 - \frac{1}{t_{2\beta}} \right) - \sum _l 2 m_l^2  Y_6 \left( -Y_6 - \frac{1}{t_{2\beta}} \right) \Big] \Delta   \label{eq:renormalizationConditionm12Sq}
\end{align}\end{mdframed}
where the sum indices $u$, $d$ and $l$ indicate a summation over all
up-type and down-type quarks and charged leptons,
respectively, and 
\begin{equation}
	\Delta \equiv \frac{1}{\varepsilon } - \gamma _E + \ln (4\pi ) + \ln \left( \frac{\mu ^2}{\mu _R^2} \right) ~.
\end{equation}
Here, $\gamma _E$ is the Euler-Mascheroni
constant, $\varepsilon$ the dimensional shift when switching
from 4 physical to $D=4-2\varepsilon$ space-time dimensions in the framework of
dimensional regularization
\cite{Wilson1971,Wilson1972,Ashmore1972,Bollini:1972ui,THOOFT1972189} and $\mu$ is the
mass-dimensional 't Hooft scale which cancels in the calculation of
the decay amplitudes.
The result in \eqref{eq:renormalizationConditionm12Sq} is in
agreement with the formula presented in \cite{Kanemura:2015mxa}. 
Since $m_{12}^2$ is $\overline{\text{MS}}$
  renormalized, the user has to specify in the input file the scale at
which the parameter is understood to be
given.\footnote{All $\overline{\text{MS}}$ input
    parameters are understood to be given at the same scale so that in
  the input file there is only one entry for its specification.} Just as for the
$\overline{\text{MS}}$ renormalized mixing angles, the automatic
parameter conversion routine adapts $m_{12}^2$ to the scale at which
the EW one-loop corrected decay widths are evaluated in case the two
scales differ.

\subsection{Electroweak Decay Processes at LO and NLO}
\label{sec:decayProcessesAtLOandNLO}
\begin{figure}[tb]
\centering
\includegraphics[width=12.5cm, trim=0cm 0cm 0cm 0.8cm, clip]{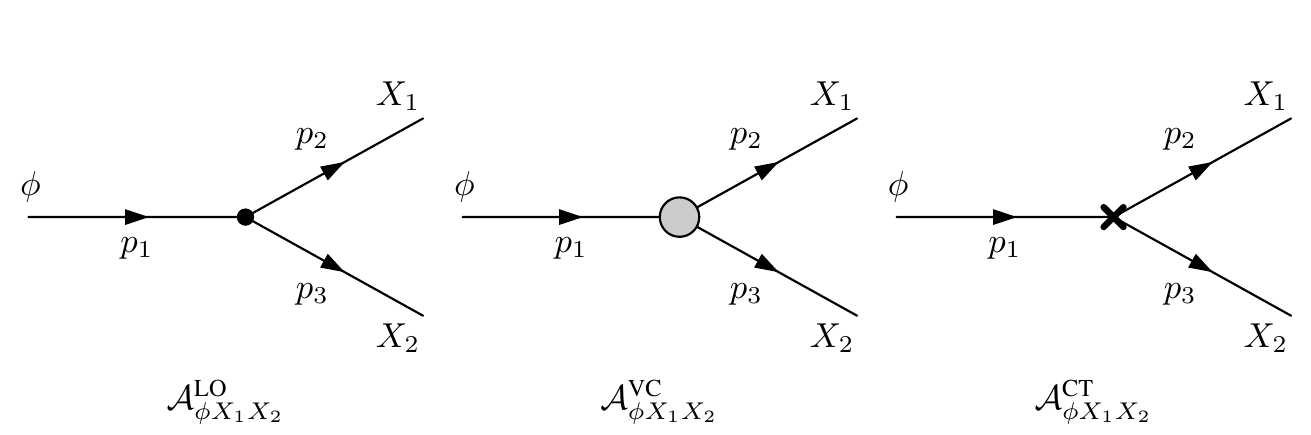}
    \caption{Decay amplitudes at LO and NLO. The LO decay amplitude
      $\mathcal{A}^\text{LO} _{\phi X _1 X _2}$ simply
      consists of the trilinear coupling of the three particles $\phi
      _1$, $X _1$ and $X _2$, while the one-loop amplitude is 
      given by the sum of the genuine vertex corrections
      $\mathcal{A}^\text{VC} _{\phi X _1 X _2}$, indicated by
      a grey blob, and the vertex counterterm $\mathcal{A}^\text{CT}
      _{\phi X _1 X _2}$ which also includes all WFRCs
      necessary to render the NLO amplitude UV-finite. We do not show
      corrections on the external legs since in the decays we
      consider, they vanish either due to OS renormalization
      conditions or due to Slavnov-Taylor identities. In the case of
      the alternative tadpole scheme, the vertex corrections
      $\mathcal{A}^\text{VC} _{\phi X _1 X _2}$ also in
      general contain tadpole diagrams.} 
\label{fig:decayHiggsParticles}
\end{figure}

Figure~\ref{fig:decayHiggsParticles} shows the topologies that
contribute to the tree-level and one-loop corrected decay of a scalar
particle $\phi$ with four-momentum $p_1$ into two other particles
$X _1$ and $X _2$ with four-momenta $p_2$ and $p_3$,
respectively. We emphasize that for the EW corrections, we restrict ourselves to
OS decays, {\it i.e.}~we demand 
\begin{equation}
p _1^2 \ge (p_2 + p_3)^2 
\end{equation}
with $p_i^2 = m_i^2$ ($i=1,2,3$) where $m_i$ denote the masses of the
three particles. Moreover, we do not calculate EW corrections to
loop-induced Higgs decays, which are of two-loop order. In particular,
we do not provide EW corrections to Higgs boson decays into two-gluon,
two-photon or $Z\gamma$ final states. 
Note, however, that the decay widths implemented in
{\texttt{HDECAY}} include also loop-induced decay widths as well as
off-shell decays into heavy-quark, 
massive gauge boson, neutral Higgs pair as well as Higgs and gauge boson final
states. We come back to this point in Sec.\,\ref{sec:connectionHDECAY}.   
The LO and NLO partial decay widths were calculated by first
generating all Feynman diagrams and the corresponding amplitudes for
all decay modes that exist for the 2HDM, shown topologically in
\figref{fig:decayHiggsParticles}, with help of the tool
{\texttt{FeynArts 3.9}} \cite{Hahn:2000kx}. To that end, we used the
2HDM model file that is implemented in {\texttt{FeynArts}}, but
modified the Yukawa couplings to implement all four 2HDM
types. Diagrams that account for NLO corrections on the external legs
were not calculated since for all decay modes that we considered, they
either vanish due to OS renormalization conditions or due to
Slavnov-Taylor identities. All amplitudes were then calculated
analytically with {\texttt{FeynCalc 8.2.0}} \cite{MERTIG1991345,
  Shtabovenko:2016sxi}, together with all self-energy amplitudes
needed for the CTs. For the numerical evaluation of all loop integrals involved in the analytic expression of the one-loop amplitudes, {\texttt{2HDECAY}} links {\texttt{LoopTools 2.14}} \cite{HAHN1999153}.

The LO partial decay width is obtained from the LO amplitude
$\mathcal{A}^\text{LO} _{\phi X _1 X _2}$, while the NLO
amplitude is given by the sum of all amplitudes stemming from 
the vertex correction and the necessary CTs as defined in
Sec.\,\ref{sec:renormalization2HDM}, 
\begin{equation}
\mathcal{A}^\text{1loop} _{\phi X _1 X _2} \equiv \mathcal{A}^\text{VC} _{\phi X _1 X _2} + \mathcal{A}^\text{CT} _{\phi X _1 X _2} ~.
\end{equation}
By introducing the K\"{a}ll\'en phase space function 
\begin{equation}
	\lambda (x,y,z) \equiv \sqrt{x^2 + y^2 + z^2 - 2xy - 2xz - 2yz} 
\end{equation}
the LO and NLO partial decay widths can be cast into the form
\begin{align}
\Gamma ^\text{LO} _{\phi X _1 X _2} &= S \frac{\lambda (m_1^2 , m_2^2 , m_3^2 )}{16\pi m_1^3} \sum _\text{d.o.f.} \left| \mathcal{A} _{\phi X _1 X _2}^\text{LO} \right| ^2  \label{eq:decayWidthLO} \\
\Gamma ^\text{NLO} _{\phi X _1 X _2} &= \Gamma ^\text{LO} _{\phi X _1 X _2} + S \frac{\lambda (m_1^2 , m_2^2 , m_3^2 )}{8\pi m_1^3} \sum _\text{d.o.f.} \text{Re} \left[ \left( \mathcal{A} _{\phi X _1 X _2}^\text{LO} \right) ^{*} \mathcal{A} _{\phi X _1 X _2}^\text{1loop} \right] + \Gamma _{\phi X _1 X _2 + \gamma}  \label{eq:decayWidthNLO}
\end{align}
where the symmetry factor $S$ accounts for identical particles in the
final state and the sum extends over all degrees of freedom of the
final-state particles, {\it i.e.}~over spins or polarizations. The partial
decay width $\Gamma _{\phi X _1 X _2 + \gamma}$ accounts for
real corrections that are necessary for removing IR divergences in all
decays that involve charged particles in the initial or final
state. For this, we implemented the results given in
\cite{Goodsell:2017pdq} for generic one-loop two-body partial decay
widths. Since the involved integrals are analytically
solvable for two-body decays \cite{Denner:1991kt}, 
the IR corrections that are implemented in {\texttt{2HDECAY}} are
given in analytic form as well and do not require numerical
integration. Additionally, since the implemented integrals account for
the full phase-space of the radiated photon, {\it i.e.} both the ``hard''
and ``soft'' parts, our results do not depend on arbitrary cuts in the
photon phase-space. 

In the following, we present all decay channels for which the
EW corrections were calculated at one-loop order: 
\begin{itemize}
	\item $h/H/A \to f\bar{f}$ ~ ($f=u,d,c,s,t,b,e, \mu ,\tau $) 
	\item $h/H \to VV$ ~ ($V=W^\pm ,Z$) 
	\item $h/H \to VS$ ~ ($V=Z, W^\pm$, $S=A, H^\pm$) 
	\item $h/H \to SS$ ~ ($S = A, H^\pm$) 
	\item $H \to hh$ 
	\item $H^\pm \to VS$ ~ ($V=W^\pm$, $S=h,H,A$) 
	\item $H^+ \to f\bar{f}$ ~ ($f=u,c,t, \nu _e , \nu _\mu ,
          \nu _\tau $ , $\bar{f} = \bar{d}, \bar{s}, \bar{b}, e^+ ,
          \mu ^+ , \tau ^+ $)  
	\item $A \to VS$ ~ ($V=Z,W^\pm$, $S=h,H,H^\pm$) 
\end{itemize}
All analytic results of these decay processes are stored in
subdirectories of {\texttt{2HDECAY}}. For a consistent connection with
{\texttt{HDECAY}}, {\it cf.}\,also Sec.\,\ref{sec:connectionHDECAY}, not all
of these decay processes are used for the calculation of the decay
widths and branching ratios, however. Decays containing pairs of
first-generation fermions are neglected, {\it i.e.}\,in {\texttt{2HDECAY}},
the EW corrections of the following processes are not used
for the calculation of the partial decay widths and branching ratios: 
$h/H/A \to f\bar{f}$ ($f=u,d,e $) and $H^+ \to f\bar{f}$
($f\bar{f}=u\bar{d}, \nu _e e^+  $). The reason is
  that they are overwhelmed by the Dalitz decays $\Phi \to f\bar{f}^{(')}
  \gamma$ ($\Phi=h,H,A,H^\pm$) that are induced {\it
    e.g.}~by off-shell $\gamma^* \to f\bar{f}$ splitting.

\subsection{Link to HDECAY, Calculated Higher-Order Corrections and Caveats}
\label{sec:connectionHDECAY}
The EW one-loop corrections to the Higgs decays in the 2HDM derived in
this work are combined with 
{\texttt{HDECAY}} version 6.52 \cite{DJOUADI199856,
  Djouadi:2018xqq}\footnote{The program code for
    {\texttt{HDECAY}} can be downloaded from the URL
    \href{http://tiger.web.psi.ch/hdecay/}{http://tiger.web.psi.ch/hdecay/}.}
  in form of the new tool {\texttt{2HDECAY}}. The Fortran code {\texttt{HDECAY}}
provides the LO and QCD corrected decay widths. 
As outlined in 
Sec.\,\ref{sec:renormalizationGaugeSector} the EW corrections use 
$\alpha _\text{em}$ at the $Z$ boson mass scale as input parameter
instead of $G_F$ as used in {\tt HDECAY}. For a
  consistent combination of the EW corrected decay widths with the {\tt HDECAY}
  implementation in the $G_F$ scheme we would have to convert between
  the $\{\alpha_{\text{em}}, m_W, m_Z\}$ and the $\{ G_F, m_W, m_Z\}$
  scheme including 2HDM higher-order corrections in the conversion
  formulae. Since these conversion 
  formulae are not implemented yet, we chose a
  pragmatic approximate solution: 
%

In the configuration of {\texttt{2HDECAY}} with {\texttt{OMIT ELW2=0}}
being set ({\it cf.}~the input file format described in
Sec.\,\ref{sec:InputFileFormat}), the EW corrections to the decay
widths are calculated automatically. This setting
  also overwrites the value that the user chooses for the input
  {\texttt{2HDM}}. If {\it e.g.}~the user does not choose the 2HDM by
  setting {\texttt{2HDM=0}} but at the same time chooses {\texttt{OMIT
      ELW2=0}} in order 
  to calculate the EW corrections, then a warning is printed and
  {\texttt{2HDM=1}} is automatically set internally. In
  this configuration, the value of 
$G_F$ given in the input file of {\texttt{2HDECAY}} is ignored by the
part of the program that calculates the EW
corrections. Instead, $\alpha _\text{em} (m_Z^2)$, given in line 26 of
the input file, is taken as independent input. This $\alpha _\text{em}
(m_Z^2)$ is used for the calculation of all electroweak
corrections. Subsequently, for the consistent combination with the
decay widths of {\texttt{HDECAY}} computed in terms of the Fermi
constant $G_F$, the latter decay widths are adapted to the input
scheme of the EW corrections by rescaling the {\texttt{HDECAY}} decay
widths with $G_F^\text{calc}/G_F$, where $G_F^\text{calc}$ 
is calculated by means of the tree-level relation
\eqref{eq:definitionFermiConstant} as a function of $\alpha_\text{em}
(m_Z^2)$. We expect the differences between the observables within
these two schemes to be small.  

On the other hand, if {\texttt{OMIT ELW2=1}} is set, no 
EW corrections are computed and {\texttt{2HDECAY}}
reduces to the original program code {\texttt{HDECAY}}, including
(where applicable)
the QCD corrections in the decay widths, the off-shell decays and the
loop-induced decays. In this case, the value of
$G_F$ given in line 27 of the input file is used as input parameter
instead of being calculated through the input value of $\alpha
_\text{em} (m_Z^2)$, and no rescaling with $G_F^\text{calc}$ is
performed. We note in particular that therefore the QCD corrected
decay widths, printed out separately by {\texttt{2HDECAY}}, will be
different in the two input options {\texttt{OMIT ELW2=0}} and {\texttt{OMIT ELW2=1}}.

Another comment is at order in view of the fact that we
implemented EW corrections to OS decays only, while
{\texttt{HDECAY}} also features the computation of
off-shell decays. 
More specifically, {\texttt{HDECAY}} includes off-shell decays into
final states with an off-shell top-quark $t^*$, {\it i.e.}~$\phi \to t^*
\bar{t}$ ($\phi=h,H,A$), $H^+ \to t^* + \bar{d},\bar{s},\bar{b}$, into gauge and Higgs
boson final states with an off-shell gauge boson, $h/H \to Z^* A, A
\to Z^* h/H, \phi \to H^- W^{+*}, H^+ \to \phi W^{+*}$, and into
neutral Higgs pairs with one off-shell Higgs boson that is assumed to
predominantly decay into the $b\bar{b}$ final state, $h/H \to AA^*$,
$H \to hh^*$. The top quark total width within the 2HDM, required for the off-shell
decays with top final states, is calculated internally in {\tt HDECAY}.
In {\texttt{2HDECAY}}, we combine the EW 
and QCD corrections in such a way that {\texttt{HDECAY}} still
computes the decay widths of off-shell decays, while the electroweak corrections
are added only to OS decay channels. It
is important to keep this restriction in mind when performing the
calculation for large varieties of input data. If {\it e.g.}~the lighter
Higgs boson $h$ is chosen to be the SM-like Higgs boson, then the OS
decay $h \rightarrow 
W^+ W^-$ would be kinematically forbidden while the heavier Higgs boson 
decay $H \rightarrow W^+ W^-$ might be OS. In such
cases, {\texttt{2HDECAY}} calculates the EW NLO corrections only for the latter decay
channel, while the LO (and QCD decay widths where applicable) are calculated
for both. The same is true for any other decay channel for which we 
implemented EW corrections but which are off-shell in certain
input scenarios. Note, that the NLO EW corrections for the off-shell decays 
into the massive gauge boson final states have been provided for the
2HDM in \cite{Altenkamp:2017ldc,Denner2018,Altenkamp:2017kxk}. For
the SM, the combination of {\texttt{HDECAY}} and {\texttt{Prophecy4f}}
\cite{Bredenstein:2006rh,Bredenstein:2006nk,Bredenstein:2006ha} 
provides the decay widths including EW corrected off-shell decays into
these final states. In a similar way, a combination of {\texttt{2HDECAY}} and 
{\texttt{Prophecy4f}} with the 2HDM decays may be envisaged in future.

For the combination of the QCD and EW corrections finally, we assume
that these corrections factorize. We denote by $\delta^{\text{QCD}}$
and $\delta^{\text{EW}}$ the relative QCD and EW corrections,
respectively. Here $\delta^{\text{QCD}}$ is normalized to the LO width
$\Gamma^{\text{HD,LO}}$, 
calculated internally by {\tt HDECAY}.  This means for example in the
case of quark pair final states that the LO width includes the running
quark mass in order to improve the perturbative
behaviour. 
The relative EW corrections $\delta^{\text{EW}}$ on the other hand are obtained by
normalization to the LO width with on-shell particle masses. With these
definitions the QCD and EW corrected decay width into a specific final
state, $\Gamma^{\text{QCD\&EW}}$, is given by
\beq
\Gamma^{\text{QCD\&EW}} = \frac{G_F^{\text{calc}}}{G_F} \Gamma^{\text{HD,LO}} 
[1+\delta^{\text{QCD}}] [1+ + \delta^{\text{EW}}] 
\equiv \frac{G_F^\text{calc}}{G_F}
\Gamma^{\text{HD,QCD}} 
[1 + \delta^{\text{EW}}]  \;.
\eeq
We have included the rescaling factor $G_F^\text{calc}/G_F$ which is
necessary for the consistent connection of our EW corrections with the
decay widths obtained from {\tt HDECAY}, as outline above.

\underline{QCD\&EW-corrected branching ratios:}
The program code will provide the branching ratios calculated
originally by {\tt HDECAY}, which, however, for {\tt OMIT ELW2=0} are
rescaled by $G_F^\text{calc}/G_F$. They include all loop decays,
off-shell decays and QCD corrections where applicable. We summarize
these branching ratios under the name 'QCD-corrected' branching ratios
and call their associated decay widths $\Gamma^{\text{HD,QCD}}$,
keeping in mind that the QCD corrections are included only where
applicable. 
Furthermore, the EW and QCD corrected branching ratios will be given
out. Here, we add the EW corrections to the decay widths calculated
internally by {\tt HDECAY} where possible, {\it i.e.}~for non-loop
induced and OS decay widths. We summarize these branching
ratios under the name 'QCD\&EW-corrected' branching ratios and call
their associated decay widths $\Gamma^{\text{QCD\&EW}}$. In
Table~\ref{tab:brs} we summarize all details and caveats on their
calculation that we described here above. All these branching ratios
are written to the output file carrying the suffix '\_BR' with its
filename, see also end of section~\ref{sec:InputFileFormat} for details.

\begin{table}[tb]
\centering
  \begin{tabular}{ c c c  }
    \hline
    {\tt IELW2=0} & QCD-corrected & QCD\&EW-corrected \\ \hline 
    on-shell and &  $\Gamma^{\text{HD,QCD}} \frac{G_F^\text{calc}}{G_F}$ & 
    $\Gamma^{\text{HD,QCD}} [1+\delta^{\text{EW}}] \frac{G_F^\text{calc}}{G_F}$ \\
    non-loop induced &  & \\ \hline
    off-shell or & $\Gamma^{\text{HD,QCD}}
    \frac{G_F^\text{calc}}{G_F}$ & $\Gamma^{\text{HD,QCD}} \frac{G_F^\text{calc}}{G_F}$  \\
    loop-induced & & \\ \hline
  \end{tabular}
    \caption{The QCD-corrected and the QCD\&EW-corrected decay widths
      as calculated by {\tt 2HDECAY} for {\tt IELW2=0}. The label QCD
      is in the sense that the QCD corrections are included where applicable.}
   \label{tab:brs}
\end{table}

\underline{NLO EW-corrected decay widths:} For ${\tt IELW2=0}$,
we additionally give out the LO and the EW-corrected NLO decay widths
as calculated by the new addition to {\tt HDECAY}. Here the LO widths
do not include any running of the quark masses in the case of quark
final states, but are obtained for OS masses. They can hence
differ quite substantially from the LO widths as calculated in the
original {\tt HDECAY} version. These LO and EW-corrected NLO widths are
computed in the $\{\alpha_{\text{em}}, m_W, m_Z\}$ scheme and therefore
obviously do not need the inclusion of the rescaling factor
$G_F^\text{calc}/G_F$. The 
decay widths are written to the output file carrying the suffix
'\_EW' with its filename.  While the widths given out here are not meant
to be applied in Higgs observables as they do not include the
important QCD corrections, the study of the NLO EW-corrected decay
widths for various renormalization schemes, as provided by {\tt 2HDECAY}, allows to
analyze the importance of the EW corrections and estimate the
remaining theoretical error due to missing higher-order EW
corrections. The decay widths can also be used for phenomenological studies like
{\it e.g.}~the comparison with the EW-corrected decay widths in the
MSSM in the limit of large supersymmetric particle masses, or the
investigation of specific 2HDM parameter regions at LO and NLO as {\it e.g.}~the
alignment limit, the non-decoupling limit or the wrong-sign limit.

\underline{Caveats:} We would like to point out to
  the user that it can happen that the EW-corrected decay widths become
  negative because of too large negative EW corrections compared to
  the LO width. There can be several reasons for this: $(i)$ The LO width
  may be very small in parts of the parameter space due to 
  suppressed couplings. For example the decay of the heavy Higgs boson
  $H$ into massive vector bosons is very small in the region where the lighter $h$
  becomes SM-like and takes over almost the whole coupling to massive
  gauge bosons. If the NLO EW width is not suppressed by the same
  power of the relevant coupling or if at NLO there are cancellations
  between the various terms that remove the
  suppression, the NLO width can largely exceed the LO width. $(ii)$ 
The EW corrections are artificially enhanced due to a badly chosen renormalization
scheme, {\it
  cf.}~Refs.~\cite{Krause:2016oke,Krause:2016xku,Krause:2017mal} for
investigations on this 
subject. The choice of a different renormalization scheme may cure
this problem, but of course raises also the question for the remaining
theoretical error due to missing higher-order corrections. 
$(iii)$ The EW corrections are parametrically enhanced due to
involved couplings that are large, because of small coupling
parameters in the denominator or due to light particles in the loop, see also
Refs.~\cite{Krause:2016oke,Krause:2016xku,Krause:2017mal} for discussions. This
would call for the resummation of EW corrections 
beyond NLO to improve the behaviour. It is obvious that the EW
corrections should not be trusted in case of extremely large positive
or negative corrections and rather be discarded, in particular in the
comparison with experimental observables, unless some of the suggested
measures are taken to improve the behaviour.
    
\subsection{Parameter Conversion}
\label{sec:ParameterConversion}
Through the higher-order corrections the decay widths depend on the
renormalization scale. In {\texttt{2HDECAY}} the user can choose this
scale, called $\mu_{\text{out}}$ in the following, in the input file. It can either
chosen to be a fixed scale or the mass of the decaying Higgs
boson. Input parameters in the $\overline{\mbox{MS}}$ scheme depend
explicitly on the renormalization scale $\mu_R$. This scale also has to be
given by the user in the input scale, and is called $\mu_R$ in the
following. The value of the scale becomes particularly important when the values of
$\mu_R$ and $\mu_{\text{out}}$ differ. In this case the $\overline{\mbox{MS}}$
parameters have to be evolved from the scale $\mu_R$ to the scale $\mu_{\text{out}}$. This
applies for $m_{12}^2$ which is always understood to be an
$\overline{\mbox{MS}}$ parameter, and for $\alpha$ and $\beta$ in case
they are chosen to be $\overline{\mbox{MS}}$ renormalized.
{\texttt{2HDECAY}} internally converts the $\overline{\mbox{MS}}$
parameters from $\mu_R$ to $\mu_{\text{out}}$ by means of a linear
approximation, applying the formula 
\begin{equation}
	\varphi  \left( \{ \mu _\text{out} \} \right) \approx \varphi
        \left( \{ \mu _R \} \right) + \ln \left( \frac{\mu _\text{out}
            ^2}{\mu _R ^2} \right) \delta \varphi^\text{div} \left( \{
          \varphi \} \right)  \label{eq:scalechange}
\end{equation}
where $\varphi$ and $\delta \varphi$ denote the $\overline{\mbox{MS}}$
parameters ($\alpha$ and $\beta$, if chosen as such, $m_{12}^2$) and
their respective counterterms. The index 'div' means that only the
divergent part of the counterterm, {\it i.e.}~the terms proportional
to $1/\varepsilon$ (or equivalently $\Delta$), is taken. \s 

In addition, a parameter conversion has to be performed, when the
chosen renormalization scheme of the input parameter differs from the
renormalization scheme at which the EW corrected decay widths are
chosen to be evaluated. {\texttt{2HDECAY}} performs this conversion
automatically which is necessary for a consistent interpretation of
the results. The renormalization schemes implemented in {\texttt{2HDECAY}} differ solely in their definition of the scalar mixing angle CTs, while the defnition of all
other CTs is fixed. Therefore, the values of $\alpha$ and $\beta$ must
be converted when switching from one renormalization scheme to
another. For this conversion, we follow the linearized approach described in Ref.\,\cite{Altenkamp:2017ldc}. Since the bare mixing angles are independent of the
renormalization scheme, their values $\varphi_i$ in a different renormalization scheme
are given by the values $\varphi_{\text{ref}}$ in the input scheme (called reference scheme in the following) and the corresponding counterterms $\delta \varphi_{\text{ref}}$ and $\delta \varphi_i$ in the reference and the other renormalization scheme, respectively, as
\begin{equation}
	\varphi _i \left( \{ \mu _\text{out} \} \right) \approx \varphi _\text{ref}  \left( \{ \mu _R \} \right) + \delta \varphi _\text{ref} \left( \{ \varphi _\text{ref} , \mu _R \} \right) - \delta \varphi _i \left( \{ \varphi _\text{ref} , \mu _\text{out} \} \right) ~. \label{eq:convertedParameterValues}
\end{equation}
Note, that Eq.~(\ref{eq:convertedParameterValues}) also contains the
dependence on the scales $\mu_R$ and $\mu_{\text{out}}$ introduced
above. They are relevant in case $\alpha$ and $\beta$ are 
understood as $\overline{\mbox{MS}}$ parameters and additionally
depend on the renormalization scale, at which they are defined.
The relation Eq.~(\ref{eq:convertedParameterValues}) holds
approximately up to higher-order terms, as the CTs 
involved in this equation are all evaluated with the mixing angles
given in the reference scheme. 

\section{Program Description}
\label{sec:programDescriptionMain}
In the following, we describe the system requirements needed for
compiling and running {\texttt{2HDECAY}}, the installation procedure
and the usage of the program. Additionally, we describe the input and
output file formats in detail. 
\subsection{System Requirements}
The \texttt{Python/FORTRAN} program code {\texttt{2HDECAY}} was
developed under {\texttt{Windows 10}} and {\texttt{openSUSE Leap
    15.0}}. The supported operating systems are: 
\begin{itemize}
	\item {\texttt{Windows 7}} and {\texttt{Windows 10}} (tested
          with {\texttt{Cygwin 2.10.0}}) 
	\item {\texttt{Linux}} (tested with {\texttt{openSUSE Leap 15.0}})
	\item {\texttt{macOS}} (tested with {\texttt{macOS Sierra 10.12}})
\end{itemize}
In order to compile and run {\texttt{2HDECAY}} on {\texttt{Windows}}, you need to install
{\texttt{Cygwin}} first (together with the packages {\texttt{cURL}},
{\texttt{find}}, {\texttt{gcc}}, {\texttt{g++}} and
{\texttt{gfortran}}, which also are required to be installed on {\texttt{Linux}}
and {\texttt{{macOS}}). For the compilation,
the {\texttt{GNU C}} compilers {\texttt{gcc}} (tested with versions
{\texttt{6.4.0}} and {\texttt{7.3.1}}), {\texttt{g++}} and the {\texttt{FORTRAN}}
compiler {\texttt{gfortran}} are required. Additionally, an up-to-date
version of {\texttt{Python 2}} or {\texttt{Python 3}} is required
(tested with versions {\texttt{2.7.14}} and {\texttt{3.5.0}}). For an
optimal performance of {\texttt{2HDECAY}}, we recommend that the
program is installed on a solid state drive (SSD) with high reading
and writing speeds. 
\subsection{License}
{\texttt{2HDECAY}} is released under the GNU General Public License
(GPL) ({\texttt{GNU GPL-3.0-or-later}}). {\texttt{2HDECAY}} is free
software, which means that anyone can redistribute it and/or modify it
under the terms of the GNU GPL as published by the Free Software
Foundation, either version 3 of the License, or any later
version. {\texttt{2HDECAY}} is distributed without any warranty. A
copy of the GNU GPL is included in the {\texttt{LICENSE.md}} file in
the root directory of {\texttt{2HDECAY}}. 
\subsection{Download}
\label{sec:Download}
The latest version of the program as well as a short quick-start
documentation is given at
\href{https://github.com/marcel-krause/2HDECAY}{https://github.com/marcel-krause/2HDECAY}. To
obtain the code either the repository is cloned or the zip archive is
downloaded and unzipped to a directory of the user's choice, which
here and in the following will be referred to as
{\texttt{\$2HDECAY}}. The main folder of {\texttt{2HDECAY}} consists
of several subfolders:  

\begin{description}
\item[{\texttt{BuildingBlocks}}] Contains the analytic electroweak
   one-loop corrections for all decays considered, as well as the
   real corrections and CTs needed to render the decay widths UV- and
   IR-finite. 
\item[{\texttt{Documentation}}] Contains this documentation. 
\item[{\texttt{HDECAY}}] This subfolder contains a modified version
   of
   {\texttt{HDECAY}} 6.52~\cite{DJOUADI199856,Djouadi:2018xqq},
   needed for the computation of the LO and (where applicable) QCD
   corrected decay widths. 
   {\tt HDECAY} also provides off-shell
   decay widths and the loop-induced decay widths into gluon and
   photon pair final states and into $Z\gamma$. {\tt HDECAY} is
   furthermore used for the computation of the branching ratios.
 \item[{\texttt{Input}}] In this subfolder, at least one
   or more input files can be stored that shall be used for the computation. The
   format of the input file is explained in
   Sec.\,\ref{sec:InputFileFormat}. In the Github repository, the
   {\texttt{Input}} folder contains an exemplary input file which is
   printed in App.\,\ref{sec:AppendixInputFile}. 
\item[{\texttt{Results}}] All results of a successful run of
  {\texttt{2HDECAY}} are stored as output files in this subfolder
  under the same name as the corresponding input files in the
  {\texttt{Input}} folder, but with the file extension {\texttt{.in}}
  replaced by {\texttt{.out}} and a suffix ``\_BR'' and ``\_EW'' for
  the branching ratios and electroweak partial decay widths,
  respectively. In the Github repository, the 
  {\texttt{Results}} folder contains two exemplary output files which 
  are given in App.\,\ref{sec:AppendixOutputFile}. 
\end{description}

The main folder {\texttt{\$2HDECAY}} itself also contains several
files: 
\begin{description}
\item[{\texttt{2HDECAY.py}}] Main program file of
  {\texttt{2HDECAY}}. It serves as a wrapper file for calling 
  {\texttt{HDECAY}} in order to convert the charm and bottom quark
  masses from the $\overline{\text{MS}}$ input values to the
  corresponding OS values and to calculate the LO widths, QCD
  corrections, off-shell and loop-induced decays, the branching ratios as well as
  {\texttt{electroweakCorrections}} for the calculation of
  the EW one-loop corrections.
\item[{\texttt{Changelog.md}}] Documentation of all changes made in
  the program since version \newline {\texttt{2HDECAY\,1.0.0}}. 
\item[{\texttt{CommonFunctions.py}}] Function library of
  {\texttt{2HDECAY}}, providing functions frequently used in the
  different files of the program. 
\item[{\texttt{Config.py}}] Main configuration file. If
  {\texttt{LoopTools}} is not installed automatically by the installer
  of {\texttt{2HDECAY}}, the paths to the {\texttt{LoopTools}}
  executables and libraries have to be set manually in this file.
\item[{\texttt{constants.F90}}] Library for all constants
  used in {\texttt{2HDECAY}}. 
\item[{\texttt{counterterms.F90}}] Definition of all fundamental CTs
  necessary for the EW one-loop renormalization of the Higgs boson 
  decays. The CTs defined in this file require the analytic results
  saved in the {\texttt{BuildingBlocks}} subfolder. 
\item[{\texttt{electroweakCorrections.F90}}] Main file for the
    calculation of the EW one-loop corrections to the Higgs boson 
    decays. It combines the EW one-loop corrections to the decay
    widths with the necessary CTs and IR corrections and calculates
    the EW contributions to the tree-level decay widths that 
    are then combined with the QCD corrections in {\texttt{HDECAY}}. 
\item[{\texttt{getParameters.F90}}] Routine to 
  read in the input values
    given by the user in the input files that are needed by
    {\texttt{2HDECAY}}.  
\item[{\texttt{LICENSE.md}}] Contains the full GNU General Public
  License ({\texttt{GNU GPL-3.0-or-later}}) agreement under which
  {\texttt{2HDECAY}} is published.  
\item[{\texttt{README.md}}] Provides an overview over basic
  information about the program as well as a quick-start guide. 
\item[{\texttt{setup.py}}] Main setup and installation file of
    {\texttt{2HDECAY}}. For a guided installation, this file should be
    called after downloading the program. 
\end{description}
\subsection{Installation}
\label{sec:Installation}
We highly recommend to use the automatic installation script
{\texttt{setup.py}} that is part of the {\texttt{2HDECAY}}
download. The script guides the user through the installation and asks
what components should be installed. For an installation under
{\texttt{Windows}}, the user should open the configuration file
{\texttt{\$2HDECAY/Config.py}} and check that the path to the
{\texttt{Cygwin}} executable in line 36 is set correctly before
starting the installation. In order to initiate the installation, the 
user navigates to the {\texttt{\$2HDECAY}} folder and executes the
following in the command-line shell: 
\begin{lstlisting}[numbers=none,language=bash,frame=single,backgroundcolor=\color{mygray}]
python setup.py
\end{lstlisting}
The script first asks the user if {\texttt{LoopTools}} should be
downloaded and installed. By entering {\texttt{y}}, the installer
downloads the {\texttt{LoopTools}} version that is specified in the
{\texttt{\$2HDECAY/Config.py}} file in line 37 and starts the
installation automatically. {\texttt{LoopTools}} is then installed in
a subdirectory of {\texttt{2HDECAY}}. Further information about the
installation of the program can be found in \cite{HAHN1999153}. 

If the user already has a working version of {\texttt{LoopTools}} on the system,
this step of the installation can be skipped. In this case, the user
has to open the file {\texttt{\$2HDECAY/Config.py}} in an editor and
change the lines 33-35 to the absolute path of the
{\texttt{LoopTools}} root directory and to the {\texttt{LoopTools}}
executables and libraries on the system. Additionally, line 32 has
to be changed to  
\begin{lstlisting}[language=Python,frame=single,backgroundcolor=\color{mygray},numbers=none]
useRelativeLoopToolsPath = False
\end{lstlisting}
This step is important if {\texttt{LoopTools}} is not installed
automatically with the install script, since otherwise,
{\texttt{2HDECAY}} will not be able to find the necessary executables
and libraries for the calculation of the EW one-loop
corrections.

As soon as {\texttt{LoopTools}} is installed (or alternatively, as
soon as paths to the {\texttt{LoopTools}} libraries and executables on
the user's system are being set manually in {\texttt{\$2HDECAY/Config.py}}),
the installation script asks whether it should automatically create
the makefile and the main EW corrections file
{\texttt{electroweakCorrections.F90}} and whether the program shall be
compiled. For an automatic installation, the user should type
{\texttt{y}} for all these requests to compile the main program as
well as to compile the modified version of {\texttt{HDECAY}} that
is included in {\texttt{2HDECAY}. The compilation may take several 
minutes to finish. At the end of the installation the
  user has the choice to 'make clean' the installation. This is optional. 

In order to test if the installation was successful, the user can type
\begin{lstlisting}[numbers=none,language=bash,frame=single,backgroundcolor=\color{mygray}]
python 2HDECAY.py
\end{lstlisting}
in the command-line shell, which runs the main program. The exemplary
input file provided by the default {\texttt{2HDECAY}} version is used for the
calculation. In the command window, the output of several steps of the
computation should be printed, but no errors. If the installation was
successful, {\texttt{2HDECAY}} terminates with no errors and the
existing output files in {\texttt{\$2HDECAY/Results}} are overwritten by
the newly created ones, which, however, are equivalent to the exemplary
output files that are provided with the program.

\subsection{Input File Format}
\label{sec:InputFileFormat}
\begin{table}[tb]
\centering
  \begin{tabular}{ c c c c }
    \hline
    Line & Input name & Allowed values and meaning \\ \hline
    \makecell[tc]{6} & \makecell[tc]{{\texttt{OMIT ELW2}}} &
\makecell[tl]{0: electroweak corrections (2HDM) are calculated \\ 1: electroweak
    corrections (2HDM) are neglected} \\  
    \makecell[tc]{9} & \makecell[tc]{{\texttt{2HDM}}} &
\makecell[tl]{0: considered model is not the 2HDM \\ 1: considered
    model is the 2HDM } \\ 
    \makecell[tc]{56} & \makecell[tc]{{\texttt{PARAM}}} & \makecell[tl]{1: 2HDM Higgs masses and $\alpha$ (lines 66-70) are given as input \\ 2: 2HDM potential parameters (lines 72-76) are given as input} \\ 
    \makecell[tc]{57} & \makecell[tc]{{\texttt{TYPE}}} &
\makecell[tl]{1: 2HDM type I \\ 2: 2HDM type II \\ 
3: 2HDM lepton-specific \\ 4: 2HDM flipped} \\ 
    \makecell[tc]{58} & \makecell[tc]{{\texttt{RENSCHEM}}} &
\makecell[tl]{0: all renormalization schemes are calculated \\ 1-17: only the chosen scheme ({\it cf.}~Tab.\,\ref{tab:2HDECAYImplementedSchemes}) is calculated} \\ 
    \makecell[tc]{59} & \makecell[tc]{ {\texttt{REFSCHEM}} } &
\makecell[tl]{1-17: the input values of
    $\alpha$, $\beta$ and $m_{12}^2$
    (\textit{cf.}\,Tab.\,\ref{tab:2HDECAYInputValues}) are given in
    the \\ \hspace*{0.35cm} chosen reference
    scheme and at the scale $\mu _R$ given by {\texttt{INSCALE}} in 
\\ \hspace*{0.35cm} case of $\overline{\mbox{MS}}$
    parameters; the values of $\alpha$, $\beta$ and $m_{12}^2$ in all other
\\ \hspace*{0.35cm} schemes and 
    at the scale $\mu_{\text{out}}$ at which the decays are calculated,
\\ \hspace*{0.35cm} are evaluated using
    Eqs.~(\ref{eq:scalechange}) and (\ref{eq:convertedParameterValues})} \\ 
    \hline
  \end{tabular}
    \caption{Input parameters for the basic control of
      {\texttt{2HDECAY}}. The line number corresponds to the line of
      the input file where the input value can be found. In order to
      calculate the EW corrections for the 2HDM, the input parameter
      {\texttt{OMIT ELW2}} has to be set to 0. In this case, the given
      input value of {\texttt{2HDM}} is ignored and {\texttt{2HDM=1}}
      is set automatically, independent of the chosen input value. All
      input values presented in this table have to be entered as integer values.}  
   \label{tab:2HDECAYControlInputs}
\end{table}
The format of the input file is adopted from {\texttt{HDECAY}}
\cite{DJOUADI199856, Djouadi:2018xqq}, with minor modifications to
account for the EW corrections that are implemented. The
file has to be stored as a text-only file in UTF-8 
format. Since {\texttt{2HDECAY}} is a program designed for the calculation of
higher-order corrections solely for the 2HDM, only a subset of input
parameters in comparison to the original {\texttt{HDECAY}} input file
is actually used ({\it e.g.}~SUSY-related input parameters are not needed
for {\texttt{2HDECAY}}). The input file nevertheless contains the full
set of input parameters from {\texttt{HDECAY}} to make
{\texttt{2HDECAY}} fully backwards-compatible,
{\it i.e.}\,{\texttt{HDECAY\,6.52}} is fully contained in
{\texttt{2HDECAY}}. The input file contains two classes of 
input parameters. The first class are input values that control the
main flow of the program ({\it e.g.\,}whether corrections for the SM or the
2HDM are calculated). The control parameters relevant for
{\texttt{2HDECAY}} are shown in Tab.\,\ref{tab:2HDECAYControlInputs},
together with their line numbers in the input file, their allowed
values and the meaning of the input values. In order to choose the
2HDM as the model that is considered, the input value {\texttt{2HDM =
    1}} has to be chosen. By setting {\texttt{OMIT ELW2 = 0}}, the
EW and QCD corrections are calculated for the 2HDM, whereas
for {\texttt{OMIT ELW2 = 1}}, only the QCD corrections are
calculated. The latter choice corresponds to the corrections for the
2HDM that are already implemented in
{\texttt{HDECAY\,6.52}}. If the user sets {\texttt{OMIT ELW2 = 0}} in
the input file, then {\texttt{2HDM = 1}} is automatically set
internally, independent of the input value of {\texttt{2HDM}} that the
user provides. The input value {\texttt{PARAM}} 
determines which parametrization of the Higgs sector shall be
used. For {\texttt{PARAM = 1}}, the Higgs boson masses and mixing angle
$\alpha$ are chosen as input, while for {\texttt{PARAM =
    2}}, the Higgs potential parameters $\lambda _i$ are used as
input. As described at the end of Sec.\,\ref{sec:setupOfModel}, 
however, it should be noted that the EW corrections in
{\texttt{2HDECAY}} are in both cases parametrized through the Higgs
masses and mixing angle. Hence, if {\texttt{PARAM = 2}} is chosen, the
masses and mixing angle are calculated as functions of $\lambda _i$ by
means of
Eqs.\,(\ref{eq:parameterTransformationInteractionToMass1})-(\ref{eq:parameterTransformationInteractionToMass5}). The
input value {\texttt{TYPE}} sets the type of the 2HDM, as described in
Sec.\,\ref{sec:setupOfModel}, and {\texttt{RENSCHEM}} determines the
renormalization schemes that are used for the calculation. By setting
{\texttt{RENSCHEM = 0}}, the EW corrections to the Higgs boson 
decays are calculated for all 17 implemented renormalization
schemes. This allows for analyses of the renormalization scheme dependence
and for an estimate of the effects of missing higher-order EW 
corrections, but this setting has the caveat of increasing 
the computation time and output file size rather significantly. A
specific integer value of {\texttt{RENSCHEME}} between 1 and 17 sets
the renormalization scheme to the chosen one. An overview of all
implemented schemes and their identifier values between 1 and 17 is
presented in Tab.\,\ref{tab:2HDECAYImplementedSchemes}. 
As discussed in
  Sec.\,\ref{sec:ParameterConversion}, the consistent comparison of
  partial decay widths calculated in different renormalization schemes
  requires the conversion of the input parameters between these
  schemes. By setting {\texttt{REFSCHEM}} to a value between 1 and 17,
  the input parameters for $\alpha$ and $\beta$ ({\it cf.}
  Tab.\,\ref{tab:2HDECAYInputValues}) are understood as input parameters in
  the chosen reference scheme and the automatic parameter conversion
  is activated. The input value of the $\overline{\mbox{MS}}$
  parameter $m_{12}^2$ is given at the input scale $\mu _R$. The same
  applies for the input values of $\alpha$ and $\beta$ when they are
  chosen to be $\overline{\text{MS}}$ renormalized. The values of
  $\alpha$, $\beta$ and $m_{12}^2$ in all other renormalization
  schemes and at all other scales $\mu _\text{out}$ are then
  calculated using Eqs.~(\ref{eq:scalechange}) and
  (\ref{eq:convertedParameterValues}). The automatic parameter
  conversion requires the input parameters to be given in the mass
  basis of \eqref{eq:inputSetMassBase}, \textit{i.e.}\,for the
  automatic parameter conversion to be active, it is necessary to set
  {\texttt{PARAM = 1}}. If instead {\texttt{PARAM = 2}} is set, then
  {\texttt{REFSCHEM = 0}} is set automatically internally so that the
  automatic parameter conversion is deactivated. In this case, a
  warning is printed in the console. All input  
values of the first class must be entered as integers. 
\begin{table}[tb]
\centering
  \begin{tabular}{ c c c c }
    \hline
    Line & Input name & Name in Sec.\,\ref{sec:EWQCD2HDMMain} & Allowed values and meaning \\ \hline
    \makecell[tc]{18} & \makecell[tc]{{\texttt{ALS(MZ)}}} & $\alpha_s (m_Z)$ & \makecell[tl]{strong coupling constant (at $m_Z$)} \\ 
    \makecell[tc]{19} & \makecell[tc]{{\texttt{MSBAR(2)}}} & $m_s (2\,\text{GeV})$ & \makecell[tl]{$s$-quark $\overline{\text{MS}}$ mass at 2 GeV in GeV} \\ 
    \makecell[tc]{20} & \makecell[tc]{{\texttt{MCBAR(3)}}} & $m_c (3\,\text{GeV})$ & \makecell[tl]{$c$-quark $\overline{\text{MS}}$ mass at 3 GeV in GeV} \\ 
    \makecell[tc]{21} & \makecell[tc]{{\texttt{MBBAR(MB)}}} & $m_b (m_b)$ & \makecell[tl]{$b$-quark $\overline{\text{MS}}$ mass at $m_b$ in GeV} \\ 
    \makecell[tc]{22} & \makecell[tc]{{\texttt{MT}}} & $m_t$ & \makecell[tl]{$t$-quark pole mass in GeV} \\ 
    \makecell[tc]{23} & \makecell[tc]{{\texttt{MTAU}}} & $m_\tau $ & \makecell[tl]{$\tau$-lepton pole mass in GeV} \\ 
    \makecell[tc]{24} & \makecell[tc]{{\texttt{MMUON}}} & $m_\mu $ & \makecell[tl]{$\mu $-lepton pole mass in GeV} \\ 
    \makecell[tc]{25} & \makecell[tc]{{\texttt{1/ALPHA}}} & $\alpha _\text{em} ^{-1} (0)$ & \makecell[tl]{inverse fine-structure constant (Thomson limit)} \\ 
    \makecell[tc]{26} & \makecell[tc]{{\texttt{ALPHAMZ}}} & $\alpha _\text{em} (m_Z)$ & \makecell[tl]{fine-structure constant (at $m_Z$)} \\ 
    \makecell[tc]{29} & \makecell[tc]{{\texttt{GAMW}}} & $\Gamma _W$ & \makecell[tl]{partial decay width of the $W$ boson } \\ 
    \makecell[tc]{30} & \makecell[tc]{{\texttt{GAMZ}}} & $\Gamma _Z$ & \makecell[tl]{partial decay width of the $Z$ boson } \\ 
    \makecell[tc]{31} & \makecell[tc]{{\texttt{MZ}}} & $m_Z$ &
\makecell[tl]{$Z$ boson on-shell mass in GeV} \\ 
    \makecell[tc]{32} & \makecell[tc]{{\texttt{MW}}} & $m_W$ &
\makecell[tl]{$W$ boson on-shell mass in GeV} \\ 
    \makecell[tc]{33-41} & \makecell[tc]{{\texttt{Vij}}} & $V_{ij}$ & \makecell[tl]{CKM matrix elements ($i\in \{ u,c,t\}$ , $j\in \{ d,s,b\} $) } \\ 
    \makecell[tc]{61} & \makecell[tc]{{\texttt{TGBET2HDM}}} & $t_\beta $ & \makecell[tl]{ratio of the VEVs in the 2HDM} \\ 
    \makecell[tc]{62} & \makecell[tc]{{\texttt{M\_12\textasciicircum
2}}} & $m_{12}^2 $ & \makecell[tl]{squared soft-$\mathbb{Z}_2$-breaking scale in GeV$^2$} \\ 
    \makecell[tc]{63} & \makecell[tc]{{\texttt{INSCALE}}} & $\mu_R $
 & \makecell[tl]{renormalization scale for $\overline{\text{MS}}$ inputs in GeV} \\ 
    \makecell[tc]{64} & \makecell[tc]{ {\texttt{OUTSCALE}} } & $\mu_\text{out} $
 & \makecell[tl]{ renormalization scale for the evaluation of the } \\ 
      &   &   & \makecell[tl]{ partial decay widths in GeV or in terms of {\texttt{MIN}} } \\ 
    \makecell[tc]{66} & \makecell[tc]{{\texttt{ALPHA\_H}}} & $\alpha $ & \makecell[tl]{CP-even Higgs mixing angle in radians} \\ 
    \makecell[tc]{67} & \makecell[tc]{{\texttt{MHL}}} & $m_h $ &
\makecell[tl]{light CP-even Higgs boson mass in GeV} \\
    \makecell[tc]{68} & \makecell[tc]{{\texttt{MHH}}} & $m_H $ &
\makecell[tl]{heavy CP-even Higgs boson mass in GeV} \\
    \makecell[tc]{69} & \makecell[tc]{{\texttt{MHA}}} & $m_A $ &
\makecell[tl]{CP-odd Higgs boson mass in GeV} \\
    \makecell[tc]{70} & \makecell[tc]{{\texttt{MH+-}}} & $m_{H^\pm } $
& \makecell[tl]{charged Higgs boson mass in GeV} \\
    \makecell[tc]{72-76} & \makecell[tc]{{\texttt{LAMBDAi}}} &
$\lambda_i $ & \makecell[tl]{Higgs potential parameters 
[see Eq.~(\ref{eq:scalarPotential})]} \\
    \hline
  \end{tabular}
    \caption{Shown are all relevant physical input parameters of
      {\texttt{2HDECAY}} that are necessary for the calculation of the
      QCD and EW corrections. The
      line number corresponds to the line of the input file where the input value
      can be found. Depending on the chosen value of {\texttt{PARAM}}
      ({\it cf.}~Tab.\,\ref{tab:2HDECAYControlInputs}), either the Higgs
      masses and mixing angle $\alpha$ (lines 66-70) or the 2HDM
      potential parameters (lines 72-76) are chosen as input, but
      never both simultaneously. The value {\texttt{OUTSCALE}} is
      entered either as a double-precision number or as 
      {\texttt{MIN}}, representing the mass scale of the decaying
      Higgs boson. All other input values presented in this table are
      entered as double-precision numbers.} 
   \label{tab:2HDECAYInputValues}
\end{table}

The second class of input values in the input file are the physical
input parameters shown in Tab.\,\ref{tab:2HDECAYInputValues}, together
with their line numbers in the input file, their allowed input values
and the meaning of the input values. This is the full set of input
parameters needed for the calculation of the electroweak and QCD
corrections. All other input parameters present in the input file that
are not shown in Tab.\,\ref{tab:2HDECAYInputValues} are neglected for
the calculation of the QCD and EW corrections in the
2HDM. We want to emphasize again that depending on the
choice of {{\texttt{PARAM}} ({\it cf.}~Tab.\,\ref{tab:2HDECAYControlInputs}),
  either the Higgs masses and mixing angle $\alpha$ or the Higgs
  potential parameters $\lambda _i$ are chosen as independent input,
  but never both simultaneously, {\it i.e.}~if {{\texttt{PARAM = 1}} is
    chosen, then the input values for $\lambda _i$ are ignored, while
    for {\texttt{PARAM = 2}}, the input values of the Higgs masses
      and $\alpha$ are ignored and instead calculated by means of
      Eqs.\,(\ref{eq:parameterTransformationInteractionToMass1})-(\ref{eq:parameterTransformationInteractionToMass5}). 
All input values of the second class are entered in
      {\texttt{FORTRAN}} double-precision format, {\it i.e.}~valid input
      formats are {\it e.g.}~{\texttt{MT = 1.732e+02}} or {\texttt{MHH =
          258.401D0}}. Since $m_{12}^2$ and, in case of a chosen $\overline{\text{MS}}$ scheme, $\alpha$ and $\beta$ depend on the renormalization scale $\mu _R$ at which these parameters are given, the calculation of the partial decay widths depends on this scale. Moreover, since the partial decay widths are evaluated at the (potentially different) renormalization scale $\mu _\text{out}$, the decay widths and branching ratios depend on this scale as well. In order to avoid artificially large corrections, both scales should be chosen appropriately. The input value
{\texttt{INSCALE}} of $\mu _R$, \textit{i.e.}\,the scale at which all $\overline{\text{MS}}$ parameters are defined, is entered as a double-precision number. The input value
{\texttt{OUTSCALE}} of $\mu _\text{out}$, \textit{i.e.}\,the renormalization scale at which the partial decay widths are evaluated, can be entered either as a
double-precision number or it can be expressed in terms of the mass
scale {\texttt{MIN}} of the decaying Higgs boson, {\it i.e.}\,setting {\texttt{OUTSCALE=MIN}} sets $\mu _R = m_1$ for each decay
channel, where $m_1$ is the mass of the decaying Higgs boson in the
respective channel. Note finally, that the input
  masses for the $W$ and $Z$ gauge bosons must be the on-shell
  values for consistency with the renormalization conditions applied 
  in the EW corrections. 
\begin{table}[tb]
\centering
  \begin{tabular}{ c c c c c }
    \hline
    Input ID & Tadpole scheme & $\delta \alpha$ & $\delta \beta$ & Gauge-par.-indep. $\Gamma$ \\ \hline
    \makecell[tc]{1} & \makecell[tc]{standard} & \makecell[tc]{KOSY} & \makecell[tc]{KOSY (odd)} & \makecell[tc]{\xmark } \\ 
    \makecell[tc]{2} & \makecell[tc]{standard} & \makecell[tc]{KOSY} & \makecell[tc]{KOSY (charged)} & \makecell[tc]{\xmark } \\ 
    \makecell[tc]{3} & \makecell[tc]{alternative (FJ)} & \makecell[tc]{KOSY} & \makecell[tc]{KOSY (odd)} & \makecell[tc]{\xmark } \\ 
    \makecell[tc]{4} & \makecell[tc]{alternative (FJ)} & \makecell[tc]{KOSY} & \makecell[tc]{KOSY (charged)} & \makecell[tc]{\xmark } \\ 
    \makecell[tc]{5} & \makecell[tc]{alternative (FJ)} & \makecell[tc]{$p_{*}$-pinched} & \makecell[tc]{$p_{*}$-pinched (odd)} & \makecell[tc]{\cmark } \\ 
    \makecell[tc]{6} & \makecell[tc]{alternative (FJ)} & \makecell[tc]{$p_{*}$-pinched} & \makecell[tc]{$p_{*}$-pinched (charged)} & \makecell[tc]{\cmark } \\ 
    \makecell[tc]{7} & \makecell[tc]{alternative (FJ)} & \makecell[tc]{OS-pinched} & \makecell[tc]{OS-pinched (odd)} & \makecell[tc]{\cmark } \\ 
    \makecell[tc]{8} & \makecell[tc]{alternative (FJ)} & \makecell[tc]{OS-pinched} & \makecell[tc]{OS-pinched (charged)} & \makecell[tc]{\cmark } \\ 
    \makecell[tc]{9} & \makecell[tc]{alternative (FJ)} & \makecell[tc]{proc.-dep. 1} & \makecell[tc]{proc.-dep. 1} & \makecell[tc]{\cmark } \\ 
    \makecell[tc]{10} & \makecell[tc]{alternative (FJ)} & \makecell[tc]{proc.-dep. 2} & \makecell[tc]{proc.-dep. 2} & \makecell[tc]{\cmark } \\ 
    \makecell[tc]{11} & \makecell[tc]{alternative (FJ)} & \makecell[tc]{proc.-dep. 3} & \makecell[tc]{proc.-dep. 3} & \makecell[tc]{\cmark } \\ 
    \makecell[tc]{12} & \makecell[tc]{alternative (FJ)} & \makecell[tc]{OS1} & \makecell[tc]{OS1} & \makecell[tc]{\cmark } \\ 
    \makecell[tc]{13} & \makecell[tc]{alternative (FJ)} & \makecell[tc]{OS2} & \makecell[tc]{OS2} & \makecell[tc]{\cmark } \\ 
    \makecell[tc]{14} & \makecell[tc]{alternative (FJ)} & \makecell[tc]{OS12} & \makecell[tc]{OS12} & \makecell[tc]{\cmark } \\ 
    \makecell[tc]{15} & \makecell[tc]{alternative (FJ)} & \makecell[tc]{BFMS} & \makecell[tc]{BFMS} & \makecell[tc]{\cmark } \\ 
    \makecell[tc]{16} & \makecell[tc]{standard} & \makecell[tc]{ $\overline{\text{MS}}$ } & \makecell[tc]{ $\overline{\text{MS}}$ } & \makecell[tc]{\xmark } \\ 
    \makecell[tc]{17} & \makecell[tc]{alternative (FJ)} & \makecell[tc]{ $\overline{\text{MS}}$ } & \makecell[tc]{ $\overline{\text{MS}}$ } & \makecell[tc]{\cmark } \\ 
    \hline
  \end{tabular}
    \caption{Overview over all renormalization schemes for the mixing
      angles $\alpha$ and $\beta$ that are implemented in
      {\texttt{2HDECAY}}. By setting {\texttt{RENSCHEM}} in the input
      file, {\it cf.}~Tab.\,\ref{tab:2HDECAYControlInputs}, equal to the
      Input ID the renormalization scheme is chosen. In case of 0 the
      results for all renormalization schemes are given out. The
      definition of the CTs $\delta \alpha$ and $\delta \beta$ in each
      scheme is explained in
      Sec.\,\ref{sec:renormalizationMixingAngles}. The crosses and
      check marks in the column for gauge independence indicate
      whether the chosen scheme in general yields explicitly gauge-independent
      partial decay widths or not.} 
   \label{tab:2HDECAYImplementedSchemes}
\end{table}
The amount of input files that can be stored in the input folder is
not limited. The input files can have arbitrary non-empty names and
filename extensions\footnote{On some systems, certain filename
  extensions should be avoided when naming the input files, as they
  are reserved for certain types of files ({\it e.g.}~under {\texttt{Windows}},
  the {\texttt{.exe}} file extension is automatically connected to
  executables by the operating system, which can under certain
  circumstances lead to runtime problems when trying to read the
  file). Choosing text file extensions like {\texttt{.in}},
  {\texttt{.out}}, {\texttt{.dat}} or {\texttt{.txt}} should in
  general be unproblematic.}. The output files are saved in the
{\texttt{\$2HDECAY/Results}} subfolder under the same name as the
corresponding input files, but with their filename extension replaced
by {\texttt{.out}}. 
For each input file, two output files are generated. The output file
containing the branching ratios is indicated by the filename suffix
'\_BR', while the output file containing the electroweak 
partial decay widths is indicated by the filename suffix '\_EW'.

\subsection{Structure of the Program}
As briefly mentioned in Sec.\,\ref{sec:Download}, the main program
{\texttt{2HDECAY}} combines the already existing QCD corrections from
{\texttt{HDECAY}} with the full EW one-loop
corrections. Depicted in \figref{fig:flowchart2HDECAY} is the
flowchart of {\texttt{2HDECAY}} which shows how the QCD and
EW corrections are combined by the main wrapper file
{\texttt{2HDECAY.py}}. 
\begin{figure}[tb]
\centering
\includegraphics[width=13.8cm]{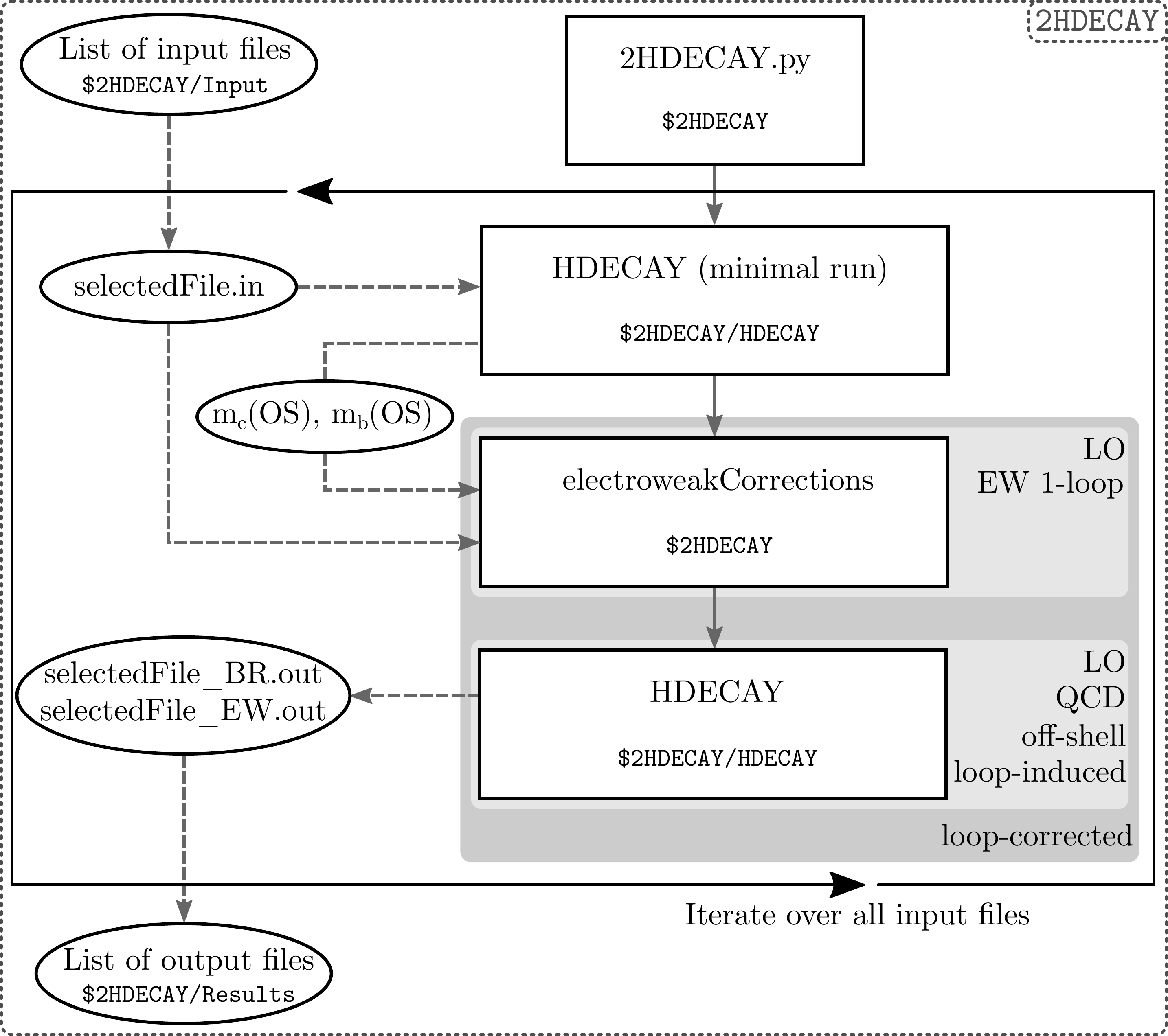}
\caption{Flowchart of {\texttt{2HDECAY}}. The main wrapper file
  {\texttt{2HDECAY.py}} generates a list of input files, provided by the
  user in the subfolder {\texttt{\$2HDECAY/Input}}, and iterates over
  the list. For each selected input file in the list, the wrapper
  calls {\tt HDECAY} and the subprogram {\tt
    electroweakCorrections}. The computed branching ratios including
  the EW and QCD corrections as described in the text are written 
to the output file with suffix '\_BR', the calculated LO and NLO EW-corrected
partial decay widths are given out in the output file with suffix
'\_EW'. For further details, we refer to the text.}
\label{fig:flowchart2HDECAY}
\end{figure}
First, the wrapper file generates a list of all input files that the
user provides in {\texttt{\$2HDECAY/Input}}. The user can provide an
arbitrary non-zero amount of input files with arbitrary filenames, as
described in Sec.\,\ref{sec:InputFileFormat}. For any input file in
the list, the wrapper file first calls {\texttt{HDECAY}} in a
so-called minimal run, technically by calling {\texttt{HDECAY}} in the
subfolder {\texttt{\$2HDECAY/HDECAY}} with an additional flag ``1'': 
\begin{lstlisting}[numbers=none,language=bash,frame=single,backgroundcolor=\color{mygray}]
run 1
\end{lstlisting}
With this flag, HDECAY reads the selected input file from the input
file list and uses the input values only to convert the
$\overline{\text{MS}}$ values of the $c$- and $b$-quark masses, as
given in the input file, to the corresponding pole masses, but no other computations are performed at
this step.

The wrapper file then calls the subprogram
{\texttt{electroweakCorrections}}, which reads the selected input file
as well as the OS values of the quark masses. With these input values,
the full EW one-loop corrections are calculated for all decays that
are kinematically allowed, as described in
Sec.\,\ref{sec:decayProcessesAtLOandNLO}, and the value of $G_F^{\text{calc}}$
at the $Z$ mass is calculated, as described in
Sec.\,\ref{sec:connectionHDECAY}. Subsequently, a temporary new input
file is created, which consists of a copy of the selected input file
with the calculated OS quark masses, the calculated value of
$G_F^{\text{calc}}$ and all EW corrections being appended. 

Lastly, the wrapper file calls {\texttt{HDECAY}} without the minimal
flag. In this configuration, {\texttt{HDECAY}} reads the temporary
input file and calculates the LO widths and QCD corrections to the
decays. Moreover, the program calculates off-shell decay widths 
as well as the loop-induced decays to final-state pairs
of gluons or photons and $Z \gamma$. Furthermore, the branching ratios
are calculated by {\texttt{HDECAY}}. The results of these computations
are consistently combined with the electroweak corrections, as 
described in Sec.\,\ref{sec:connectionHDECAY}. The results are saved in an
output file in the {\texttt{\$2HDECAY/Results}} subfolder.  

The wrapper file repeats these steps for each file in the input file
list until the end of the list is reached. 
\subsection{Usage}
Before running the program, the user should check that all input files
for which the computation shall be performed are stored in the
subfolder {\texttt{\$2HDECAY/Input}}. The input files have to be
formatted exactly as described in Sec.\,\ref{sec:InputFileFormat} or
otherwise the input values are not read in correctly and the program
might crash with a segmentation error. The exemplary input file
printed in App.\,\ref{sec:AppendixInputFile} that is part of the
{\texttt{2HDECAY}} repository can be used as a template for generating 
other input files in order to avoid formatting problems. 

The user should check the output subfolder
{\texttt{\$2HDECAY/Results}} for any output files of previous runs of
{\texttt{2HDECAY}}. These previously created output files are
overwritten if in a new run input files with the same names as the
already stored output files are used. Hence, the user is advised to
create backups of the output files before starting a new run of
{\texttt{2HDECAY}}. 

In order to run the program, open a terminal, navigate to the
{\texttt{\$2HDECAY}} folder and execute the following command:
\begin{lstlisting}[numbers=none,language=bash,frame=single,backgroundcolor=\color{mygray}]
python 2HDECAY.py
\end{lstlisting}
If {\texttt{2HDECAY}} was installed correctly according to
Sec.\,\ref{sec:Installation} and if the input files have the correct
format, the program should now compute the EW and/or QCD
corrections according to the flowchart shown in
\figref{fig:flowchart2HDECAY}. Several intermediate results and
information about the computation are printed in the terminal. As
soon as the computation for all input files is done,
{\texttt{2HDECAY}} is terminated and the resulting output files can be
found in the {\texttt{\$2HDECAY/Results}} subfolder. 

\subsection{Output File Format}
\label{sec:OutputFileFormat}
For each input file, two output files with the suffixes '\_QCD'
and '\_EW' for the branching ratios and electroweak partial decay
widths, respectively, are generated in an SLHA format, as described in 
Sec.~\ref{sec:connectionHDECAY}. The SLHA output 
format \cite{Skands:2003cj,Allanach:2008qq,Mahmoudi:2010iz} in its strict and original
sense has only been designed for supersymmetric models. We have
modified the format to account for the EW corrections that are implemented in
{\texttt{2HDECAY}} in the 2HDM. As a reference for the following description, 
exemplary output files are given in
App.\,\ref{sec:AppendixOutputFile}. These modified SLHA output files are
only generated if {\texttt{OMIT ELW2=0}} is set in the input file,
{\it i.e.}\,only if the electroweak corrections to the 2HDM decays are
taken into account. In the following we describe the changes that we
have applied. 

The first block {\texttt{BLOCK DCINFO}} contains basic information
about the program itself, while the subsequent three blocks
{\texttt{SMINPUTS}}, {\texttt{2HDMINPUTS}} and {\texttt{VCKMIN}}
contain the input parameters used for the calculation that were
already described in Sec.\,\ref{sec:InputFileFormat}. As explained in
Sec.\,\ref{sec:connectionHDECAY}, the value of $G_F$ printed in the
output file is not necessarily the same as the one given in the input
file if {\texttt{OMIT ELW2=0}} is set, since in this case, $G_F$ is
calculated from the input value $\alpha _\text{em} (m_Z^2)$
instead, and this value is then given out. These four blocks are
given out in both output files.

In the output file containing the branching ratios, indicated by the
suffix '\_BR', subsequently two blocks follow for
each Higgs boson ($h,H,A$ and 
$H^\pm$). They are called {\texttt{DECAY QCD}} and {\texttt{DECAY
    QCD\&EW}}.
The block {\texttt{DECAY QCD}} contains the total decay
width, the mixing angles $\alpha$, $\beta$,
  the $\overline{\mbox{MS}}$ parameter
  $m_{12}^2$\footnote{Note that they differ from
      the input values if $\mu_R \ne \mu_{\text{out}}$ or if the
      reference/input scheme is different from the renormalization scheme in
    which the decays are evaluated.}, and the 
branching ratios of the decays of the respective Higgs boson, as
implemented in {\tt HDECAY}. These are in 
particular the LO (loop-induced for the $gg$, $\gamma\gamma$ and
$Z\gamma$ final states) decay widths including the relevant and state-of-the
art QCD corrections where applicable ({\it
  cf.}~\cite{DJOUADI199856,Djouadi:2018xqq} for further details). 
For decays into heavy quarks, massive vector bosons, neutral Higgs
pairs as well as gauge and Higgs boson final states
also off-shell decays are computed if necessary. We want to emphasize
again that the partial and total decay widths differ from the ones of the original {\tt
  HDECAY} version if  {\texttt{OMIT ELW2=0}} is set, as for
consistency with the computed EW corrections in this case the {\tt
  HDECAY} decay widths are rescaled by $G_F^\text{calc}/G_F$, as
explained in Sec.~\ref{sec:connectionHDECAY}. If {\texttt{OMIT
  ELW2=1}} is set, no EW corrections are computed and the {\tt HDECAY}
decay widths are computed with $G_F$ as in the original {\tt HDECAY} version.

The block {\texttt{DECAY
    QCD\&EW}} contains the total decay width, the
mixing angles $\alpha$, $\beta$, the $\overline{\mbox{MS}}$
parameter $m_{12}^2$, and the
branching ratios of the respective Higgs boson including both the QCD
corrections (provided by {\tt 
  HDECAY}) and the EW corrections (computed by {\tt 2HDECAY}) }to the
LO decay widths. Note that the 
LO decay widths are also computed by {\tt 2HDECAY}. As an additional
cross-check, we internally compare the respective {\tt HDECAY} LO decay
width (rescaled by $G_F^\text{calc}/G_F$ and calculated with OS masses
for this comparison) with the one computed by
{\tt 2HDECAY}. If they differ (which they should not), a warning is
printed on the screen. As described in
Sec.\,\ref{sec:connectionHDECAY}, we emphasize again that the EW corrections are
calculated and included only for OS decay channels that are
kinematically allowed and for non-loop-induced decays. Therefore, some
of the branching ratios given out may be QCD-, but not
EW-corrected. The total decay width given out in this block is 
the sum of all accordingly computed partial decay widths. 

The last block at the end of the file with the branching ratios 
contains the QCD-corrected branching ratios of the top-quark calculated in the 2HDM.
It is required for the computation of the Higgs decays into final states
with an off-shell top.

In the output file with the EW corrected NLO decay widths, indicated
by the suffix '\_EW', the first four blocks
described above are instead followed by the two blocks {\texttt{LO DECAY WIDTH}} and
{\texttt{NLO DECAY WIDTH}} for each Higgs boson ($h,H,A$ and 
$H^\pm$). In these blocks, the partial decay widths at 
LO and including the one-loop EW corrections are given out,
respectively, together with the
mixing angles $\alpha$, $\beta$ and the $\overline{\mbox{MS}}$
parameter $m_{12}^2$. These values of the widths
are particularly useful for studies of  
the relative size of the EW corrections and for studying the
renormalization scheme dependence of the EW corrections. This allows
for a rough estimate of the remaining theoretical error due to missing higher-order
EW corrections. Since the EW corrections are calculated only for OS
decays and additionally only for decays that are not loop-induced,
these two blocks do not contain all final states written out in the
blocks {\texttt{DECAY QCD}} and {\texttt{DECAY
    QCD\&EW}}. Hence, depending on the input values that are
chosen, it can happen that the two blocks {\texttt{DECAY QCD}} and {\texttt{DECAY
    QCD\&EW}} contain decays that are not printed out in the blocks
{\texttt{LO DECAY WIDTH}} and {\texttt{NLO DECAY WIDTH}}, since for the calculation of 
the branching ratios, off-shell and loop-induced decays are considered
by {\texttt{HDECAY}} as well.  

\section{Summary}
\label{sec:summary}
We have presented the program package {\texttt{2HDECAY}} for the 
calculation of the Higgs boson decays in the 2HDM. The tool computes the
NLO EW corrections to all 2HDM Higgs boson decays into OS final states
that are not loop-induced. The user can choose among 17 different
renormalization schemes that have been specified in the manual. 
They are based on different renormalization schemes for the mixing
angles $\alpha $ and $\beta$, an $\overline{\text{MS}}$ condition for
the soft-$\mathbb{Z}_2$-breaking scale $m_{12}^2$ and an OS scheme for
all other counterterms and wave function renormalization constants of
the 2HDM necessary for calculating the EW corrections.
The EW corrections are combined with the state-of-the-art QCD
corrections obtained from {\texttt{HDECAY}}. The EW\&QCD-corrected
total decay widths and branching ratios are given out in an
SLHA-inspired output file format. Moreover, the tool provides
separately an SLHA-inspired output for the LO and  
EW NLO partial decay widths to all OS and non-loop-induced
decays. This separate output enables {\it e.g.}~an
efficient analysis of the size of the EW corrections in the 2HDM or
the comparison with the relative 
EW corrections in the MSSM as a SUSY benchmark model. The
implementation of several different renormalization schemes 
additionally allows for the investigation of the numerical effects of
the different schemes and an estimate of the residual theoretical
uncertainty due to missing higher-order EW
corrections. For a consistent estimate of this error,
  an automatic parameter conversion routine is implemented, 
performing the automatic conversion of the input
values of $\alpha$, $\beta$ and $m_{12}^2$ from a reference scheme to
all other renormalization schemes that are implemented, as well as
from the $\overline{\text{MS}}$ input renormalization scale $\mu _R$
to the renormalization scale $\mu _\text{out}$ at which the partial
decay widths are evaluated. Being fast, our new 
tool enables efficient phenomenological studies of the 2HDM Higgs
sector at high precision. The latter is necessary to reveal indirect
new physics effects in the Higgs sector and to identify the true
underlying model in case of the discovery of additional Higgs bosons. This
brings us closer to our goal of understanding electroweak symmetry
breaking and deciphering the physics puzzle in fundamental particle
physics.  

\subsection*{Acknowledgments}
The authors thank David Lopez-Val and Jonas M\"{u}ller for
independently cross-checking some of the analytic results derived for
this work. The authors express gratitude to David Lopez-Val for his
endeavors on debugging the early alpha versions of {\texttt{2HDECAY}}
and to Stefan Liebler and Florian Staub for helpful discussions concerning the real corrections to the decays. The authors thank Ansgar Denner, Stefan Dittmaier and Jean-Nicolas Lang for helpful discussions and for providing the analytic results of their mixing angle counterterms to us for the implementation in {\texttt{2HDECAY}}. MK
and MM acknowledge financial support from the DFG project “Precision
Calculations in the Higgs Sector - Paving the Way to the New Physics
Landscape” (ID: MU 3138/1-1). 

\begin{appendix}

\section{Exemplary Input File}
\label{sec:AppendixInputFile}
In the following, we present an exemplary input file
{\texttt{2hdecay.in}} as it is included in the subfolder
{\texttt{\$2HDECAY/Input}} in the {\texttt{2HDECAY}} repository. The
first integer in each line represents the line number and is not part
of the actual input file, but printed here for convenience. The
meaning of the input parameters is specified in
Sec.\,\ref{sec:InputFileFormat}. In comparison to the input file
format of the unmodified {\texttt{HDECAY}}
program\cite{DJOUADI199856,Djouadi:2018xqq}, the lines 6, 26, 28, 58, 59, 63 and 64 are new, but the rest of the input file format is unchanged. We
want to emphasize again that the value {\texttt{GFCALC}} in the input
file is overwritten by the program and thus not an input value that
is provided by the user, but it is calculated by {\texttt{2HDECAY}}
internally. The sample 2HDM parameter point has been checked
  against all relevant theoretical and experimental constraints. In
  particular it features a SM-like Higgs boson with a mass of
  125.09~GeV which is given by the lightest CP-even neutral Higgs
  boson $h$. For details on the applied constraints, we refer to
  Refs.~\cite{Basler:2016obg,Muhlleitner:2017dkd}. 

\begin{lstlisting}
SLHAIN   = 1
SLHAOUT  = 1
COUPVAR  = 1
HIGGS    = 5
OMIT ELW = 1
OMIT ELW2= 0
SM4      = 0
FERMPHOB = 0
2HDM     = 1
MODEL    = 1
TGBET    = 5.07403e+01
MABEG    = 4.67967e+02
MAEND    = 4.67967e+02
NMA      = 1
********************* hMSSM (MODEL = 10) *********************************
MHL      = 125.D0
**************************************************************************
ALS(MZ)  = 1.18000e-01
MSBAR(2) = 9.50000e-02
MCBAR(3) = 0.98600e+00
MBBAR(MB)= 4.18000e+00
MT       = 1.73200e+02
MTAU     = 1.77682e+00
MMUON    = 1.056583715e-01
1/ALPHA  = 1.37036e+02
ALPHAMZ  = 7.754222173973729e-03
GF       = 1.1663787e-05
GFCALC   = 0.000000000
GAMW     = 2.08500e+00
GAMZ     = 2.49520e+00
MZ       = 9.11876e+01
MW       = 8.0385e+01
VTB      = 9.9910e-01
VTS      = 4.040e-02
VTD      = 8.67e-03
VCB      = 4.12e-02
VCS      = 9.7344e-01
VCD      = 2.252e-01
VUB      = 3.51e-03
VUS      = 2.2534e-01
VUD      = 9.7427e-01
********************* 4TH GENERATION *************************************
  SCENARIO FOR ELW. CORRECTIONS TO H -> GG (EVERYTHING IN GEV):
  GG_ELW = 1: MTP = 500    MBP = 450    MNUP = 375    MEP = 450
  GG_ELW = 2: MBP = MNUP = MEP = 600    MTP = MBP+50*(1+LOG(M_H/115)/5)

GG_ELW   = 1
MTP      = 500.D0
MBP      = 450.D0
MNUP     = 375.D0
MEP      = 450.D0
************************** 2 Higgs Doublet Model *************************
  TYPE: 1 (I), 2 (II), 3 (lepton-specific), 4 (flipped)
  PARAM: 1 (masses), 2 (lambda_i)

PARAM    = 1
TYPE     = 1
RENSCHEM = 7
REFSCHEM = 5
********************
TGBET2HDM= 4.23635D0
M_12^2   = 28505.5D0
INSCALE  = 125.09D0
OUTSCALE = MIN
******************** PARAM=1:
ALPHA_H  = -0.189345D0
MHL      = 125.09D0
MHH      = 381.767D0
MHA      = 350.665D0
MH+-     = 414.114D0
******************** PARAM=2:
LAMBDA1  = 6.368674377530086700D0
LAMBDA2  = 0.235570240072350970D0
LAMBDA3  = 1.780416490847621700D0
LAMBDA4  = -1.52623758540479430D0
LAMBDA5  = 0.074592764717552856D0
**************************************************************************
SUSYSCALE= 2.22449e+03
MU       = -1.86701e+03
M2       = -2.39071e+02
MGLUINO  = 7.32754e+02
MSL1     = 1.49552e+03
MER1     = 1.62210e+03
MQL1     = 9.30379e+01
MUR1     = 2.77029e+03
MDR1     = 1.76481e+03
MSL      = 1.97714e+03
MER      = 9.29678e+02
MSQ      = 2.68124e+03
MUR      = 1.85939e+03
MDR      = 2.28235e+03
AL       = -4.62984e+03
AU       = 5.31164e+03
AD       = 2.54430e+03
ON-SHELL = 0
ON-SH-WZ = 0
IPOLE    = 0
OFF-SUSY = 0
INDIDEC  = 0
NF-GG    = 5
IGOLD    = 0
MPLANCK  = 2.40000e+18
MGOLD    = 1.00000e-13
******************* VARIATION OF HIGGS COUPLINGS *************************
ELWK     = 1
CW       = 1.D0
CZ       = 1.D0
Ctau     = 1.D0
Cmu      = 1.D0
Ct       = 1.D0
Cb       = 1.D0
Cc       = 1.D0
Cs       = 1.D0
Cgaga    = 0.D0
Cgg      = 0.D0
CZga     = 0.D0
********************* 4TH GENERATION *************************************
Ctp      = 0.D0
Cbp      = 0.D0
Cnup     = 0.D0
Cep      = 0.D0
\end{lstlisting}

\section{Exemplary Output Files}
\label{sec:AppendixOutputFile}
In the following, we present exemplary output files
  {\texttt{2hdecay\_BR.out}} and {\texttt{2hdecay\_EW.out}} as they
  are generated from the sample input file
    {\texttt{2hdecay.in}} and included in the subfolder \\
{\texttt{\$2HDECAY/Results}} in the
  {\texttt{2HDECAY}} repository. The suffixes ``\_BR'' and ``\_EW''
  stand for the branching ratios and electroweak partial decay widths,
  respectively. The first integer in each line represents the line
number and is not part of the actual output file, but printed here for
convenience. The output file format is explained in detail in
Sec.\,\ref{sec:OutputFileFormat}. The exemplary output file was
generated for a specific choice of the renormalization scheme, {\it i.e.}~we
have set {\texttt{RENSCHEM = 7}} in line 58 of the input file,
{\it cf.}~App.\,\ref{sec:AppendixInputFile}. For {\texttt{RENSCHEM = 0}},
the output file becomes considerably longer, since the electroweak
corrections are calculated for all 17 implemented renormalization
schemes. We chose {\texttt{REFSCHEM = 5}} and 
 {\texttt{INSCALE = 125.09D0}}. This means that the input values for 
$\alpha$ and $\beta$ are understood to be given in the renormalization
scheme 5 and the scale at which $\alpha$, $\beta$ and the
$\overline{\mbox{MS}}$ parameter $m_{12}^2$ are defined is equal to 125.09~GeV.

\subsection{Exemplary Output File for the Branching Ratios}
The exemplary output file {\texttt{2hdecay\_BR.out}} contains the
branching ratios without and with the electroweak corrections. The
content of the file is presented in the following. 
\begin{lstlisting}
#
BLOCK DCINFO  # Decay Program information
     1   2HDECAY     # decay calculator
     2   1.1.0       # version number
#
BLOCK SMINPUTS  # Standard Model inputs
         2     1.19596488E-05   # G_F [GeV^-2]
         3     1.18000000E-01   # alpha_S(M_Z)^MSbar
         4     9.11876000E+01   # M_Z on-shell mass
         5     4.18000000E+00   # mb(mb)^MSbar
         6     1.73200000E+02   # mt pole mass
         7     1.77682000E+00   # mtau pole mass
         8     4.84141297E+00   # mb pole mass
         9     1.43141297E+00   # mc pole mass
        10     1.05658372E-01   # muon mass
        11     8.03850000E+01   # M_W on-shell mass
        12     2.08500000E+00   # W boson total width
        13     2.49520000E+00   # Z boson total width
#
BLOCK 2HDMINPUTS  # 2HDM inputs of reference scheme
         1                  1   # 2HDM parametrization
         2                  1   # 2HDM type
         3     4.23635000E+00   # tan(beta)
         4     2.85055000E+04   # M_12^2
         5    -1.89345000E-01   # alpha
         6     1.25090000E+02   # M_h
         7     3.81767000E+02   # M_H
         8     3.50665000E+02   # M_A
         9     4.14114000E+02   # M_CH
        10     6.53022117E+00   # LAMBDA1 calc. from masses/alpha
        11     2.41545678E-01   # LAMBDA2 calc. from masses/alpha
        12     1.82557826E+00   # LAMBDA3 calc. from masses/alpha
        13    -1.56495189E+00   # LAMBDA4 calc. from masses/alpha
        14     7.64848730E-02   # LAMBDA5 calc. from masses/alpha
        15                  7   # renormalization scheme EW corrs
        16                  5   # reference ren. scheme EW corrs
        17     125.09D0         # ren. scale of MSbar parameters
        18     MIN              # ren. scale at which decay is evaluated
#
BLOCK VCKMIN  # CKM mixing
         1     9.99100000E-01   # V_tb
         2     4.04000000E-02   # V_ts
         3     8.67000000E-03   # V_td
         4     4.12000000E-02   # V_cb
         5     9.73440000E-01   # V_cs
         6     2.25200000E-01   # V_cd
         7     3.51000000E-03   # V_ub
         8     2.25340000E-01   # V_us
         9     9.74270000E-01   # V_ud
#
#            PDG    Width QCD Only
DECAY QCD    25     4.22730978E-03   # h decays with QCD corrections only
                       7                # Renormalization Scheme Number
      -0.18809815E+00   # Corresponding mixing angle alpha
       0.43048897E+01   # Corresponding tan(beta)
       0.28505500E+05   # Corresponding m_12^2
#          BR         NDA         ID1       ID2
     5.93838905E-01    2           5        -5   # BR(h -> b       bb       )
     6.38675293E-02    2         -15        15   # BR(h -> tau+    tau-     )
     2.26112814E-04    2         -13        13   # BR(h -> mu+     mu-      )
     2.24253909E-04    2           3        -3   # BR(h -> s       sb       )
     2.90727883E-02    2           4        -4   # BR(h -> c       cb       )
     7.76755623E-02    2          21        21   # BR(h -> g       g        )
     2.19865876E-03    2          22        22   # BR(h -> gam     gam      )
     1.54187689E-03    2          22        23   # BR(h -> Z       gam      )
     2.05662673E-01    2          24       -24   # BR(h -> W+      W-       )
     2.56916399E-02    2          23        23   # BR(h -> Z       Z        )
#
#               PDG    Width QCD and EW
DECAY QCD&EW    25     4.10575180E-03   # h decays with QCD and EW corrections
                       7                # Renormalization Scheme Number
      -0.18809815E+00   # Corresponding mixing angle alpha
       0.43048897E+01   # Corresponding tan(beta)
       0.28505500E+05   # Corresponding m_12^2
#          BR         NDA         ID1       ID2
     5.85412930E-01    2           5        -5   # BR(h -> b       bb     )
     6.30251890E-02    2         -15        15   # BR(h -> tau+    tau-   )
     2.18258112E-04    2         -13        13   # BR(h -> mu+     mu-    )
     2.25768642E-04    2           3        -3   # BR(h -> s       sb     )
     2.90873264E-02    2           4        -4   # BR(h -> c       cb     )
     7.99752836E-02    2          21        21   # BR(h -> g       g      )
     2.26375391E-03    2          22        22   # BR(h -> gam     gam    )
     1.58752686E-03    2          22        23   # BR(h -> Z       gam    )
     2.11751677E-01    2          24       -24   # BR(h -> W+      W-     )
     2.64522859E-02    2          23        23   # BR(h -> Z       Z      )
#
#            PDG    Width QCD Only
DECAY QCD    35     1.82054627E-01   # H decays with QCD corrections only
                       7                # Renormalization Scheme Number
      -0.18809815E+00   # Corresponding mixing angle alpha
       0.43048897E+01   # Corresponding tan(beta)
       0.28286295E+05   # Corresponding m_12^2
#          BR         NDA         ID1       ID2
     1.23718494E-03    2           5        -5   # BR(H -> b       bb     )
     1.64167073E-04    2         -15        15   # BR(H -> tau+    tau-   )
     5.80581591E-07    2         -13        13   # BR(H -> mu+     mu-    )
     4.66487784E-07    2           3        -3   # BR(H -> s       sb     )
     6.05366336E-05    2           4        -4   # BR(H -> c       cb     )
     5.39857087E-01    2           6        -6   # BR(H -> t       tb     )
     4.22983124E-03    2          21        21   # BR(H -> g       g      )
     1.81583473E-05    2          22        22   # BR(H -> gam     gam    )
     1.06915278E-05    2          22        23   # BR(H -> Z       gam    )
     1.27161194E-01    2          24       -24   # BR(H -> W+      W-     )
     5.90077292E-02    2          23        23   # BR(H -> Z       Z      )
     1.19027442E-10    2          36        36   # BR(H -> A       A      )
     2.68081410E-01    2          25        25   # BR(H -> h       h      )
     1.70962359E-04    2          23        36   # BR(H -> Z       A      )
#
#               PDG    Width QCD and EW
DECAY QCD&EW    35     2.02874707E-01   # H decays with QCD and EW corrections
                       7                # Renormalization Scheme Number
      -0.18809815E+00   # Corresponding mixing angle alpha
       0.43048897E+01   # Corresponding tan(beta)
       0.28286295E+05   # Corresponding m_12^2
#          BR         NDA         ID1       ID2
     1.05832016E-03    2           5        -5   # BR(H -> b       bb     )
     1.33470074E-04    2         -15        15   # BR(H -> tau+    tau-   )
     4.61132083E-07    2         -13        13   # BR(H -> mu+     mu-    )
     3.90544832E-07    2           3        -3   # BR(H -> s       sb     )
     5.01923622E-05    2           4        -4   # BR(H -> c       cb     )
     4.75365783E-01    2           6        -6   # BR(H -> t       tb     )
     3.79574350E-03    2          21        21   # BR(H -> g       g      )
     1.62948412E-05    2          22        22   # BR(H -> gam     gam    )
     9.59430645E-06    2          22        23   # BR(H -> Z       gam    )
     1.19103021E-01    2          24       -24   # BR(H -> W+      W-     )
     6.65098085E-02    2          23        23   # BR(H -> Z       Z      )
     1.06812214E-10    2          36        36   # BR(H -> A       A      )
     3.33803504E-01    2          25        25   # BR(H -> h       h      )
     1.53417294E-04    2          23        36   # BR(H -> Z       A      )
#
#            PDG    Width QCD Only
DECAY QCD    36     4.12723113E-01   # A decays with QCD corrections only
                       7                # Renormalization Scheme Number
      -0.18809815E+00   # Corresponding mixing angle alpha
       0.43048897E+01   # Corresponding tan(beta)
       0.28302990E+05   # Corresponding m_12^2
#          BR         NDA         ID1       ID2
     7.27392605E-04    2           5        -5   # BR(A -> b       bb     )
     9.74019639E-05    2         -15        15   # BR(A -> tau+    tau-   )
     3.44437869E-07    2         -13        13   # BR(A -> mu+     mu-    )
     2.60982854E-07    2           3        -3   # BR(A -> s       sb     )
     3.67755809E-05    2           4        -4   # BR(A -> c       cb     )
     9.62219843E-01    2           6        -6   # BR(A -> t       tb     )
     9.63661208E-03    2          21        21   # BR(A -> g       g      )
     3.97340998E-05    2          22        22   # BR(A -> gam     gam    )
     6.75641349E-06    2          22        23   # BR(A -> Z       gam    )
     2.72348787E-02    2          23        25   # BR(A -> Z       h      )
#
#               PDG    Width QCD and EW
DECAY QCD&EW    36     4.10006463E-01   # A decays with QCD and EW corrections
                       7                # Renormalization Scheme Number
      -0.18809815E+00   # Corresponding mixing angle alpha
       0.43048897E+01   # Corresponding tan(beta)
       0.28302990E+05   # Corresponding m_12^2
#          BR         NDA         ID1       ID2
     6.84138168E-04    2           5        -5   # BR(A -> b       bb     )
     8.64209164E-05    2         -15        15   # BR(A -> tau+    tau-   )
     2.98361191E-07    2         -13        13   # BR(A -> mu+     mu-    )
     2.38527876E-07    2           3        -3   # BR(A -> s       sb     )
     3.32826226E-05    2           4        -4   # BR(A -> c       cb     )
     9.70984653E-01    2           6        -6   # BR(A -> t       tb     )
     9.70046304E-03    2          21        21   # BR(A -> g       g      )
     3.99973728E-05    2          22        22   # BR(A -> gam     gam    )
     6.80118062E-06    2          22        23   # BR(A -> Z       gam    )
     1.84637072E-02    2          23        25   # BR(A -> Z       h      )
#
#            PDG    Width QCD Only
DECAY QCD    37     9.52122446E-01   # H+ decays with QCD corrections only
                       7                # Renormalization Scheme Number
      -0.18809815E+00   # Corresponding mixing angle alpha
       0.43048897E+01   # Corresponding tan(beta)
       0.28270317E+05   # Corresponding m_12^2
#          BR         NDA         ID1       ID2
     6.38478540E-07    2           4        -5   # BR(H+ -> c       bb    )
     4.98617485E-05    2         -15        16   # BR(H+ -> tau+    nu_t  au )
     1.76321144E-07    2         -13        14   # BR(H+ -> mu+     nu_m  u  )
     4.41813442E-09    2           2        -5   # BR(H+ -> u       bb    )
     6.73510166E-09    2           2        -3   # BR(H+ -> u       sb    )
     8.84349820E-07    2           4        -1   # BR(H+ -> c       db    )
     1.66493576E-05    2           4        -3   # BR(H+ -> c       sb    )
     9.70305009E-01    2           6        -5   # BR(H+ -> t       bb    )
     1.58487563E-03    2           6        -3   # BR(H+ -> t       sb    )
     7.29902662E-05    2           6        -1   # BR(H+ -> t       db    )
     2.60001008E-02    2          24        25   # BR(H+ -> W+      h     )
     5.03002930E-05    2          24        35   # BR(H+ -> W+      H     )
     1.91850311E-03    2          24        36   # BR(H+ -> W+      A     )
#
#               PDG    Width QCD and EW
DECAY QCD&EW    37     9.15926373E-01   # H+ decays with QCD and EW corrections
                       7                # Renormalization Scheme Number
      -0.18809815E+00   # Corresponding mixing angle alpha
       0.43048897E+01   # Corresponding tan(beta)
       0.28270317E+05   # Corresponding m_12^2
#          BR         NDA         ID1       ID2
     5.95091084E-07    2           4        -5   # BR(H+ -> c       bb     )
     4.71994255E-05    2         -15        16   # BR(H+ -> tau+    nu_tau )
     1.63076322E-07    2         -13        14   # BR(H+ -> mu+     nu_mu  )
     3.64995618E-09    2           2        -5   # BR(H+ -> u       bb     )
     6.55853493E-09    2           2        -3   # BR(H+ -> u       sb     )
     8.55031171E-07    2           4        -1   # BR(H+ -> c       db     )
     1.60986318E-05    2           4        -3   # BR(H+ -> c       sb     )
     9.66117646E-01    2           6        -5   # BR(H+ -> t       bb     )
     1.58023685E-03    2           6        -3   # BR(H+ -> t       sb     )
     7.31159911E-05    2           6        -1   # BR(H+ -> t       db     )
     3.01174725E-02    2          24        25   # BR(H+ -> W+      h      )
     5.22880871E-05    2          24        35   # BR(H+ -> W+      H      )
     1.99431955E-03    2          24        36   # BR(H+ -> W+      A      )
#
#         PDG            Width
DECAY         6     1.36958730E+00   # top decays
#          BR         NDA         ID1       ID2
     1.00000000E+00    2           5        24   # BR(t -> b       W+    )
\end{lstlisting}

In the following, we make some comments on the output
  files that partly pick up hints and caveats made in the main text of
the manual. 
As can be inferred from the output, we give for the decays of each
Higgs boson the values of $\alpha$, $\beta$ and $m_{12}^2$. These
values change from the input values and for each Higgs boson as we
have to perform the parameter conversion from the input reference
scheme 5 to the renormalization scheme 7 and because we use for the
loop corrected widths the renormalization scale given by the mass of
the decaying Higgs boson, since we set 
{\texttt{OUTSCALE = MIN}} while the input values for these parameters 
are understood to be given at the mass of the SM-like Higgs boson.
 Furthermore notice that indeed the branching ratios of
the lightest CP-even Higgs boson $h$ are SM-like. All branching ratios
presented in the blocks {\texttt{DECAY QCD}} can be compared to the
ones generated by the program code {\texttt{HDECAY}} version 6.52. The
user will notice that the partial widths related to the branching
ratios generated by {\texttt{2HDECAY}} and {\texttt{HDECAY}},
respectively, differ due to 
the rescaling factor $G_F^\text{calc}/G_F =
1.026327$, which is applied in {\texttt{2HDECAY}}
for the consistent 
combination of the EW-corrected decay widths with the decay widths generated by
{\texttt{HDECAY}}. Be aware that the rescaling factor
  appears in the loop induced decay into $Z\gamma$ and in the off-shell decays non-linearly. This is why also the branching
ratios given here differ from the ones generated by
{\texttt{HDECAY}}6.52.
The comparison furthermore shows an additional difference
between the decay widths for the heavy CP-even Higgs boson $H$ into 
massive vector bosons, $\Gamma (H \to VV)$ ($V=W,Z$), of around
2-3\%. The reason is that {\texttt{HDECAY}}
throughout computes these 
decay widths using the double off-shell formula while
{\texttt{2HDECAY}} uses the on-shell formula for Higgs boson masses
above the threshold. Let us also note some phenomenological features of the chosen
parameter point. The $H$ boson with a mass of 382~GeV is heavy enough
to decay on-shell into $WW$ and $ZZ$,  
and also into the 2-Higgs boson final state $hh$. It decays off-shell into
$AA$ and the gauge plus Higgs boson final state $ZA$ with branching
ratios of ${\cal O}(10^{-10})$ and ${\cal O}(10^{-4})$,
respectively. The pseudoscalar with a mass of 351~GeV decays on-shell
into the gauge plus Higgs boson final state $Zh$ with a branching ratio
at the per cent level. The charged Higgs boson has a mass of 414~GeV
allowing it to decay on-shell in the gauge plus Higgs boson final
state $W^+ h$ with a branching ratio at the per cent level. It decays
off-shell into the final states $W^+ H$ and $W^+ A$ with branching
ratios of ${\cal O}(10^{-5})$ and ${\cal O}(10^{-3})$,
respectively.

\subsection{Exemplary Output File for the Electroweak Partial Decay Widths} 
The exemplary output file {\texttt{2hdecay\_EW.out}} contains the LO
and electroweak NLO partial decay widths. The content of the file is
presented in the following. 
\begin{lstlisting}
#
BLOCK DCINFO  # Decay Program information
     1   2HDECAY     # decay calculator
     2   1.1.0       # version number
#
BLOCK SMINPUTS  # Standard Model inputs
         2     1.19596488E-05   # G_F [GeV^-2]
         3     1.18000000E-01   # alpha_S(M_Z)^MSbar
         4     9.11876000E+01   # M_Z on-shell mass
         5     4.18000000E+00   # mb(mb)^MSbar
         6     1.73200000E+02   # mt pole mass
         7     1.77682000E+00   # mtau pole mass
         8     4.84141297E+00   # mb pole mass
         9     1.43141297E+00   # mc pole mass
        10     1.05658372E-01   # muon mass
        11     8.03850000E+01   # M_W on-shell mass
        12     2.08500000E+00   # W boson total width
        13     2.49520000E+00   # Z boson total width
#
BLOCK 2HDMINPUTS  # 2HDM inputs of reference scheme
         1                  1   # 2HDM parametrization
         2                  1   # 2HDM type
         3     4.23635000E+00   # tan(beta)
         4     2.85055000E+04   # M_12^2
         5    -1.89345000E-01   # alpha
         6     1.25090000E+02   # M_h
         7     3.81767000E+02   # M_H
         8     3.50665000E+02   # M_A
         9     4.14114000E+02   # M_CH
        10     6.53022117E+00   # LAMBDA1 calc. from masses/alpha
        11     2.41545678E-01   # LAMBDA2 calc. from masses/alpha
        12     1.82557826E+00   # LAMBDA3 calc. from masses/alpha
        13    -1.56495189E+00   # LAMBDA4 calc. from masses/alpha
        14     7.64848730E-02   # LAMBDA5 calc. from masses/alpha
        15                  7   # renormalization scheme EW corrs
        16                  5   # reference ren. scheme EW corrs
        17     125.09D0         # input scale
        18     MIN              # ren. scale at which decay is evaluated
#
BLOCK VCKMIN  # CKM mixing
         1     9.99100000E-01   # V_tb
         2     4.04000000E-02   # V_ts
         3     8.67000000E-03   # V_td
         4     4.12000000E-02   # V_cb
         5     9.73440000E-01   # V_cs
         6     2.25200000E-01   # V_cd
         7     3.51000000E-03   # V_ub
         8     2.25340000E-01   # V_us
         9     9.74270000E-01   # V_ud
#
#                 PDG
LO DECAY WIDTH    25   # h non-zero LO EW decay widths of on-shell and non-loop induced decays
                    7   # Renormalization Scheme Number
      -0.18809815E+00   # Corresponding mixing angle alpha
       0.43048897E+01   # Corresponding tan(beta)
       0.28505500E+05   # Corresponding m_12^2
#         WIDTH       NDA         ID1       ID2
     5.96669359E-03    2           5        -5   # GAM(h -> b       bb   )
     2.69987831E-04    2         -15        15   # GAM(h -> tau+    tau- )
     9.55848909E-07    2         -13        13   # GAM(h -> mu+     mu-  )
     2.31819649E-06    2           3        -3   # GAM(h -> s       sb   )
     5.25887869E-04    2           4        -4   # GAM(h -> c       cb   )
#
#                 PDG
NLO DECAY WIDTH    25   # h non-zero NLO EW decay widths of on-shell and non-loop induced decays
                    7   # Renormalization Scheme Number
      -0.18809815E+00   # Corresponding mixing angle alpha
       0.43048897E+01   # Corresponding tan(beta)
       0.28505500E+05   # Corresponding m_12^2
#         WIDTH       NDA         ID1       ID2
     5.71289204E-03    2           5        -5   # GAM(h -> b       bb    )
     2.58765783E-04    2         -15        15   # GAM(h -> tau+    tau-  )
     8.96113636E-07    2         -13        13   # GAM(h -> mu+     mu-   )
     2.26674392E-06    2           3        -3   # GAM(h -> s       sb    )
     5.11021167E-04    2           4        -4   # GAM(h -> c       cb    )
#
#                 PDG
LO DECAY WIDTH    35   # H non-zero LO EW decay widths of on-shell and non-loop induced decays
                    7   # Renormalization Scheme Number
      -0.18809815E+00   # Corresponding mixing angle alpha
       0.43048897E+01   # Corresponding tan(beta)
       0.28286295E+05   # Corresponding m_12^2
#         WIDTH       NDA         ID1       ID2
     6.65125616E-04    2           5        -5   # GAM(H -> b       bb   )
     2.98873753E-05    2         -15        15   # GAM(H -> tau+    tau- )
     1.05697565E-07    2         -13        13   # GAM(H -> mu+     mu-  )
     2.56345478E-07    2           3        -3   # GAM(H -> s       sb   )
     5.81931534E-05    2           4        -4   # GAM(H -> c       cb   )
     6.32882453E-02    2           6        -6   # GAM(H -> t       tb   )
     2.31502837E-02    2          24       -24   # GAM(H -> W+      W-   )
     1.07426301E-02    2          23        23   # GAM(H -> Z       Z    )
     4.88054612E-02    2          25        25   # GAM(H -> h       h    )
#
#                 PDG
NLO DECAY WIDTH    35   # H non-zero NLO EW decay widths of on-shell and non-loop induced decays
                    7   # Renormalization Scheme Number
      -0.18809815E+00   # Corresponding mixing angle alpha
       0.43048897E+01   # Corresponding tan(beta)
       0.28286295E+05   # Corresponding m_12^2
#         WIDTH       NDA         ID1       ID2
     6.34033642E-04    2           5        -5   # GAM(H -> b       bb    )
     2.70777021E-05    2         -15        15   # GAM(H -> tau+    tau-  )
     9.35520362E-08    2         -13        13   # GAM(H -> mu+     mu-   )
     2.39156652E-07    2           3        -3   # GAM(H -> s       sb    )
     5.37672029E-05    2           4        -4   # GAM(H -> c       cb    )
     6.21009650E-02    2           6        -6   # GAM(H -> t       tb    )
     2.41629904E-02    2          24       -24   # GAM(H -> W+      W-    )
     1.34931579E-02    2          23        23   # GAM(H -> Z       Z     )
     6.77202880E-02    2          25        25   # GAM(H -> h       h     )
#
#                 PDG
LO DECAY WIDTH    36   # A non-zero LO decay widths of on-shell and non-loop induced decays
                    7   # Renormalization Scheme Number
      -0.18809815E+00   # Corresponding mixing angle alpha
       0.43048897E+01   # Corresponding tan(beta)
       0.28302990E+05   # Corresponding m_12^2
#         WIDTH       NDA         ID1       ID2
     8.95079946E-04    2           5        -5   # GAM(A -> b       bb   )
     4.02000418E-05    2         -15        15   # GAM(A -> tau+    tau- )
     1.42157470E-07    2         -13        13   # GAM(A -> mu+     mu-  )
     3.44770691E-07    2           3        -3   # GAM(A -> s       sb   )
     7.82705947E-05    2           4        -4   # GAM(A -> c       cb   )
     1.78189744E-01    2           6        -6   # GAM(A -> t       tb   )
     1.12404639E-02    2          23        25   # GAM(A -> Z       h    )
#
#                 PDG
NLO DECAY WIDTH    36   # A non-zero NLO EW decay widths of on-shell and non-loop induced decays
                    7   # Renormalization Scheme Number
      -0.18809815E+00   # Corresponding mixing angle alpha
       0.43048897E+01   # Corresponding tan(beta)
       0.28302990E+05   # Corresponding m_12^2
#         WIDTH       NDA         ID1       ID2
     8.36312673E-04    2           5        -5   # GAM(A -> b       bb    )
     3.54331343E-05    2         -15        15   # GAM(A -> tau+    tau-  )
     1.22330017E-07    2         -13        13   # GAM(A -> mu+     mu-   )
     3.13032490E-07    2           3        -3   # GAM(A -> s       sb    )
     7.03701609E-05    2           4        -4   # GAM(A -> c       cb    )
     1.78629290E-01    2           6        -6   # GAM(A -> t       tb    )
     7.57023930E-03    2          23        25   # GAM(A -> Z       h     )
#
#                 PDG
LO DECAY WIDTH    37   # H+ non-zero LO decay widths of on-shell and non-loop induced decays
                    7   # Renormalization Scheme Number
      -0.18809815E+00   # Corresponding mixing angle alpha
       0.43048897E+01   # Corresponding tan(beta)
       0.28270317E+05   # Corresponding m_12^2
#         WIDTH       NDA         ID1       ID2
     1.95134769E-06    2           4        -5   # GAM(H+ -> c       bb   )
     4.74744900E-05    2         -15        16   # GAM(H+ -> tau+    nu_tau)
     1.67879319E-07    2         -13        14   # GAM(H+ -> mu+     nu_mu)
     1.30241820E-08    2           2        -5   # GAM(H+ -> u       bb   )
     2.06744724E-08    2           2        -3   # GAM(H+ -> u       sb   )
     4.68777754E-06    2           4        -1   # GAM(H+ -> c       db   )
     8.79746167E-05    2           4        -3   # GAM(H+ -> c       sb   )
     9.20580564E-01    2           6        -5   # GAM(H+ -> t       bb   )
     1.50367679E-03    2           6        -3   # GAM(H+ -> t       sb   )
     6.92515961E-05    2           6        -1   # GAM(H+ -> t       db   )
     2.47552795E-02    2          24        25   # GAM(H+ -> W+      h    )
#
#                 PDG
NLO DECAY WIDTH    37   # H+ non-zero NLO EW decay widths of on-shell and non-loop induced decays
                    7   # Renormalization Scheme Number
      -0.18809815E+00   # Corresponding mixing angle alpha
       0.43048897E+01   # Corresponding tan(beta)
       0.28270317E+05   # Corresponding m_12^2
#         WIDTH       NDA         ID1       ID2
     1.74960317E-06    2           4        -5   # GAM(H+ -> c       bb     )
     4.32311986E-05    2         -15        16   # GAM(H+ -> tau+    nu_tau )
     1.49365904E-07    2         -13        14   # GAM(H+ -> mu+     nu_mu  )
     1.03506338E-08    2           2        -5   # GAM(H+ -> u       bb     )
     1.93671125E-08    2           2        -3   # GAM(H+ -> u       sb     )
     4.36006145E-06    2           4        -1   # GAM(H+ -> c       db     )
     8.18307685E-05    2           4        -3   # GAM(H+ -> c       sb     )
     8.81761847E-01    2           6        -5   # GAM(H+ -> t       bb     )
     1.44227891E-03    2           6        -3   # GAM(H+ -> t       sb     )
     6.67336642E-05    2           6        -1   # GAM(H+ -> t       db     )
     2.75853874E-02    2          24        25   # GAM(H+ -> W+      h      )
\end{lstlisting} 

The inspection of the output file shows that the EW
  corrections reduce the $h$ decay widths, and the 
  relative NLO EW corrections,
  $\Delta^{\text{EW}} = (\Gamma^{\text{EW}} -
  \Gamma^{\text{LO}})/\Gamma^{\text{LO}}$, range between -6.3 and
  -2.2\% for the decays $\Gamma(h \to \mu^+\mu^-)$ and $\Gamma(h\to
  s\bar{s})$, respectively. Regarding $H$, the corrections can both 
  enhance and reduce the decay widths. The relative corrections range
  between -11.5 and 27.7\% for the decays $\Gamma(H\to \mu^+ \mu^-)$
  and $\Gamma(H \to hh)$, respectively. The relative corrections to
  the $A$ decay widths vary between -31.2 and 0.3\% for the decays $\Gamma(A\to Zh)$
  and $\Gamma(A \to t\bar{t})$, respectively. And those for the $H^\pm$
  decays between -20.6 and 11.1\% for the decays $\Gamma(H^+ \to u\bar{b})$
  and $\Gamma(H^+ \to W^+h)$, respectively. The EW corrections (for the
  renormalization scheme number 7) of the
  chosen parameter point can hence be sizeable. Finally, note also
  that LO and NLO EW-corrected decay widths are given out for
  on-shell and non-loop induced decays only.

\end{appendix}



\end{document}